\documentclass[prd,aps,showpacs,twocolumn,superscriptaddress,floatfix,nofootinbib]{revtex4-1}
\usepackage[utf8]{inputenc}
\usepackage{bm}
\usepackage{bbold}
\usepackage{mathtools}
\usepackage[toc,page]{appendix}
\usepackage{dsfont}
\usepackage{combelow}
\usepackage[colorlinks=true,linkcolor=blue,citecolor=blue, urlcolor=blue]{hyperref} 
\usepackage[a4paper,top=2.4cm,bottom=2.4cm,right=1.0cm,bindingoffset=-1.2cm]{geometry}
\usepackage{tikz}
\usepackage{graphicx}
\usepackage{subfigure}
\usepackage{scalerel,amssymb}
\DeclareMathAlphabet{\pazocal}{OMS}{zplm}{m}{n}
\usepackage{ulem}

\renewcommand{\d}{\mathrm{d}}
\newcommand{\epn}{\frac{\d E_\perp^0}{\d \eta}}
\newcommand{\epni}{{\left.\d E_\perp^0\right/\d \eta}}

\newcommand{\xT}{{\mathbf{x}_\perp}}

\newcommand{\vT}{\mathbf{v_\perp}}

\newcommand{\trento}{\textsc{trento}}

\begin{document}
\title{Collective dynamics in heavy and light-ion collisions -- I) Kinetic Theory vs. Hydrodynamics}
\author{Victor E. Ambru\cb{s}}
\affiliation{Department of Physics, West University of Timi\cb{s}oara, \\
Bd.~Vasile P\^arvan 4, Timi\cb{s}oara 300223, Romania}
\author{S.~Schlichting}
\affiliation{Fakultät für Physik, Universität Bielefeld, D-33615 Bielefeld, Germany}
\author{C.~Werthmann}
\email{clemens.werthmann@ugent.be}
\affiliation{Fakultät für Physik, Universität Bielefeld, D-33615 Bielefeld, Germany}
\affiliation{Incubator of Scientific Excellence-Centre for Simulations of Superdense Fluids, University of Wrocław, pl. Maxa Borna 9, 50-204 Wrocław, Poland}
\affiliation{Department of Physics and Astronomy, Ghent University, 9000 Ghent, Belgium}
\date{\today}

\begin{abstract} 
High-energy nuclear collisions exhibit collective flow, which emerges as a dynamical response of the Quark-Gluon Plasma (QGP) to the initial state geometry of the collision. 
Collective flow in heavy-ion collisions is usually
described within multi-stage evolution models, which employ a viscous relativistic  hydrodynamic description of the space-time evolution of the QGP. By comparing event-by-event simulations in kinetic theory and viscous hydrodynamics in OO, AuAu and PbPb collisions at RHIC and LHC energies, we quantify to what extent a macroscopic hydrodynamic description can accurately describe the development of collective flow and to what extent collective flow in small systems, such as OO, is sensitive to the non-equilibrium evolution of the QGP beyond hydrodynamics.
\end{abstract}

\maketitle

\tableofcontents

\section{Introduction}\label{sec:intro}
Since a theoretical description of the space-time dynamics of high-energy heavy-ion collisions from the underlying microscopic theory of QCD is not possible, the standard model of heavy-ion collisions~\cite{Heinz:2013wva} is based on effective macroscopic descriptions of QCD, which  -- exploiting a separation of time scales in the reaction dynamics -- are combined into multi-stage evolution models~(see e.g. Refs.~\cite{Shen:2014vra,McDonald:2016vlt,Putschke:2019yrg,Nijs:2020roc}).
Generally, such models describe the energy deposition during the initial collision, the pre-equilibrium dynamics of the QGP, the viscous hydrodynamic evolution of the QGP, the transition from the QGP to the hadronic phase as well as hadronic re-scatterings and resonance decays in the final state of the collision. By now it is well established that such multi-stage evolution models can provide an excellent theoretical description of soft physics observables, including particle spectra and collective flow, in collisions of large nuclei and significant efforts have been invested in recent years to determine model parameters and physical properties of the QGP from statistical comparisons to RHIC and LHC measurements~\cite{Putschke:2019yrg,JETSCAPE:2020mzn,Nijs:2020roc}. 

Besides this approach, various transport models such as BAMPS~\cite{Xu:2004mz,Xu:2007jv}, AMPT~\cite{Lin:2004en} or PHSD~\cite{Bratkovskaya:2011wp,Linnyk:2015rco} based on effective kinetic descriptions, have also attempted to provide an alternative description of collective flow, and generally also fare well in comparisons to experimental measurements in heavy-ion collisions. While less firmly rooted in QCD, one key advantage of these models is that they are based on kinetic theory, which faithfully interpolates between a non-interacting system at very small opacity and ideal hydrodynamics in the large opacity limit~\cite{Ambrus:2021fej}. 

Despite the long-standing phenomenological success of such hydrodynamic models of the QGP, a theoretical justification for the applicability of viscous hydrodynamics has only emerged in recent years, where dedicated studies of the early time pre-equilibrium dynamics~\cite{Heller:2016rtz,Spalinski:2017mel,Romatschke:2017acs,Du:2020dvp,Du:2020zqg} 
have indeed demonstrated the applicability of viscous hydrodynamics on time scales $\sim 1$ fm/c in heavy-ion collisions and provided new theoretical tools to describe the pre-equilibrium evolution of the QGP~\cite{Kurkela:2018wud,Kurkela:2018gitrep,Liyanage:2022nua}.
Nevertheless, it is strongly debated whether a hydrodynamic description provides a quantitatively reliable description of collisions of lighter ions, e.g. pPb, pAu, OO, NeNe, where the lifetime of the QGP is much shorter and, moreover, it is subject to a rapid transverse expansion. Despite the fact that mutli-stage evolution models such as IPGlasma+MUSIC+UrQMD~\cite{Mantysaari:2017cni} or 3DGlauber+MUSIC+UrQMD~\cite{Zhao:2022ugy} or core-corona simulation frameworks like in EPOS~\cite{Werner:2010ss,Werner:2013ipa}
can indeed describe the collective flow experimentally observed in pp, pPb and p/d/He$^3$+Au collisions at RHIC and LHC, there are significant uncertainties primarily due to the poorly constrained initial state geometry in such systems~\cite{Schenke:2014zha,Schenke:2021mxx,Demirci:2021kya}, but also related to the influence of the early-time pre-equilibrium dynamics~\cite{Schenke:2015aqa,McLerran:2015sva,Schenke:2022mjv} and other non-hydrodynamic excitations~\cite{Kurkela:2019kip}, and further efforts are required to quantify the (in)applicability of hydrodynamics in small systems.

Several recent works~\cite{Kurkela:2019kip,Kurkela:2020wwb,Ambrus:2022koq,Ambrus:2022qya} have started to address this question by systematically comparing the microscopic evolution in kinetic theory to the effective macroscopic description in hydrodynamics. Clearly, the most important conclusion so far is that 
the collective flow response to the initial state increases monotonically with increasing opacity, however it can only be quantitatively described using viscous hydrodynamics, when the opacity is sufficiently large for the QGP to equilibrate to a sufficient degree, before the onset of the transverse expansion. Depending on the desired level of accuracy and the actual values of QCD transport coefficients, these studies~\cite{Ambrus:2022qya,Ambrus:2022koq} suggest that OO collisions at RHIC and LHC may fall right into the regime where hydrodynamics becomes increasingly inaccurate, and it is therefore particularly interesting to explore the dynamics of these systems.

So far, the aforementioned systematic studies~\cite{Ambrus:2022qya,Ambrus:2022koq} were based on a simple average initial state geometry. In this work, we perform for the first time a systematic comparison of the collective flow response in kinetic theory and hydrodynamics based on event-by-event simulations of OO, AuAu and PbPb collisions. While in this paper we focus on a theoretical comparison of kinetic theory and hydrodynamics in event-by-event simulations, our companion paper~\cite{Ambrus:2024eqa} exploits these results to develop a phenomenological approach to quantify the degree of hydrodynamic behavior in heavy-ion collisions.

We start by introducing our theoretical setup as well as the relevant observables in Sec.~\ref{sec:setup}. We characterize the most important properties of the initial state in Sec.~\ref{sec:geometry}, and subsequently present our simulation results for elliptic flow in kinetic theory and hydrodynamics in Sec.~\ref{sec:elliptic}. Since these results are based on conformal kinetic theory/hydrodynamics, we further assess the impact of a non-conformal equation of state in Sec.~\ref{sec:non-conformal}. We conclude with Sec.~\ref{sec:conclusion}. The appendices include a discussion of elliptic flow results at fixed final transverse energy in App.~\ref{app:scaling_details}, an assessment of conceptual difficulties with the non-conformal setup in App.~\ref{app:switching_problems} and details on one of the numerical simulation codes in App.~\ref{app:LOebe}. The raw data for all plots presented in this work are publicly available~\cite{werthmann_2025_14849750}.

\section{Setup}\label{sec:setup}

\subsection{Kinetic theory}\label{sec:kinetic_theory}
Within the framework of kinetic theory, the dynamics of the system is governed by the Boltzmann equation, which describes the time evolution of the phase space distribution $f$ of (quasi-)particles in the system. We follow our previous works~\cite{Ambrus:2021fej,Ambrus:2022koq,Ambrus:2022qya} and effectively consider only a single average distribution of massless bosons. We employ the conformal relaxation time approximation (RTA) for the collision kernel, such that the microscopic evolution of the system is determined by the following equation:
\begin{align}
    p^{\mu}\partial_{\mu} f(x,p)&= -\frac{u^{\mu}(x)p_{\mu}}{\tau_R(x)} \left[f(x,p)-f_{\rm eq}(x,p)\right]\;,\label{eq:Boltzmann}
\end{align}
where the relaxation time $\tau_R$ is determined according to
\begin{align}
    \tau_R(x)&=5\frac{\eta}{s}T^{-1}(x)\label{eq:tauR}
\end{align}
from the specific shear viscosity $\eta/s$, which is constant for a conformal system. We note that the effective temperature $T(x)$ is determined by matching the local energy density $e(x)$ obtained from the energy-momentum tensor 
\begin{align}
    T^{\mu\nu}(x)=\nu_{\rm eff}\tau \int\frac{d^3p}{(2\pi)^3p^\tau} p^\mu p^\nu f(x,p)\;.
\end{align}
via Landau matching, to the conformal equation of state $e(x)=u_\mu u_\nu T^{\mu\nu} = aT^4(x)$, where $a=\nu_{\rm eff}\frac{\pi^2}{30}$ with the effective number of degrees of freedom $\nu_{\rm eff}=42.25$ chosen in accordance with lattice QCD results~\cite{HotQCD:2014kol,Borsanyi:2016ksw} at high temperatures.

We initialize the dynamics at very early times $\tau_0=10^{-6}R$, where $R$ represents the transverse size of the system [see Eq.~\eqref{eq:R}]. We assume that the longitudinal pressure vanishes, in accordance with the early time behavior of the kinetic theory attractor for Bjorken flow~\cite{Mueller:1999pi}, and numerically solve the evolution equation (\ref{eq:Boltzmann}) following the setup of our previous work~\cite{Ambrus:2021fej}, which exploits a closed set of evolution equations for energy weighted moments of the phase-space distribution. We will assume an effectively $2+1$D boost-invariant dynamics without any transverse momentum anisotropies in the initial state, such that anisotropic flow develops solely in response to the initial state geometry.

Due to the particular simplicity of this setup, the time evolution of the system then only depends on the initial state energy density profile in the transverse plane
as well as a single dimensionless opacity parameter $\hat{\gamma}$~\cite{Kurkela:2019kip,Ambrus:2021fej} 
\begin{align}
    \hat{\gamma}=\frac{1}{5 \eta/ s} 
 \left(\frac{R}{\pi a} \frac{\d E_\perp^0}{\d\eta}\right)^{1/4}\label{eq:ghat_def}
\end{align}
which collects the dependencies on the specific shear viscosity ${\eta}/{s}$, the initial transverse size $R$ and the initial transverse energy per unit rapidity $\epni$.

Clearly, this simple conformal kinetic description should not be considered as a realistic description of QCD, where most importantly the presence of a confinement scale makes QCD a non-conformal theory, with a non-trivial equation of state. Similarly, the simple relaxation to local equilibrium also does not reflect the complex interaction dynamics of full QCD. However, we note that in the context of thermalization studies, it has been found that the dynamics of the energy momentum tensor which is severely restricted by conservation laws, exhibits a rather similar behavior for different underlying microscopic theories, as discussed e.g. in \cite{Schlichting:2024uok}. Since for the purposes of this work only the evolution of the energy-momentum tensor is relevant, we expect that comparing full QCD dynamics to hydrodynamics matched to QCD will yield qualtitively similar results as comparing conformal RTA to conformal hydrodynamics matched to RTA.

\subsubsection{Linear order}\label{sec:linear_order}
Since at very small opacity $\hat{\gamma} \ll 1$ a numerical solution of the Boltzmann equation becomes impractical, it is useful to note that the Boltzmann equation permits an expansion of the solution $f$ in the interaction strength~\cite{Heiselberg:1998es,Borghini:2010hy}, i.e. a successive approximation valid for dilute systems that is of the following form 
\begin{align}
    f(x,p)=f^{(0)}(x,p)+f^{(1)}(x,p)+O(\hat{\gamma}^2)\;,
\end{align}
where in our case, $f^{(1)}$ is of linear order in $\hat{\gamma}$, i.e. the expansion parameter is the opacity, though the main idea is more general. Here, the zeroth order term is the solution of \eqref{eq:Boltzmann} in the absence of interactions, $p^\mu\partial_{\mu}f^{(0)}=0$, which is also called the free-streaming solution. The expansion is successively computed by evaluating the right hand side of Eq. \eqref{eq:Boltzmann} for the previous term in the expansion and using the result as a source term for the evolution equation of the higher order term. Since this scheme becomes more involved at higher orders, it is typically employed only to linear order. More precisely, the linear order term in our case is the solution of the following equation.
\begin{multline}
    p^{\mu}\partial_{\mu} f^{(1)}(x,p)\\
    = -\frac{u^{\mu}(x)p_{\mu}}{\tau_R[f^{(0)}](x)} \left[f^{(0)}(x,p)-f_{\rm eq}[f^{(0)}](x,p)\right]
\end{multline}
We have employed this linearized description already in previous works~\cite{Ambrus:2021fej,Ambrus:2022koq,Ambrus:2022qya} to examine the low opacity limit. The setup in this paper is the same, though some modification of the code was necessary to speed it up in order to make event-by-event simulations feasible. These are laid out in Appendix~\ref{app:LOebe}.

\subsection{Hydrodynamics}\label{sec:hydrodynamics}

Hydrodynamics is an effective macroscopic description of the system in terms of coarse-grained collective quantities: the energy density $e$ and the flow velocity $u^\mu$. Evolution equations for these quantities are obtained from energy-momentum conservation, along with so-called constitutive relations, derived in an expansion around local thermal equilibrium~\cite{Denicol:2012cn}. We follow the community standard and employ second order Mueller-Israel-Stewart type hydrodynamic evolution equations, where the shear stress tensor $\pi^{\mu\nu}$ is promoted to a dynamical quantity, such that for a conformal system the dynamics is governed by the following evolution equations.

\begin{align}
    \dot{e} &= -(e + P) \theta + \pi^{\mu\nu} \sigma_{\mu\nu} ,
    \label{eq:hydro_e}\\
 (e + P) \dot{u}^\mu &= \nabla^\mu P
 - \Delta^\mu{}_\lambda \partial_\nu \pi^{\lambda\nu} ,
 \label{eq:hydro_umu}\\
  \tau_\pi \dot{\pi}^{\langle \mu\nu \rangle} + \pi^{\mu\nu} &= 2\eta \sigma^{\mu\nu} + 
 2\tau_\pi \pi^{\langle\mu}_\lambda \omega^{\nu\rangle \lambda}- \delta_{\pi\pi} \pi^{\mu\nu} \theta
 \label{eq:hydro_pi}\\
 &\quad  - 
 \tau_{\pi\pi} \pi^{\lambda\langle \mu}\sigma^{\nu\rangle}_\lambda + \phi_7 \pi_\alpha^{\langle\mu}\pi^{\nu\rangle\alpha}\nonumber
\end{align}
In order to match the microscopic kinetic description, the hydrostatic pressure $P$ is fixed by the conformal equation of state $e=3P$. We note that ideal hydrodynamics can be recovered from the above evolution equations when the shear-stress tensor $\pi^{\mu\nu}$ vanishes identically at all times and only Eqs. \eqref{eq:hydro_e} and \eqref{eq:hydro_umu} are to be solved. In the dissipative case, all transport coefficients are chosen in agreement with the microscopic kinetic description, meaning that we compare hydrodynamics at a given specific shear viscosity $\eta/s$ to conformal kinetic theory with a relaxation time $\tau_R$ given by Eq.~\eqref{eq:tauR}, while the other coefficients in conformal RTA are determined by $\tau_R$ as~\cite{Jaiswal:2013npa,Molnar:2013lta,Ambrus:2022vif}
\begin{align}
    \tau_\pi=\tau_R\;, \quad \delta_{\pi\pi}=\frac{4}{3}\tau_R\;,\quad \tau_{\pi\pi}=\frac{10}{7}\tau_R\;,\quad\phi_7=0\;.
\end{align}

Since irrespective of the specific shear viscosity $\eta/s$, hydrodynamics always fails to properly describe the early time pre-equilibrium dynamics of the system, special attention needs to be paid to determine the initial conditions for hydrodynamics, as discussed in detail in our previous work~\cite{Ambrus:2021fej,Ambrus:2022koq}. Essentially, to properly account for the early time pre-equilibrium dynamics, one either needs to match kinetic theory at early times to hydrodynamics at later times, or perform a local re-scaling of the hydrodynamic initial conditions that compensate for the different far-from equilibrium evolution. Both approaches give comparable results within the range of applicability, as demonstrated in Ref.~\cite{Ambrus:2022koq}; however since for small systems or large viscosities, a reasonable level of equilibration may never be reached, matching kinetic theory to hydrodynamics is somewhat impractical and we will thus adopt the local re-scaling procedure of Ref.~\cite{Ambrus:2022koq}, which we briefly outline below. 

We initialize the system at very early times, $\tau_0=10^{-4}R$, on the Bjorken flow attractor of this specific hydrodynamic description; the re-scaling factor for the local energy density is determined by matching the Bjorken flow energy attractor curves of kinetic theory and hydrodynamics in the late time limit, such that hydrodynamics and kinetic theory agree when a sufficient degree of equilibration is achieved before the onset of transverse expansion.  Since we initialize hydrodynamics at such early times, we can safely neglect initial transverse flow setting $u^\mu=(1,0,0,0)$ and set the dissipative components as $\pi^\mu_\nu=\mathrm{diag}(0,\pi_d/2,\pi_d/2,-\pi_d)^\mu_\nu$, where $\pi_d/e$ is determined as the respective value of the Bjorken attractor curve evaluated at the scaling time $\tilde{w}(\xT)=\frac{\tau_0T(\tau_0,\xT)}{4\pi\eta/s}$.

Numerical solutions of the hydrodynamic evolution Eqs.~\eqref{eq:hydro_e}--\eqref{eq:hydro_pi}, are obtained using the open-source code vHLLE~\cite{Karpenko:2013wva}\footnote{Commit number \texttt{efa9e28d24d5115a8d8134852-32fb342b38380f0}.}, which we modified to implement the initialization scheme described above. For a more detailed explanation of the hydrodynamic setup, we refer to our previous work~\cite{Ambrus:2022koq}.

\subsection{Initial conditions}\label{sec:initial_conditions}
Due to the simplicity of our setup, the initial conditions for both hydrodynamics and kinetic theory are then fully determined by the initial energy density profile in the transverse plane. We employ the parametric initial state model \trento~\cite{Moreland:2014oya}  to generate ensembles of event-by-event initial profiles for OO collisions. We considered collision energies of $200\,$GeV and $7\,$TeV, following Ref.~\cite{Nijs:2021clz}\footnote{The OO collisions at the LHC are expected to run at $6.8\,$TeV.}, as well as PbPb collisions at $2.76\,$TeV and AuAu collisions at $200\,$ GeV, reflecting the respective (planned) runs at RHIC and LHC. We use the same value of the \trento ~parameters $w=0.985$ and $p=0.038$ for all four cases, which are taken as the maximum a posteriori values of a recent Bayesian analysis study of PbPb collisions at $2.76\,$TeV~\cite{Liyanage:2023nds}. Similarly, we employ the normalization factor $N=20.013$ for $2.76\,$TeV collisions and follow the strategy proposed in~\cite{Nijs:2021clz} to extrapolate its values at the other energies to be $N=9.69$ at $200\,$GeV and $N=30$ at $7\;$TeV. Inelastic nucleon-nucleon cross-sections were chosen in accordance with the \trento ~documentation as follows: $\sigma_{\rm NN}(200\,\mathrm{GeV})=4.23\,\mathrm{fm}^2$, $\sigma_{\rm NN}(2.76\,\mathrm{TeV})=6.4\,\mathrm{fm}^2$ and $\sigma_{\rm NN}(7\,\mathrm{TeV})=7.32\,\mathrm{fm}^2$.

We note that, in order to account for nucleon correlations and non-trivial nuclear structure effects, such as $\alpha$-clustering in $^{16}\mathrm{O}$, we used the \trento ~mode in which nucleon positions are randomly picked from a list of pre-generated nucleon positions and then rotated randomly. We used a list of 10000 configurations for Pb and 1820 configurations for Au from studies using variational methods to include higher order correlations~\cite{Alvioli:2009ab,Alvioli:2011sk} that were advertised in the \trento  ~documentation. For oxygen we used 6000 nucleon configurations taken from works that employed quantum Monte Carlo calculations based on two- and three-body potentials that were obtained in chiral effective field theory~\cite{Lim:2018huo}\footnote{The list of  $^{16}\mathrm{O}$ configurations is publicly available~\cite{TGlauberMC}.}, which have already been used in other works on simulations of OO collisions~\cite{Lim:2018huo,Rybczynski:2019adt,Nijs:2021clz}.

Centrality selection is performed on the basis of 64000 events, where in order to classify the centrality we employ an estimation formula for the final state particle multiplicity~\cite{Giacalone:2019ldn}, such that the selection can readily be performed on the basis of the initial state energy density profiles. We then divide the events into ten centrality classes of equal size, i.e. $0-10\%$,$10-20\%$ and so on, and randomly select 1600 events from each centrality class for which we perform the evolution in kinetic theory and hydrodynamics. Some of the properties of the initial state ensembles are discussed in Sec.~\ref{sec:geometry} in terms of distributions of three characteristic geometric observables as defined in Sec.~\ref{sec:observables_initial} for the full 64000 events of each system.

\subsection{Observables}\label{sec:observables}

\subsubsection{Initial state}\label{sec:observables_initial}

Given an initial transverse energy density profile $e(\xT)$, we characterize the geometry of the corresponding event in terms of the characteristic observables that are introduced in the following. The total energy at early times is equal to the initial transverse energy $\epni$, since longitudinal momenta (and masses) vanish:
\begin{align}
    \epn=\tau_0 \int_\xT e(\xT)\;.
\end{align}
The transverse size is quantified by the rms transverse radius $R$:
\begin{align}
    R^2=\left(\int_\xT e(\xT)\right)^{-1} \int_\xT \mathbf{x}_\perp^2 e(\xT)\;.
    \label{eq:R}
\end{align}
The initial state ellipticity $\epsilon_2$ is well established as the main geometric quantity governing the development of elliptic flow, which is given by 
\begin{align}
    e^{2i\Psi_2}\epsilon_{2}=-\left(\int_\xT \mathbf{x}_\perp^2 e(\xT) \right)^{-1} \int_{\xT} e^{2i\phi_x} \mathbf{x}_\perp^2 e(\xT)\;.\label{eq:def_e2}
\end{align}
Here, $\epsilon_2$ is the magnitude of the complex quantity on the right-hand side, and $\Psi_2$ denotes the symmetry plane angle. We are primarily interested in the magnitude $\epsilon_2$ and will not mention the phase when discussing our results. \footnote{ Technically this phase can become important to discern the orientation of the elliptic flow. The sign in Eq.~\eqref{eq:def_e2} is chosen such that elliptic flow will typically be positive when $\epsilon_2$ is positive, though at very low opacities it can be negative~\cite{Ambrus:2021fej}. However, the cases considered in this paper have large enough opacity for this not to happen.}

\subsubsection{Final state}\label{sec:observables_final}
When considering the collective flow buildup in response to the initial state geometry, 
we will focus on the elliptic flow, which is usually quantified in terms of the anisotropic flow coefficients $v_n$~of charged particles~\cite{Voloshin:1994mz,Borghini:2000sa}. However, instead of considering the particle number weighted observable $v_n$, we will examine elliptic flow in terms of the anisotropic energy flow 
\begin{align}
    e^{2i\Psi_{2,p}}\varepsilon_p=\frac{\int_\xT T^{xx}-T^{yy}+2iT^{xy}}{\int_\xT T^{xx}+T^{yy}}\;,\label{eq:epsp_def}
\end{align}
since it can be extracted directly from the energy-momentum tensor in both kinetic theory and hydrodynamics. Since $\varepsilon_p$ measures the energy flow, it can further be expected to be less sensitive to hadronization effects than a particle number weighted measurement of elliptic flow. In particular, in this work, we did not even attempt to model hadronization in any of our simulation setups. 

Besides this important difference, we follow the common procedure and quantify the event-by-event fluctuations of observables $\mathcal{O}$ by the cumulants $c_\mathcal{O}\{2k\}$, defined in the following ways:
\begin{align}
    c_\mathcal{O}\{2\}&=\langle |\mathcal{O}|^2 \rangle\label{eq:def_cumulant_2}\\
    c_\mathcal{O}\{4\}&=\langle |\mathcal{O}|^4 \rangle-2\langle |\mathcal{O}|^2 \rangle^2\\
    c_\mathcal{O}\{6\}&=\langle |\mathcal{O}|^6 \rangle-9\langle |\mathcal{O}|^4 \rangle\langle |\mathcal{O}|^2 \rangle+12\langle |\mathcal{O}|^2 \rangle^3\\
    c_\mathcal{O}\{8\}&=\langle |\mathcal{O}|^8 \rangle-16\langle |\mathcal{O}|^6 \rangle\langle |\mathcal{O}|^2 \rangle-18\langle |\mathcal{O}|^4 \rangle^2\label{eq:def_cumulant_8}\\
    &\quad + 144\langle |\mathcal{O}|^4 \rangle\langle |\mathcal{O}|^2 \rangle^2-144\langle |\mathcal{O}|^2\rangle^4\nonumber
\end{align}
where $\langle\,\cdot\,\rangle$ denotes the average over an ensemble of events, typically within a certain centrality class. We will compute these cumulants for both the final state elliptic flow $\varepsilon_p$ and the initial state ellipticity $\epsilon_2$. In the cases $\mathcal{O}=v_n$, the above definition reduces to the conventional flow cumulants $c_{v_n}\{2k\}=c_n\{2k\}$.

We are also interested in the transverse energy per unit rapidity in the final state. Similarly to elliptic flow, we want to define it in a way that allows to compute it directly from hydrodynamics without the need of invoking a particular hadronization prescription. We thus infer the transverse energy per unit rapidity via the proxy
\begin{align}
    \frac{\d E_\perp}{\d \eta}=\tau\int_\xT T^{xx}+T^{yy}\,,\label{eq:Eperp_def}
\end{align}
noting that at least in kinetic theory,  the transverse energy could in principle be more accurately computed as the integral of transverse momenta of all particles. However, in the late time limit, when particles are free-streaming again and longitudinal momenta vanish, the two definitions become identical.\footnote{Note that this is unfortunately not the case in ideal hydrodynamics, which maintains an isotropic equilibrium pressure at all times. In order to account for this difference, we multiply the result of the right-hand side of Eq.~\eqref{eq:Eperp_def} by a factor $\frac{3}{8}\pi$ for ideal hydrodynamics to obtain the transverse energy the system would have in the absence of longitudinal pressure.}

We extract all final state observables at an evolution time of $\tau=3R$, at which point the time evolution of all observables of interest has typically leveled off to its asymptotic late time value.

\section{Geometry}\label{sec:geometry}

\begin{figure*}
    \centering
    \includegraphics[width=.9\textwidth]{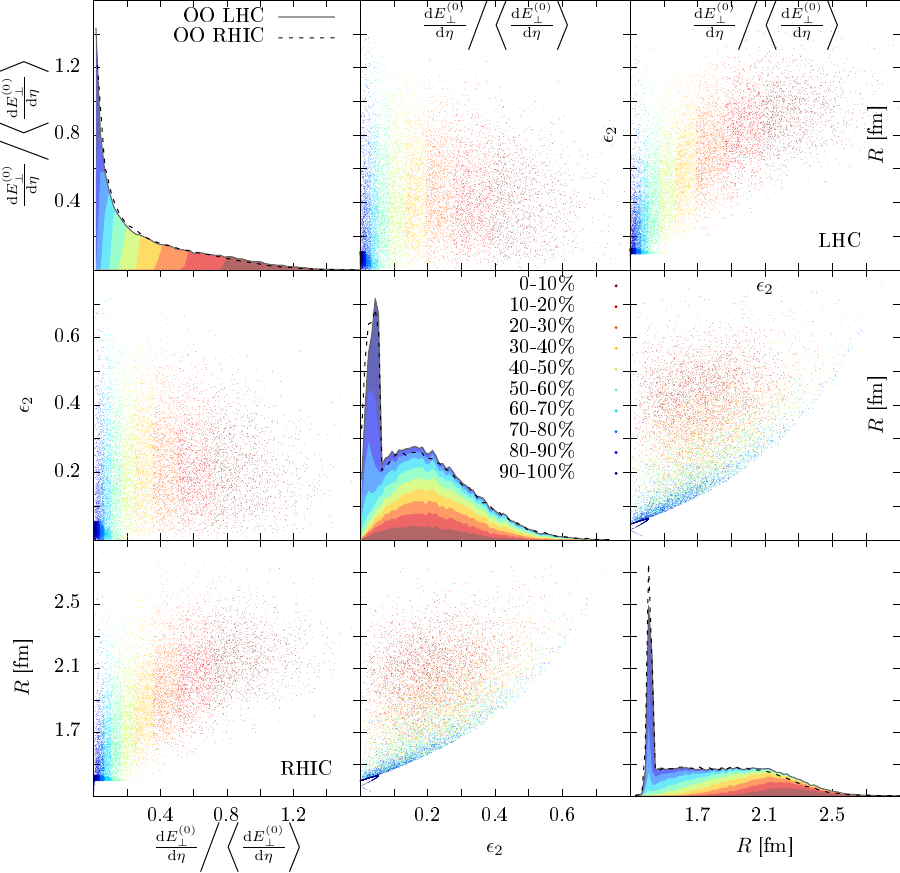}
    \caption{Distributions of transverse energy $\epni$, transverse radius $R$ and ellipticity $\epsilon_2$ in OO initial states at RHIC and LHC. Different colors show distributions for different centrality classes. Diagonal: histograms of one-dimensional distributions at LHC (gray solid) and RHIC (black dashed), where colors show centrality classes of LHC. Upper right: two-dimensional distributions for LHC as scatter plots. Lower left: the same for RHIC.
    }
    \label{fig:OOhistograms}
\end{figure*}

\begin{figure*}
    \centering
    \includegraphics[width=.9\textwidth]{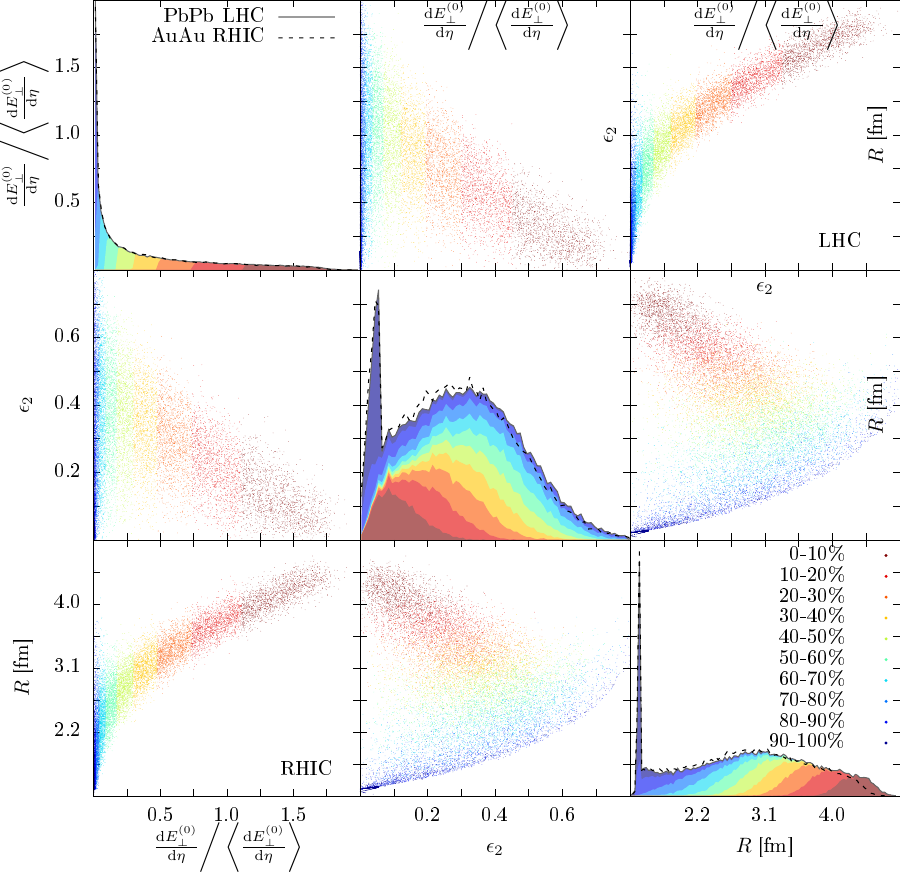}
    \caption{Distributions of transverse energy $\epni$, transverse radius $R$ and ellipticity $\epsilon_2$ in AuAu initial states at RHIC and PbPb initial states at LHC. Different colors show distributions for different centrality classes. Diagonal: histograms of one-dimensional distributions at LHC (gray solid) and RHIC (black dashed), where colors show centrality classes of LHC. Upper right: two-dimensional distributions for LHC as scatter plots. Lower left: the same for RHIC.}
    \label{fig:PbAuhistograms}
\end{figure*}

Before we investigate the collective dynamics of small and large systems, it is important to understand the initial state geometry. Since we will compare the collective flow response between heavy and light ion systems and also between RHIC and LHC, it is of particular interest how system size, eccentricity and initial state energy compare between the different systems. In order to examine this, we show in Fig.~\ref{fig:OOhistograms} one- and two-dimensional distributions of the transverse energy per unit rapidity $\epni$, the  transverse rms radius $R$ and the elliptic eccentricity $\epsilon_2$ of our initial state samples for OO collisions at RHIC and LHC. Fig.~\ref{fig:PbAuhistograms} shows the same for AuAu and PbPb collisions. The one-dimensional distributions are plotted as histograms in the panels along the diagonal, where LHC results are shown as gray solid lines and RHIC results are shown as black dashed lines. The colored area plots show the contributions from each centrality class at LHC. Two-dimensional distributions are shown as scatter plots, where different colored dots again represent events from different centrality classes. Results for LHC are shown in the upper right set of panels, while RHIC results are in the lower left set.

By comparing the one-dimensional distributions, as well as corresponding two-dimensional distributions on opposite ends of the diagonals, we observe that for OO collisions almost no difference in geometry is visible between RHIC and LHC. Hence, the most important difference concerns the overall scale of the transverse energy distribution $\langle \epni \rangle$, which was scaled out in Fig.~\ref{fig:OOhistograms}, since for our purposes this scale is addressed via the mean opacity. The distribution of transverse energy is quickly falling towards higher values. For intermediate to large values, the $\epsilon_2$-distribution follows a Bessel-Gaussian--like curve, while the $R$-distribution shows a plateau that drops towards large values. For both observables it is very noticeable that due to fluctuations their values in the most central classes are significantly more spread out in OO than in PbPb and AuAu. In particular, central OO collisions can have large ellipticities $\epsilon_2$. We further note that the peak at small $\epsilon_2$ and $R$, as well as the tail structure close to the origin in their joint distribution, are due to events with only one to a few nucleon collisions in the more peripheral centrality classes ($\gtrsim 70\%$).

When considering the correlations of the initial state observables, we see a clear correlation between transverse energy $\epni$ and transverse rms radius $R$. This is straightforwardly understood: if the system has a larger transverse size, it is also more likely to carry a larger amount of energy and vice versa. Furthermore, a correlation between the transverse rms radius $R$ and the maximum possible eccentricity $\epsilon_2$ is apparent, as in the \trento ~model with nucleon size fluctuations, a more elongated geometry will also necessarily have a large rms radius. Finally, the slight correlation between transverse energy and $\epsilon_2$ might be a result of both of the above correlations combined.

Statistical distributions of the initial state observables in PbPb and AuAu collisions show a similar behavior, but now the LHC PbPb system reaches noticeably higher transverse radii than AuAu collisions at RHIC. In general the form of the one-dimensional distributions is similar to OO collisions, except for a slight increase in the $R$ distribution at small values resulting in a wide peak at $R\approx 3\,$fm.

Differences between OO and AuAu/PbPb collisions emerge when considering the two-dimensional scatter plots, which illustrate the (anti-)correlations between initial state observables. In PbPb and AuAu collisions, the correlation between transverse energy and radius is much stronger, as the overall geometry plays a more important role compared to the fluctuations of nucleon positions. Due to this fact, one also finds that -- on top of the positive correlation of $R$ and the maximum possible $\epsilon_2$ -- there is an anti-correlation between $R$ and $\epsilon_2$ for more central collisions, which can be understood by the fact that the mean geometry dictated by the overlap of the two lead nuclei has the largest radius at maximal overlap, in which case $\epsilon_2$ is small. Similar considerations also explain the anticorrelation between $\epsilon_2$ and transverse energy $\epni$ in PbPb and AuAu collisions, which is not visible in OO collisions, where even in central events nucleon position fluctuations give rise to sizeable elliptic eccentricities $\epsilon_2$.

\section{Collective flow results}\label{sec:elliptic}

\begin{figure*}[t]
    \centering
    \includegraphics[width=.49\textwidth]{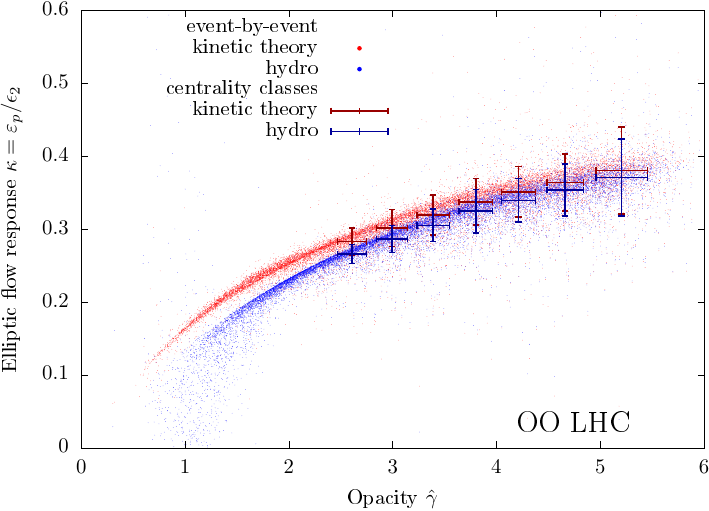}
    \includegraphics[width=.49\textwidth]{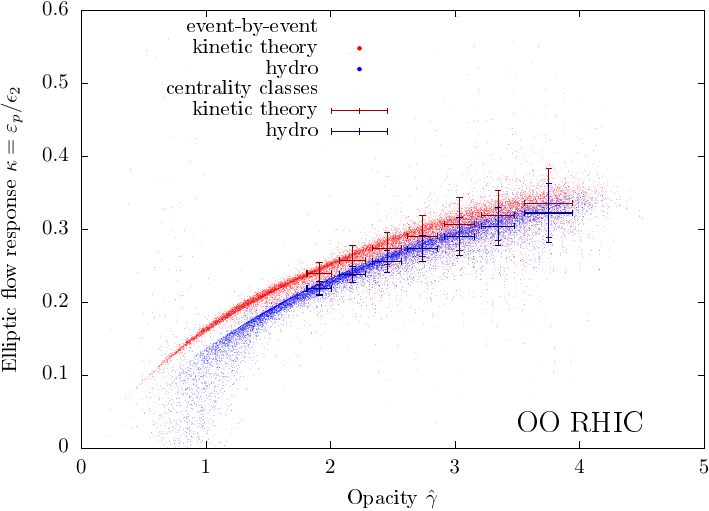}
    \includegraphics[width=.49\textwidth]{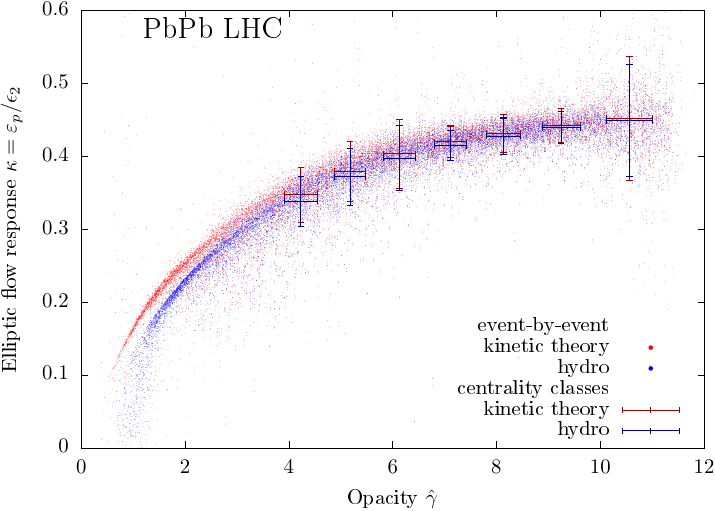}
    \includegraphics[width=.49\textwidth]{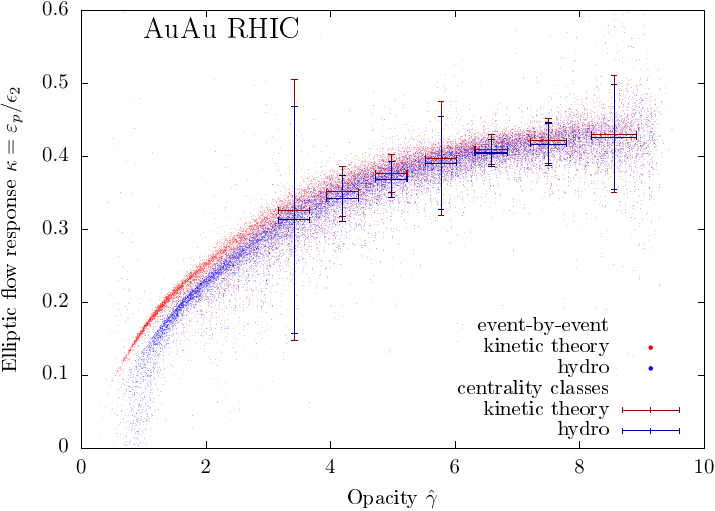}
    \caption{Point clouds of event-by-event flow responses as a function of event-by-event opacity in OO at LHC (top left) and RHIC (top right), as well as PbPb (bottom left) and AuAu (bottom right) from simulations in kinetic theory (red) and hydrodynamics (blue) at a shear viscosity of $\eta/s=0.12$. Points show the corresponding means in centrality classes of size 10\% and their error bars indicate standard deviations of the corresponding quantity within each centrality class.}
    \label{fig:point_cloud}
\end{figure*}

We will now examine the results for collective flow from the event-by-event simulations in conformal kinetic theory and viscous hydrodynamics.

\subsection{Event-by-event flow response}
Since collective flow is built up in response to the initial state geometry, it is instructive to first consider the response coefficient $\kappa=\varepsilon_p/\epsilon_2$ for the elliptic flow response. Simulation results in kinetic theory and hydrodynamics are compactly summarized in Fig.~\ref{fig:point_cloud}, where we show event-by-event results from simulations at a realistic value of $\eta/s=0.12$ as a point cloud in the plane of opacity $\hat{\gamma}$ and elliptic flow response coefficient $\kappa=\varepsilon_p/\epsilon_2$. Different panels in Fig.~\ref{fig:point_cloud} show the results obtained for each of the collision systems, and each plot also compares results from kinetic theory and hydrodynamics, shown in different colors. In addition to the individual dots for each event, points with horizontal and vertical error bars indicate the mean values and standard deviation within centrality classes of size 10\% up to 70\% centrality.

Generally, the event-by-event results show an increase in the magnitude of the flow response with increased opacity, which seems to saturate towards high opacities. When comparing results from kinetic theory and hydrodynamics, the flow response of the latter approaches that of the former from below for increasing opacity, as seen particularly well for central PbPb and AuAu collisions which have the largest opacities. 

When comparing the point clouds for different systems, one observes a very similar behavior across all systems, as the biggest difference seems to be the accessible range of opacities $\hat{\gamma}$ in each system. Clearly, each point cloud exhibits a clustering along a fixed curve in the $\kappa$-$\hat{\gamma}$ plane, with relatively small fluctuations which introduce a spread of individual events around this curve. By carefully examining this, it seems that the spread of the response coefficient $\kappa$ is somewhat bigger in more central events, where fluctuations play a central role in determining the eccentricity. This is especially true for PbPb and AuAu collisions, where the eccentricity $\epsilon_2$ is moreover very small. Hydrodynamic simulations of very peripheral events also show a larger spread of the response coefficient, especially in OO collisions where the opacities are smaller, which is not observed in kinetic theory simulations. At this point it is unclear where exactly the spread in the hydrodynamic results comes from, but as argued previously on grounds of the average flow response~\cite{Ambrus:2022koq,Ambrus:2022qya} and verified later in this work, hydrodynamics already becomes inaccurate for $\hat{\gamma}\lesssim 3$ and should certainly not be trusted in the regime where these tails lie.

\begin{figure*}
    \centering
    \includegraphics[width=.49\textwidth]{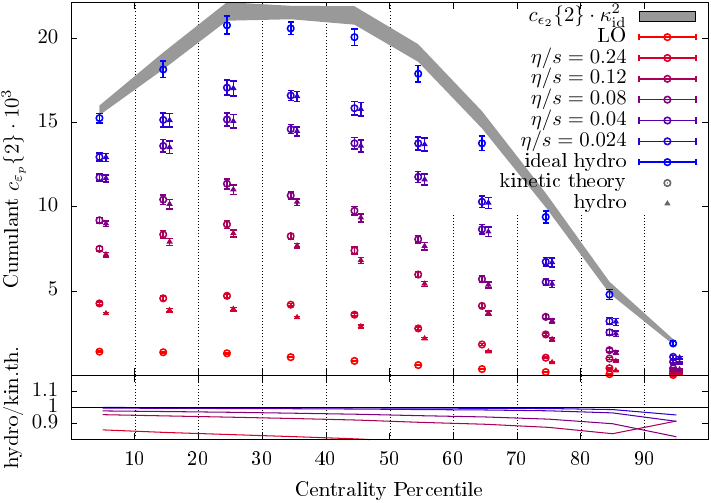}
    \includegraphics[width=.49\textwidth]{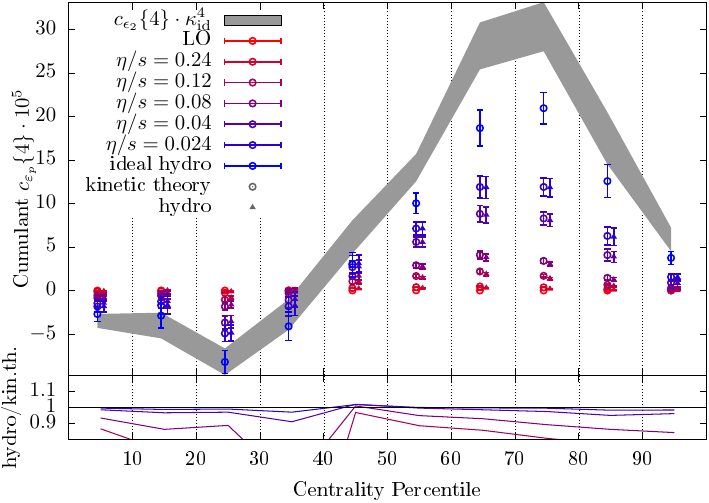}
    \includegraphics[width=.49\textwidth]{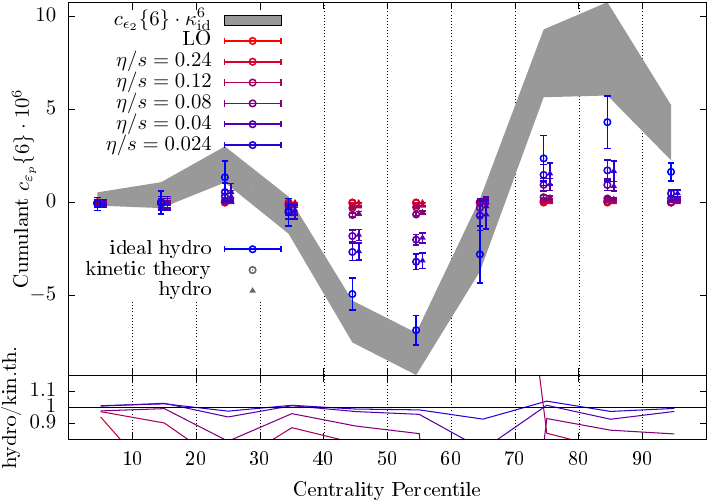}
    \includegraphics[width=.49\textwidth]{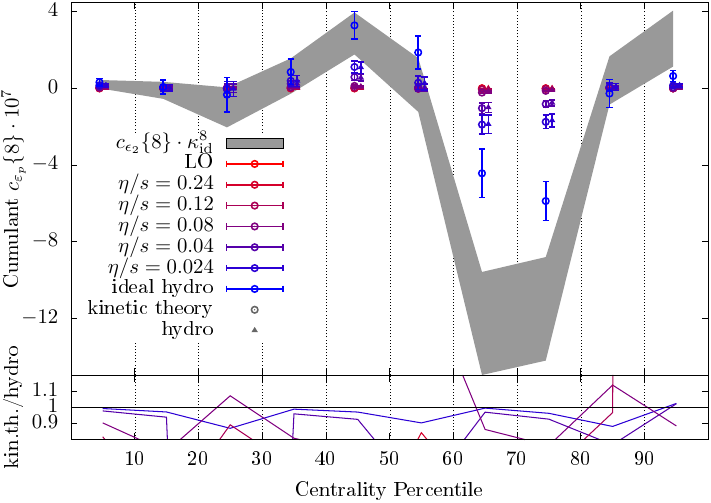}
    \caption{
    Elliptic flow cumulants $c_{\varepsilon_p}\{2\}$ (top left), $c_{\varepsilon_p}\{4\}$ (top right), $c_{\varepsilon_p}\{6\}$ (bottom left) and $c_{\varepsilon_p}\{8\}$ (bottom right) as a function of centrality for OO collisions at LHC ($\sqrt{s_{NN}}=7\;$TeV) from simulations in kinetic theory (empty circles) and hydrodynamics (filled triangles) covering a range in shear viscosity from the linear order result (red) to ideal hydrodynamics (blue) and various values of $\eta/s$ inbetween (color gradient from red to blue). Also shown are cumulants $c_{\epsilon_2}\{2k\}$ of the initial state ellipticity scaled by a power of the mean ideal hydrodynamic flow response coefficient $\kappa_{\rm id}^{2k}$ (grey bands). Colored lines in the lower panel of each plot shows the ratio of hydrodynamic to kinetic theory results for the corresponding value of $\eta/s$.}
    \label{fig:cumulants_vs_centrality_OOLHC}
\end{figure*}

\begin{figure*}
    \centering
    \includegraphics[width=.49\textwidth]{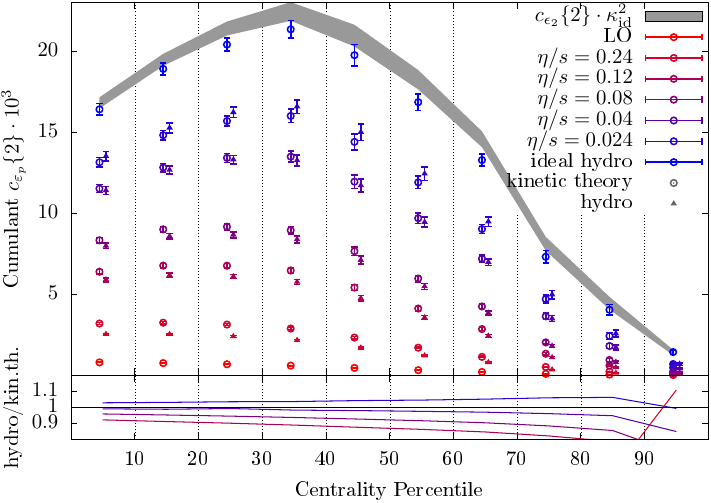}
    \includegraphics[width=.49\textwidth]{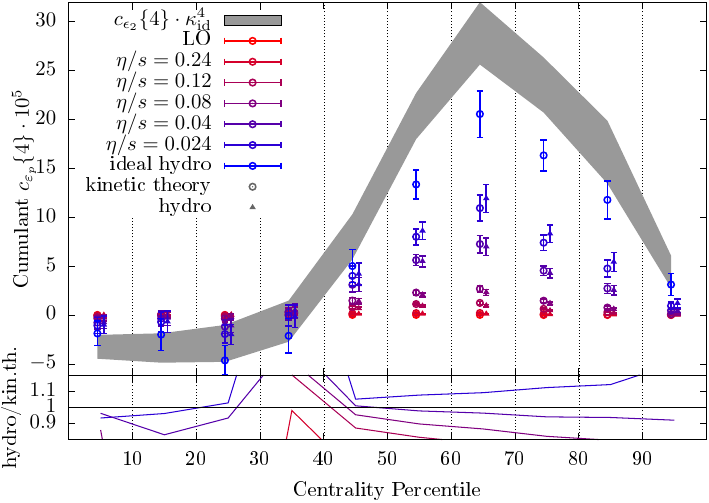}
    \includegraphics[width=.49\textwidth]{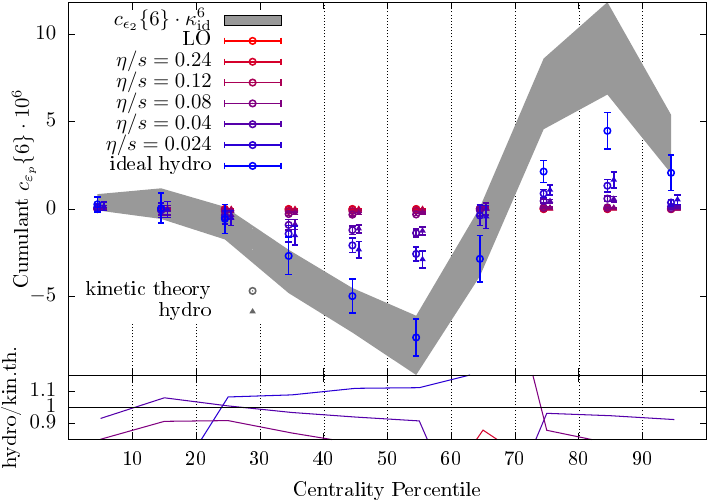}
    \includegraphics[width=.49\textwidth]{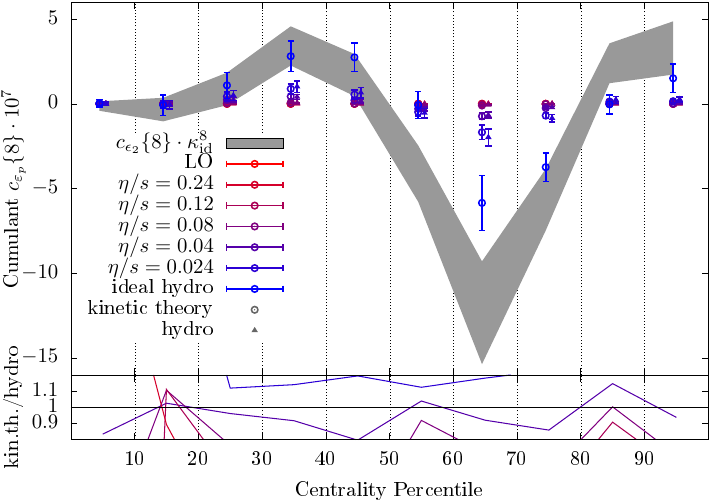}
    \caption{Same as Fig.~\ref{fig:cumulants_vs_centrality_OOLHC}, but for OO collisions at RHIC ($\sqrt{s_{NN}} = 200\;$GeV).}
    \label{fig:cumulants_vs_centrality_OORHIC}
\end{figure*}

\begin{figure*}
    \centering
    \includegraphics[width=.49\textwidth]{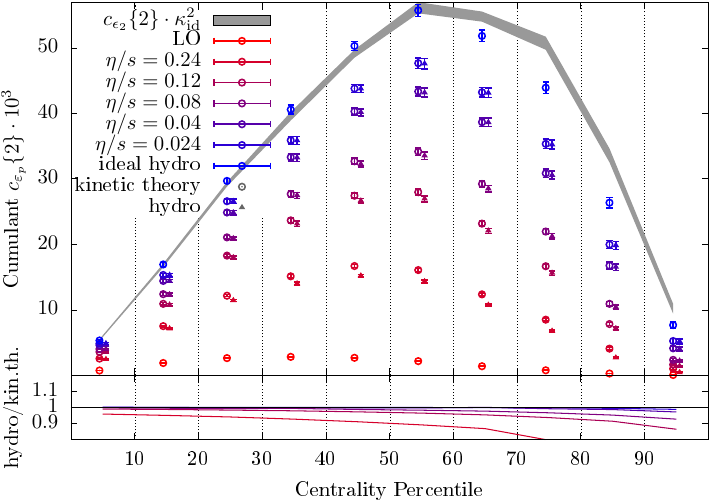}
    \includegraphics[width=.49\textwidth]{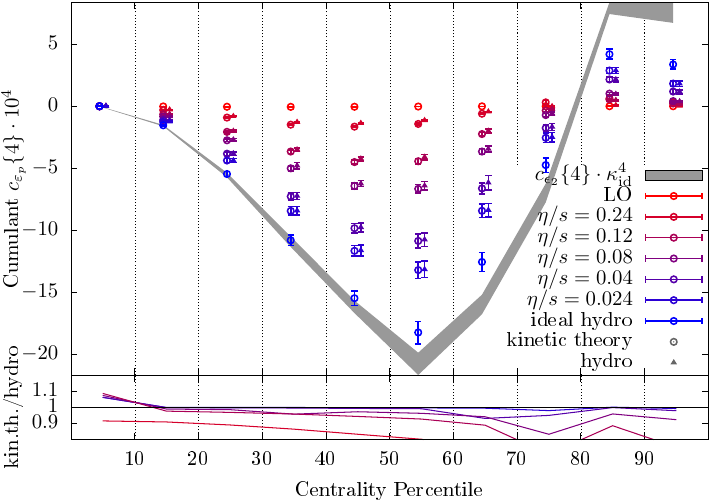}
    \includegraphics[width=.49\textwidth]{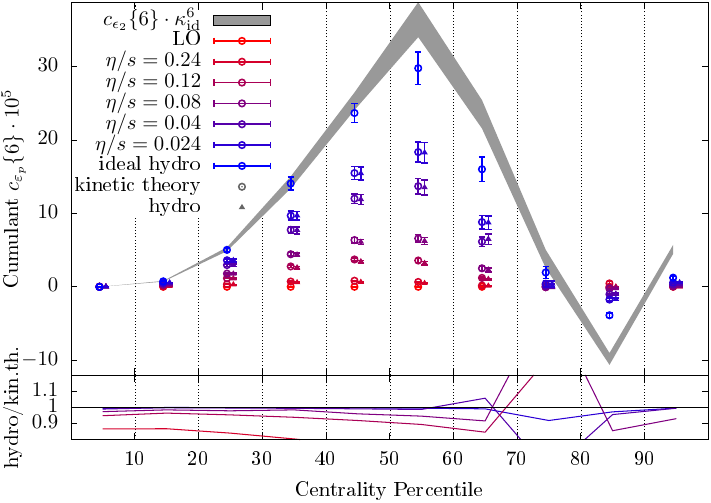}
    \includegraphics[width=.49\textwidth]{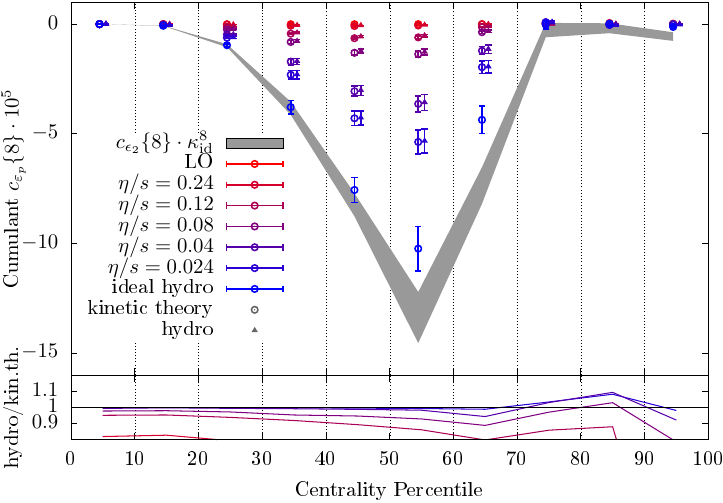}
    \caption{Same as Fig.~\ref{fig:cumulants_vs_centrality_OOLHC}, but for PbPb collisions at LHC ($\sqrt{s_{NN}} = 2.76\;$TeV).}
    \label{fig:cumulants_vs_centrality_PbPbLHC}
\end{figure*}

\begin{figure*}
    \centering
    \includegraphics[width=.49\textwidth]{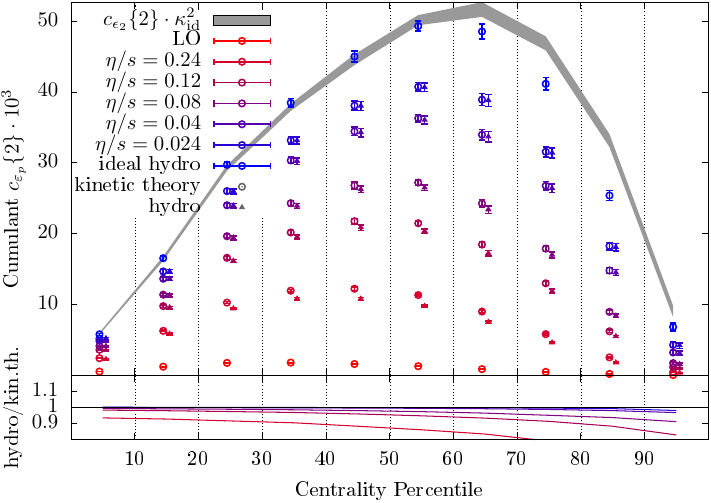}
    \includegraphics[width=.49\textwidth]{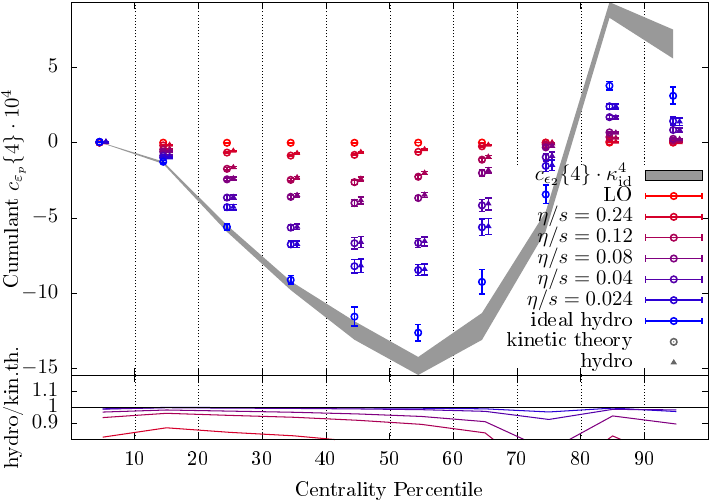}
    \includegraphics[width=.49\textwidth]{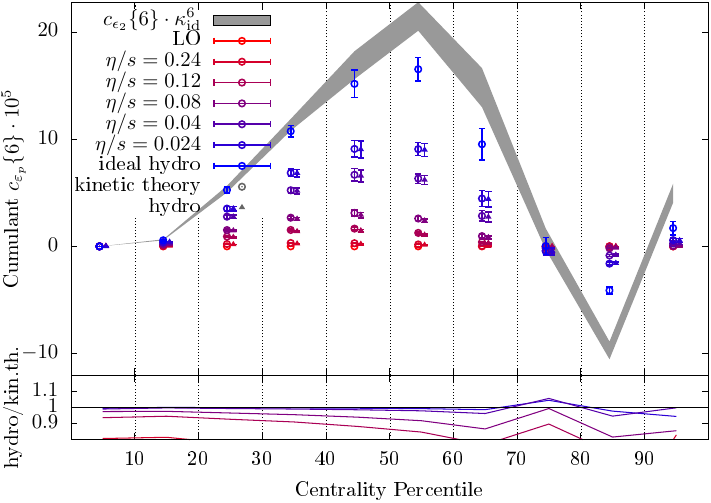}
    \includegraphics[width=.49\textwidth]{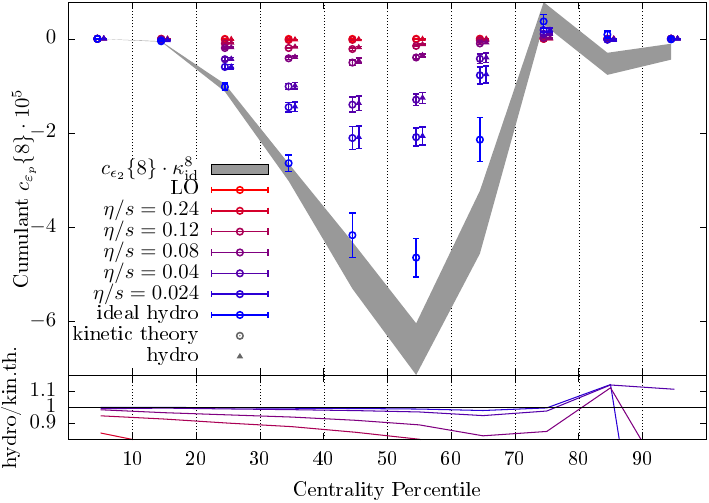}
    \caption{Same as Fig.~\ref{fig:cumulants_vs_centrality_OOLHC}, but for AuAu collisions at RHIC ($\sqrt{s_{NN}} = 200\;$GeV).}
    \label{fig:cumulants_vs_centrality_AuAuRHIC}
\end{figure*}
\subsection{Cumulants}
Now that we have established the general features of the initial state geometry and the geometric response in the different collisions systems, we continue to  characterize the statistics of final state flow by the flow cumulants $c_{\varepsilon_p}\{2k\}$, c.f. Eqs.~\eqref{eq:def_cumulant_2}--\eqref{eq:def_cumulant_8}. We computed these cumulants for each centrality class of size 10\% and obtained an error estimate via Jackknife resampling. We present our results for $c_{\varepsilon_p}\{2\}$, $c_{\varepsilon_p}\{4\}$, $c_{\varepsilon_p}\{6\}$ and $c_{\varepsilon_p}\{8\}$ separately for each collision system,  in Fig.~\ref{fig:cumulants_vs_centrality_OOLHC} for OO at LHC, in Fig.~\ref{fig:cumulants_vs_centrality_OORHIC} for OO at RHIC, in Fig.~\ref{fig:cumulants_vs_centrality_PbPbLHC} for PbPb as well as in Fig.~\ref{fig:cumulants_vs_centrality_AuAuRHIC} for AuAu. Each plot shows results for a large range of $\eta/s$ values in different colors, along with the limiting cases in the dilute limit, obtained from the leading order opacity expansion (c.f. Sec.~\ref{sec:linear_order}) and the ideal hydrodynamic result. We further compare results obtained in kinetic theory and hydrodynamics for all finite values of $\eta/s$ and depict the ratio of the two in the lower inset of each plot. We also compare the results for the final state elliptic flow $c_{\varepsilon_p}\{2k\}$ to the corresponding cumulant $c_{\epsilon_2}\{2k\}$ computed for the initial state ellipticity, which is scaled by an appropriate power of the flow response coefficient $\kappa_{\rm id}=0.547$, determined in ideal hydrodynamics (c.f. Fig.~\ref{fig:kappa_vs_gamma}).

Generally, the results in Figs.~\ref{fig:cumulants_vs_centrality_OOLHC}--\ref{fig:cumulants_vs_centrality_AuAuRHIC} allow us to compare the development of collective flow in large and small systems, the development of collective flow in kinetic theory and hydrodynamics and the development of collective flow at RHIC and LHC energies. In this paper we focus on the first two aspects, and simply note that the overall behavior of the flow cumulants shows little deviation when comparing the LHC collision system with the corresponding RHIC collision system, such that the similarity in the initial state geometry also extends to final state observables. Nevertheless, there are small but important quantitative differences in the collective flow at RHIC and LHC, which are explored further in our companion paper~\cite{Ambrus:2024eqa}.

Starting with the comparison of the small OO and large AuAu/PbPb systems, one immediately notes the characteristic differences in the shapes of the cumulant curves. In OO collisions, the initial state eccentricity is, to a large extent, driven by nucleon position fluctuations. These give rise to significant event-by-event initial ellipticity, even in central OO events, where the mean geometry is almost isotropic. This results in a large $c\{2\}$ even in central OO collisions with a relatively weak centrality dependence up to about $50-60\%$ centrality. Conversely, in PbPb or AuAu collisions, the effects of nucleon position fluctuations are suppressed by the much larger number of nucleons, and the mean geometry is dominant in determining the initial state eccentricity. Due to this fact, the centrality dependence of $c\{2\}$ is quite different, showing a significantly smaller elliptic flow in central than in mid-central to peripheral collisions. Notably, the much larger impact of nucleon position fluctuations in OO collisions also provides the reason why some of the cumulants show a different sign than expected from combinatorial considerations~\cite{Borghini:2001vi}, where typically the second and sixth order cumulant are positive while the fourth and eighth order ones are negative. For the PbPb and AuAu systems depicted in Figs.~\ref{fig:cumulants_vs_centrality_PbPbLHC} and \ref{fig:cumulants_vs_centrality_AuAuRHIC}, these expectations hold for all but the most peripheral centrality classes. However, we observe in both OO systems shown in Figs.~\ref{fig:cumulants_vs_centrality_OOLHC} and \ref{fig:cumulants_vs_centrality_OORHIC} that the fourth order cumulant is positive for mid-central to peripheral classes; the sixth order cumulant is negative for central to mid-central classes; and the eighth order cumulant is positive for both central and peripheral classes. 

Strikingly, these general observations for small and large systems hold irrespective of the opacity -- as determined by the value of $\eta/s$ -- and can be observed even in the limiting cases of the leading order opacity expansion and the ideal hydrodynamic limit. By far, the most important effect of varying the opacity is the overall increase of the magnitude of collective flow for increasing opacity (decreasing $\eta/s$), due to the stronger final state response to the initial state eccentricity. However, the opacity dependence of the final state response also has a more subtle effect on the centrality dependence of the flow cumulants. Since for any finite value of $\eta/s$, the more peripheral event classes with lower opacity also exhibit a weaker flow response (c.f. Figs.~\ref{fig:point_cloud} and~\ref{fig:kappa_vs_gamma}), the magnitude of the flow response in more peripheral centrality classes decreases faster with decreasing opacity (increasing $\eta/s$) than the one in central classes. Most notably, this effect results in a shift of the extrema of the curves towards lower centrality percentile for smaller opacities (larger $\eta/s$), which is present for all cumulants in all collision systems, but perhaps most easily seen for $c_{\varepsilon_p}\{2\}$ in PbPb or AuAu collisions. Hence, the centrality dependence of the flow cumulants only corresponds to the initial state geometry in the ideal hydrodynamic limit, where the response coefficient $\kappa(\hat{\gamma})=\varepsilon_p/\epsilon_2$ saturates to an opacity independent value $\kappa_{\rm id}$, and the behavior of the flow cumulants closely follows that of the cumulants of the initial state eccentricity. Nevertheless, it turns out that, also away from the ideal hydrodynamic limit, the centrality dependence of the flow cumulants $c_{\varepsilon_p}\{2k\}$ can be very well described by the corresponding initial state eccentricity cumulant $c_{\epsilon_2} \{2k\}$ multiplied with an appropriate power of the mean response coefficient $\kappa(\langle\hat{\gamma}\rangle)$ of each centrality class, i.e.
\begin{equation}
    c_{\varepsilon_p}\{2k\}=c_{\epsilon_2} \{2k\}\kappa(\langle\hat{\gamma}\rangle)^{2k}\;,
\end{equation}
which indicates that flow cumulants are dominated by initial state fluctuations, whereas fluctuations of the flow response give subleading contributions to the flow cumulants. We will explore this further in Sec.~\ref{sec:cumulant_ratios}, where we consider ratios of the flow cumulants to cancel the response coefficients.

Now that we have described the basic properties of the flow cumulants in small and large collision systems, we turn to a more detailed comparison of the results obtained using kinetic theory and hydrodynamics. By comparing kinetic theory (open circles) and hydrodynamic (filled triangles) results for AuAu, PbPb and OO collisions at RHIC and LHC in Figs.~\ref{fig:cumulants_vs_centrality_OOLHC}--\ref{fig:cumulants_vs_centrality_AuAuRHIC}, one generally finds that kinetic theory and hydrodynamics exhibit the same centrality dependence and the most important difference lies in the exact magnitude of the flow response. Generally, as expected for large opacity (small $\eta/s$), kinetic theory and hydrodynamics agree well with each other, in particular for the larger AuAu and PbPb systems, where, even for $\eta/s=0.24$, deviations between kinetic theory and hydrodynamics are less than $10\%$, except in very peripheral collisions. However, due to the smaller opacity reached in OO collisions, deviations between hydrodynamics and kinetic theory at the $10\%$ level emerge in mid-central collisions already for rather small viscosities $\eta/s=0.12$, where hydrodynamics systematically underpredicts the final state response to the initial state eccentricity. We can thus conclude from this analysis that collective flow in OO collisions at RHIC and LHC energies do provide some sensitivity to the underlying nonequilibrium dynamics of the QGP. However, for realistic values of $\eta/s$, the effects in central and mid-central collisions are typically only at the $10\%$ level, such that an actual discrimination between hydrodynamic and nonhydrodynamic behavior requires a very precise understanding of the initial state geometry, as in all model calculations, a larger initial eccentricity can always compensate for a weaker final state response. Due to this ambiguity, it would clearly be desirable to further separate the effects of the final state response from the initial state eccentricity, and we explore one possibility to do this in our companion paper~\cite{Ambrus:2024eqa}.

\subsection{Universality of response coefficient}

\begin{figure}
    \centering
    \includegraphics[width=.49\textwidth]{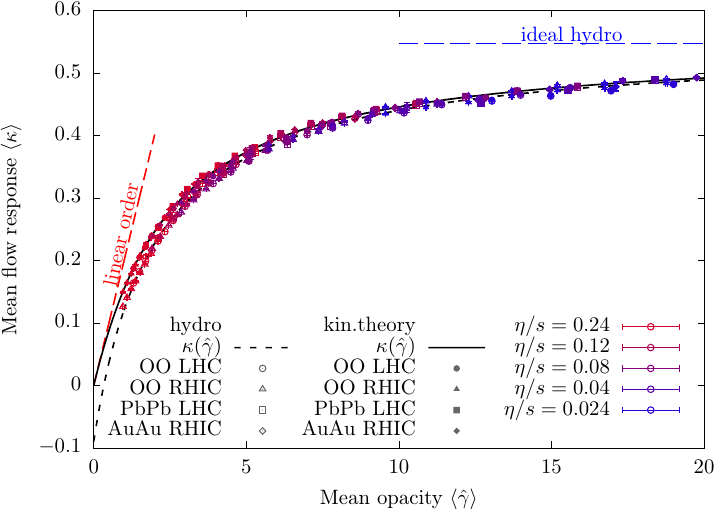}
    \caption{Mean values of the response $\kappa=\varepsilon_p/\epsilon_2$ as a function of the mean opacity from our kinetic theory (filled symbols) and hydrodynamic (open symbols) simulation results for
    centrality classes of size 10\% of OO collisions at LHC (circles) and RHIC (triangles), as well as PbPb collisions
    (squares) and AuAu collisions (diamonds), for various
    values of $\eta/s$, in color gradient from highest (red) to lowest
    (blue), together with Padé fits to the kinetic theory (solid line) 
    and hydrodynamic (dashed line) simulation data,
    with the limiting cases of opacity-linearized buildup for dilute
    systems (red dashed) and saturated response for ideal hydro-
    dynamics (blue dashed). Error bars show the standard error of
    the mean.}
    \label{fig:kappa_vs_gamma}
\end{figure}

Next, in order to scrutinize the comparison between kinetic theory and hydrodynamics and further quantify the opacity dependence of the collective flow response, we extract the dependence of the mean elliptic flow response coefficient $\langle \kappa( \hat{\gamma}) \rangle$ as a function of the mean opacity $\langle \hat{\gamma}\rangle$ for each centrality class in all collision systems, and summarize the results of this analysis in Fig.~\ref{fig:kappa_vs_gamma}. Different symbols in Fig.~\ref{fig:kappa_vs_gamma} correspond to the results for different collisions systems, with different colors indicating the results for different specific shear viscosities, in kinetic theory (full symbols) and hydrodynamics (open symbols), and we also depict the limiting behavior in ideal hydrodynamics and to leading order in the opacity expansion.

Strikingly, both kinetic theory and hydrodynamics show a universal response as a function of opacity, such that the results for different systems and viscosities collapse onto a universal scaling curve $\kappa(\hat{\gamma})$. Specifically, for kinetic theory, the response coefficient increases monotonically as a function of opacity and smoothly interpolates between a linear behavior at low opacity  
\begin{equation}
    \kappa(\hat{\gamma}\ll 1)= \kappa'_{0} \hat{\gamma}\;,
\end{equation}
and the ideal hydrodynamic limit 
\begin{equation}
    \kappa(\hat{\gamma}\gg 1)= \kappa_{\rm id}\;,
\end{equation}
where the response saturates at large opacity. Beyond these two limits, we find that the kinetic theory response can be well described with a simple Pad\'e approximant as
\begin{equation}
    \kappa(\hat{\gamma})=\frac{0.201\hat{\gamma}+0.129\hat{\gamma}^2}{1+0.892\hat{\gamma}+0.235\hat{\gamma}^2}\;,
\end{equation}
while in hydrodynamics, it is approximated by
\begin{align}
\kappa(\hat{\gamma})=\frac{-0.0896+0.271\hat{\gamma}}{1+0.496\hat{\gamma}}\;.
\end{align}

Hydrodynamics agrees with kinetic theory for large opacities, but shows visible deviations at low opacities, which become of the order of $\gtrsim 5\%$ for $\hat{\gamma}\lesssim 3$. However, it is important to point out that hydrodynamics gradually fails more and more severely to describe the kinetic theory results when considering lower and lower opacities and there is not really an abrupt breakdown. We note that all of these statements are in agreement with our previous findings obtained from studying the development of collective flow in averaged PbPb events~\cite{Ambrus:2022koq,Ambrus:2022qya}. Based on a detailed comparison of kinetic theory + hydrodynamics hybrid results to pure kinetic theory, we previously determined that hydrodynamics provides a quantitatively accurate tool to study collective flow for $\hat{\gamma}\gtrsim 3$, which is consistent with the event-by-event simulations presented in this paper.

\subsection{Cumulant ratios}\label{sec:cumulant_ratios}
Since the flow cumulants $c\{2k\}$ show a strong sensitivity to the magnitude of the final state response, it is instructive to consider ratios of flow cumulants to separate the effects of the initial state geometry from the final state response. Indeed, if we neglect event-by-event fluctuations of the response $\kappa(\hat{\gamma})$ and instead consider the average response $\langle{\kappa}\rangle$ for each centrality class (as depicted in Fig.~\ref{fig:kappa_vs_gamma}), it immediately follows that
\begin{align}
    \langle\varepsilon_p^n\rangle \approx \langle{\kappa}\rangle^n\langle\epsilon_2^n\rangle\label{eq:moment_approximation}
\end{align}
By definition, this approximation directly extends to the cumulants, such that -- to the extent that this approximation is valid -- the response coefficients can be eliminated from the cumulant ratios~\cite{Bhalerao:2011yg}
\begin{align}
    c_{\varepsilon_p}\{2k\}/c_{\varepsilon_p}\{2\}^k\approx c_{\epsilon_2}\{2k\}/c_{\epsilon_2}\{2\}^k\label{eq:cumulant_ratio}
\end{align}
We note that this observation has already been made for $v_2\{6\}/v_2\{4\}$ and $\epsilon_2\{6\}/\epsilon_2\{4\}$~\cite{Giacalone:2016eyu}, where in that case the authors mainly attribute the equivalence to the absence of a cubic response coefficient. In our discussion, a possible cubic response coefficient would be considered as a fluctuation of the response, and we refer to our companion paper~\cite{Ambrus:2024eqa} for a more detailed discussion.

\begin{figure*}
    \centering
    \includegraphics[width=.49\textwidth]{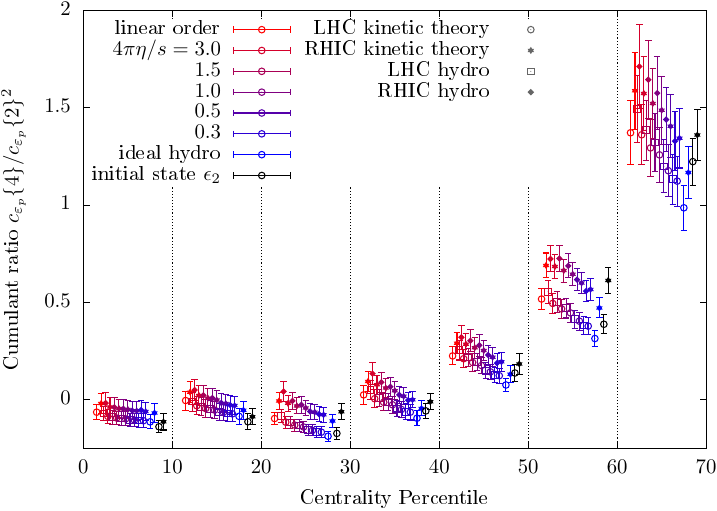}
    \includegraphics[width=.49\textwidth]{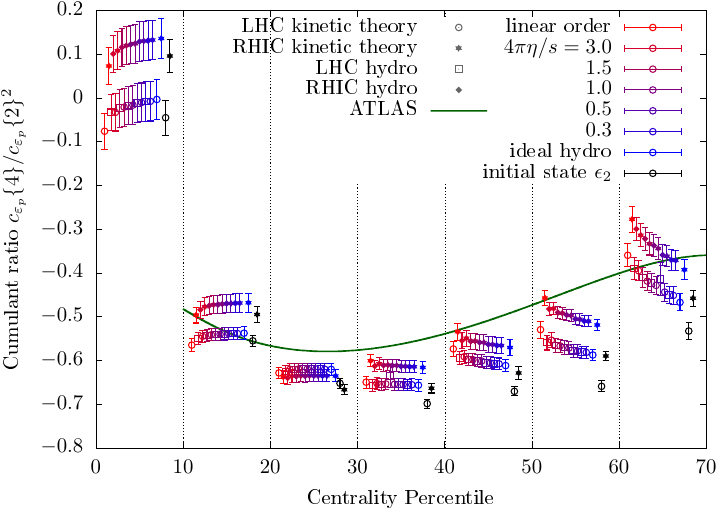}
    \includegraphics[width=.49\textwidth]{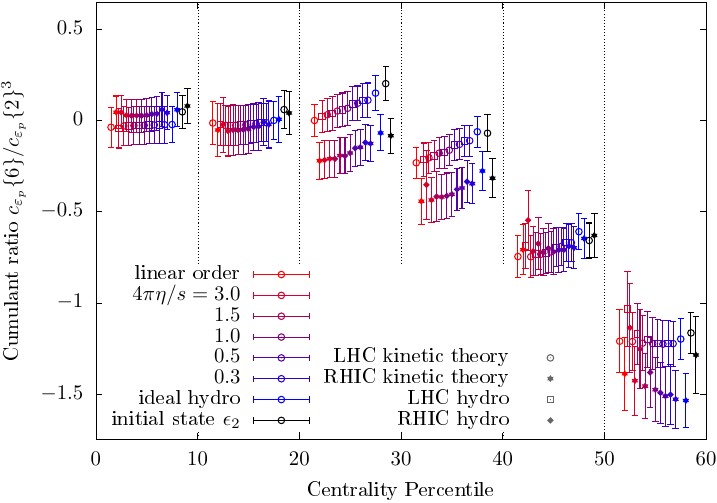}
    \includegraphics[width=.49\textwidth]{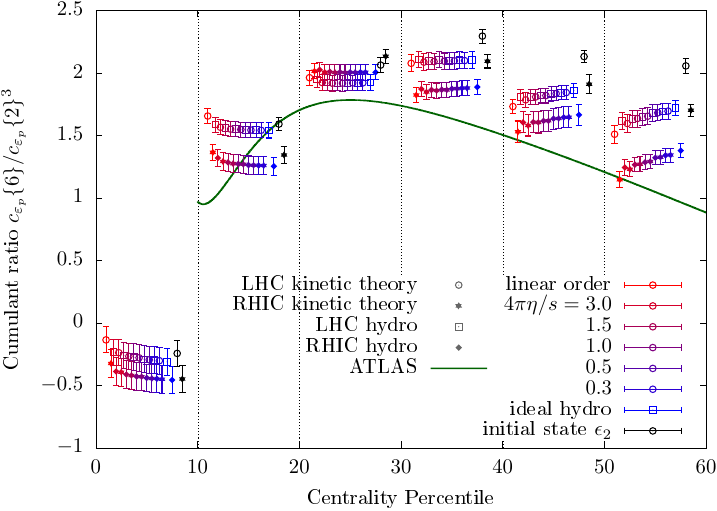}
    \includegraphics[width=.49\textwidth]{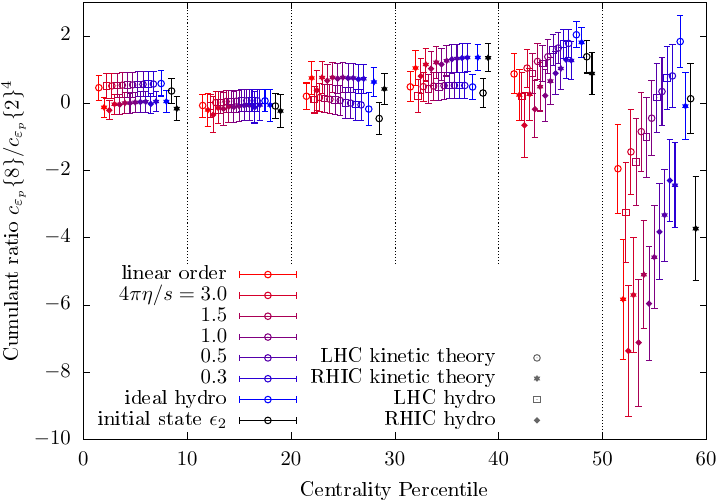}
    \includegraphics[width=.49\textwidth]{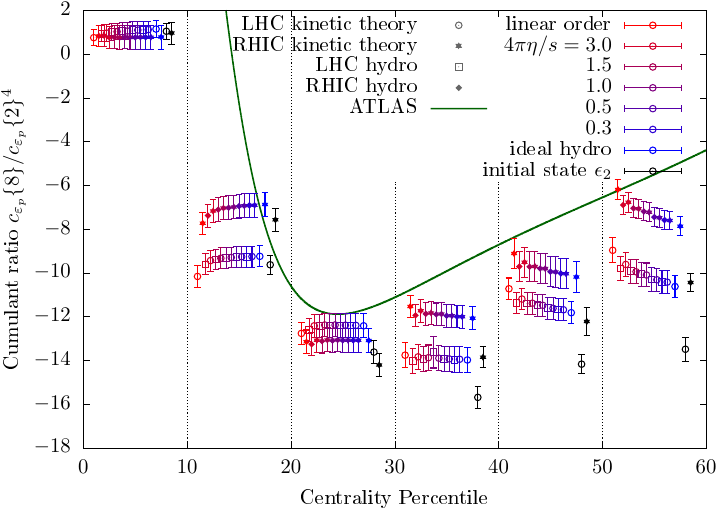}
    \caption{
    Flow cumulant ratios $c_{\varepsilon_p}\{4\}/c_{\varepsilon_p}\{2\}^2$ (top), $c_{\varepsilon_p}\{6\}/c_{\varepsilon_p}\{2\}^3$ (middle) and $c_{\varepsilon_p}\{8\}/c_{\varepsilon_p}\{2\}^4$ (bottom) for OO collisions (left) and PbPb/AuAu collisions (right) from LHC simulations in kinetic theory (open circles) and hydrodynamics (open squares) as well as RHIC simulations in kinetic theory (filled stars) and hydrodynamics (filled diamonds) for a range in shear viscosity from the linear order result (red) to ideal hydrodynamics (blue) and various values of $\eta/s$ in-between (color gradient from red to blue). We also compare to the same ratios of the initial state ellipticity (black) and a curve that was obtained by fitting to ATLAS data for the PbPb cumulants~\cite{ATLAS:2014qxy} (dark green). Error bars were obtained via jackknife resampling.}
    \label{fig:cumulant_ratio}
\end{figure*}

Nevertheless, we note that already the fact that the ideal hydrodynamic results for the cumulants in Figs.~\ref{fig:cumulants_vs_centrality_OOLHC}--\ref{fig:cumulants_vs_centrality_AuAuRHIC} closely follow the cumulants of the initial eccentricity is a priori nontrivial, and supports the hypothesis underlying  Eq.~\eqref{eq:moment_approximation}, as the mean response to the geometry is only one of several contributions to moments of the flow distribution, where others also contain event-by-event fluctuations of the flow response coefficient and their correlations with the event-by-event ellipticity. Hence, the close agreement in Figs.~\ref{fig:cumulants_vs_centrality_OOLHC}--\ref{fig:cumulants_vs_centrality_AuAuRHIC} is a validation of the assumption that those contributions are negligible and Eq.~\eqref{eq:moment_approximation} is a valid approximation. We note that we have checked that also beyond the ideal hydrodynamic limit the cumulant curves are in good agreement with the prediction from Eq.~\eqref{eq:moment_approximation}. Thus, the expectations we derived from this approximate equality should also hold.

We present a compact summary of our results for the cumulant ratios in Fig.~\ref{fig:cumulant_ratio}, which shows the ratios $c_{\varepsilon_p}\{4\}/c_{\varepsilon_p}\{2\}^2$ in the first row, $c_{\varepsilon_p}\{6\}/c_{\varepsilon_p}\{2\}^3$ in the second row and $c_{\varepsilon_p}\{8\}/c_{\varepsilon_p}\{2\}^4$ in the third row, all as a function of centrality.\footnote{We have restricted the range in centrality because results for more peripheral classes have excessively large error bars, most likely due to the fact that the denominator becomes very small. This is also the reason why we do not show ratios with a different cumulant in the denominator.} The first column of plots shows results for the OO collisions at RHIC and LHC, while the second column shows the results for AuAu collisions at RHIC and PbPb collisions at LHC. Different symbols in Fig.~\ref{fig:cumulant_ratio} indicate whether the results were obtained from the dynamical response in hydrodynamics or kinetic theory for a collision system at LHC (open symbols) or at RHIC (full symbols). Different colors correspond to the results obtained for different values of the specific shear-viscosity $\eta/s$, adopting the color coding from Figs.~\ref{fig:cumulants_vs_centrality_OOLHC}--\ref{fig:cumulants_vs_centrality_AuAuRHIC}, covering the whole range from the leading order opacity expansion to the ideal hydrodynamic limit. Black points finally show the corresponding ratios of cumulants of the initial state ellipticity for each system.

Strikingly, we find that the cumulant ratios exhibit almost no visible opacity dependence in all collision systems, indicating that indeed the final state response largely cancels from the ratio. Moreover, the ratios of flow cumulants are in close agreement with the ratios of cumulants of the initial state eccentricity, as in Eq.~\eqref{eq:cumulant_ratio}, demonstrating that these observables can indeed be used to probe the initial state geometry and constrain models of the initial state. 

Small deviations between the flow and eccentricity ratios appear in mid-central to peripheral PbPb and AuAu collisions. As alluded to above, this might be attributed to an extra cubic term on top of the linear relation between final state elliptic flow and initial state ellipticity~\cite{Noronha-Hostler:2015dbi,Giacalone:2016eyu}. 

Differences between RHIC and LHC are very minor for OO collisions, but become more pronounced when comparing PbPb collisions at LHC with AuAu collisions at RHIC, which should presumably be attributed to the fact that the collision geometries are similar but not exactly identical.

Even though, experimentally, elliptic flow cumulants have only been measured for particle weighted elliptic flow, it is nevertheless interesting to compare our results in PbPb collisions to experimental data from the ATLAS Collaboration~\cite{ATLAS:2014qxy}\footnote{ Since we were unable to obtain the original ATLAS data, we fitted sixth order polynomials to ATLAS data points for the cumulants as a function of the centrality and then took their ratio to obtain the depicted values. Since for more central collisions the cumulants become very small, our fit is no longer reliable, which is the reason that we chose to cut off the experimental curves at 10\% centrality.} shown as dark green curves in the right column of plots in Fig.~\ref{fig:cumulant_ratio}. Generally, we observe a satisfactory description of the experimental data, indicating that the initial state geometry of PbPb collisions  is reasonably well reproduced by the \trento ~initial state model.

\section{Nonconformal effects}\label{sec:non-conformal}

\begin{figure*}
    \centering
    \includegraphics[width=0.49\linewidth]{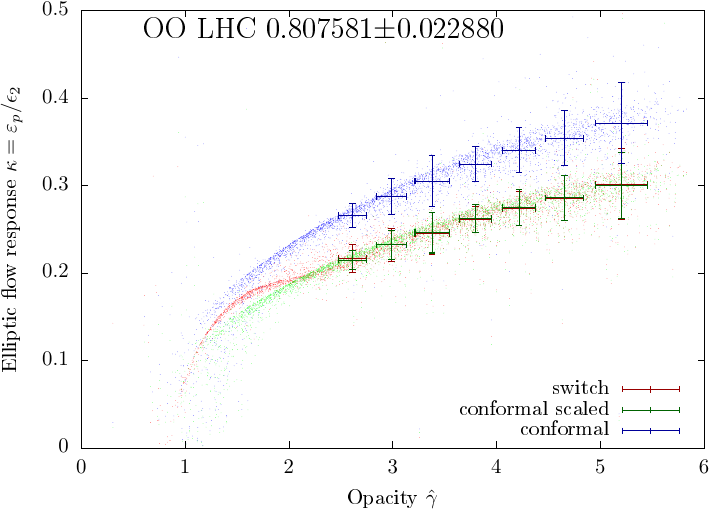}
    \includegraphics[width=0.49\linewidth]{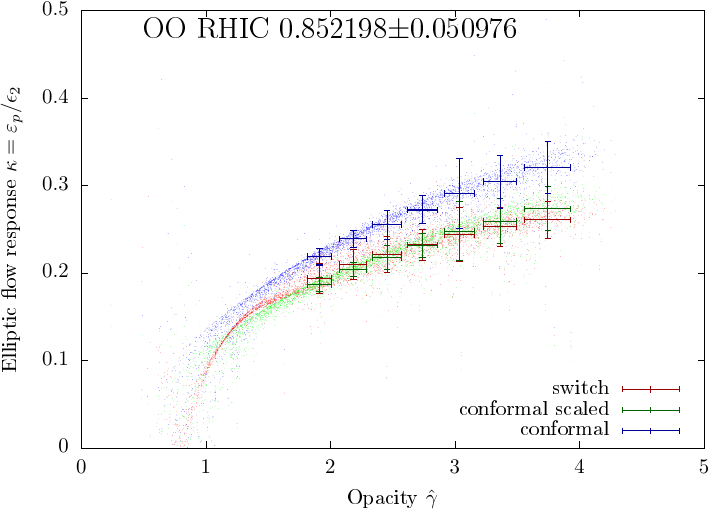}\\    
    \includegraphics[width=0.49\linewidth]{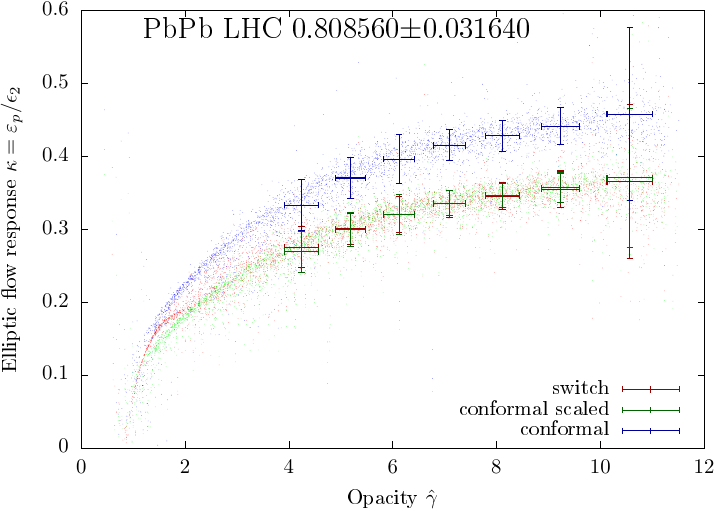}
    \includegraphics[width=0.49\linewidth]{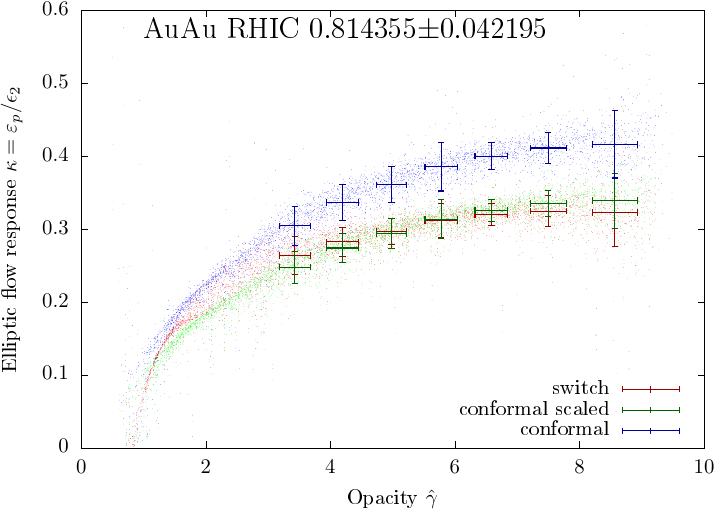}
    \caption{ 
    Point clouds of event-by-event flow responses as a function of event-by-event opacity in OO at LHC (top left) and RHIC (top right), as well as PbPb (bottom left) and AuAu (bottom right) from simulations in conformal hydrodynamics (blue) and nonconformal hydrodynamics (red) as well as conformal hydrodynamics scaled by a fitting factor (green) at a shear viscosity of $\eta/s=0.12$. Pluses show the corresponding means in centrality classes of size 10\% and their error bars indicate standard deviations of the corresponding quantity within each centrality class. The fitting factor is displayed at the top and was chosen to match conformal  with nonconformal hydrodynamics in centrality classes $<$70\%.}
    \label{fig:point_cloud_switch}
\end{figure*}

So far, we have scrutinized the development of collective flow in conformal kinetic theory and hydrodynamics with a conformal equation of state, constant specific shear viscosity and vanishing bulk viscosity, which were chosen to match the underlying kinetic description. Differences between the microscopic and macroscopic description emerge when dissipative effects become large; however, since real-world QCD is nonconformal, it is also important to assess the magnitude of nonconformal effects, which enter already at the level of ideal hydrodynamics via the nonconformal equation of state of QCD.

Clearly, extending the kinetic description to include nonconformal effects is a formidable task, which will require further dedicated efforts e.g. along the lines of~\cite{Bratkovskaya:2011wp,Sambataro:2022sns}. However, as a first step in this direction, a more straightforward way of improving the theoretical description and quantifying the magnitude of nonconformal effects is to include a more realistic equation of state in the hydrodynamic description. We thus also performed hydrodynamic simulations with a realistic QCD equation of state~\cite{Steinheimer:2010ib}, which is included in the vHLLE hydro code, and report on the results below.

\subsection{Setup for non-conformal hydrodynamics}
Before we comment on the results, some further remarks on the details of the nonconformal hydrodynamic simulations are in order. First of all, since the overall energy scale is important when considering a realistic QCD equation of state, we always rescaled the initial energy density profile by a centrality-dependent global factor to match the final state results for the transverse energy with those obtained from kinetic theory at $\eta/s=0.12$. In contrast to the conformal case, the presence of a nonvanishing trace of the energy-momentum tensor implies that the formula in Eq.~\eqref{eq:Eperp_def} has to be modified. Since, in experiment, transverse energy is extracted as the sum of transverse projections of the particles' total energies, according to
\begin{align}
    E_\perp=\sum_i E_i \sin(\theta_i)\;,
\end{align}
we define, in terms of the energy-momentum tensor, the following proxy for transverse energy in the nonconformal case:
\begin{align}
    \frac{\d E_{\perp,nc}}{\d\eta}&=\tau\int_\xT T^\mu_\mu +T^{xx}+T^{yy} \nonumber\\
    &=\tau\int_\xT T^{\tau\tau}-\tau^2T^{\eta\eta}\;.
\end{align}

Secondly, we recall that we had previously performed a local scaling of the hydrodynamic initial profile in order to correct for the difference between the conformal hydrodynamic and the conformal RTA kinetic theory early time attractors, allowing us to start the simulation at arbitrarily early times. However, with the QCD equation of state, the early time dynamics also changes in a way that cannot be predicted in a straightforward fashion. We chose to treat this issue by using the conformal equation of state for the early time evolution and then switching to the QCD equation of state at a fixed value of the evolution time in terms of the event-by-event rms radius of $\tau_{\rm switch}/R=0.1$. The switch in the equation of state is set up as follows. The equation of state describing the system is changed throughout the entire system instantaneously, once the evolution time exceeds the given threshold. In this procedure of changing the equation of state, we keep the four-velocity $u^\mu$, the local energy density $e = u_\mu u_\nu T^{\mu\nu}$, and the shear-stress tensor, $\pi^{\mu\nu} = T^{\langle \mu\nu \rangle}$ unchanged. What changes is the local thermodynamic pressure, from the conformal expression $P = e/ 3$ to $P_{\rm chiral} \equiv P_{\rm QCD}(e)$. This procedure is routinely used in more realistic frameworks when transitioning from the free-streaming pre-equilibrium model to the hydrodynamic model, in which case the difference $P - P_{\rm QCD}$ is stored as bulk viscous pressure, $\Pi$ \cite{ExTrEMe:2023nhy}. Since in our simulations, bulk viscous effects are not taken into account, we simply set $\Pi = 0$. To summarize the procedure described above, let us denote the energy-momentum tensor at position $\mathbf{x}$ and time $\tau = \tau_{\rm switch}$ before the switch via
\begin{equation}
 T^{\mu\nu}(\tau_{\rm switch}, \mathbf{x}) = e u^\mu u^\nu - P \Delta^{\mu\nu} + \pi^{\mu\nu}.
\end{equation}
After the switch, the energy momentum tensor becomes 
\begin{equation}
 T^{\mu\nu}_{\rm QCD}(\tau_{\rm switch}, \mathbf{x}) = e u^\mu u^\nu - P_{\rm QCD} \Delta^{\mu\nu} + \pi^{\mu\nu},
\end{equation}
where $P_{\rm QCD} \equiv P_{\rm QCD}(e)$ is determined from the QCD equation of state, based on the value of the energy density $e$ at point $\mathbf{x}$. Evidently this switch also comes with theoretical uncertainties, which are discussed further in App.~\ref{app:switching_problems}; however this setup allows us to initialize nonconformal hydrodynamics at the same initial time as conformal hydrodynamics with the same initial condition, up to the aforementioned global scaling factor. We then used a subset of 4000 events from the same ensemble of initial profiles, such that we can directly compare conformal and nonconformal hydrodynamic results for the same initial conditions.

\subsection{Numerical results}
In Fig.~\ref{fig:point_cloud_switch}, we show comparisons of the event-by-event flow response plotted against the event-by-event opacity at a shear viscosity of $\eta/s=0.12$ for OO at LHC in the upper left panel, OO at RHIC in the upper right panel, PbPb at LHC in the lower left panel and AuAu at RHIC in the lower right panel. Results from the conformal hydrodynamic case are plotted in blue and from the nonconformal case in red. By far the most obvious difference is an overall reduction of the magnitude of the flow response by $\sim 20\%$, which was already observed previously~\cite{Kurkela:2019kip} and can be understood as the reduced pressure resulting in weaker expansion.  In order to quantify this effect, we computed the mean of the event-by-event ratio of the nonconformal to conformal flow response for events with less than $70\%$ centrality, which is provided in the top row of each plot. When the conformal results are scaled by this ratio, we obtain the "conformal scaled points" that are plotted in green and in each case, we also show mean values of the flow response and opacity in each centrality class.

By carefully comparing the nonconformal and conformal scaled results one notices that in addition to the global scale difference, there is also a slight difference in the form of the point cloud. Nonconformal results in both LHC systems have a low opacity tail where flow results overshoot the conformal expectation. In the two RHIC systems, this behaviour is seen already for mid-central collisions, causing also the mean scale difference between the two systems to be higher than in the LHC case, such that results from central events fall below the scaled conformal expectation, while more peripheral events are above, such that there are noticeable deviations from the scaling by a constant factor. If we compute the ratio between conformal and nonconformal results using only events with opacity $\hat{\gamma}> 3$, then the factors for the LHC systems remain the same, while the RHIC factors move closer to the LHC ones, specifically they change to $0.8197(25)$ for OO and $0.8097(39)$ for AuAu. 

\begin{figure}
    \centering
    \includegraphics[width=0.99\linewidth]{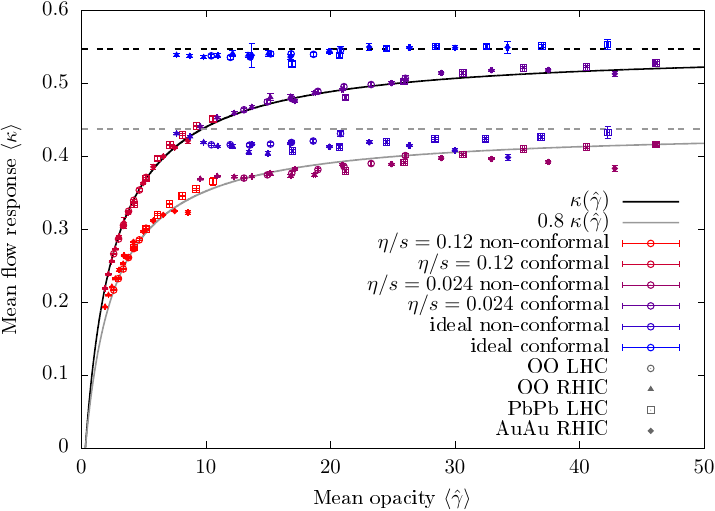}
    \caption{Mean values of the response coefficient $\kappa=\varepsilon_p/\epsilon_2$ as a function of the mean opacity (arbitrarily scaled for ideal hydrodynamics) for centrality classes of size 10\% from conformal kinetic theory and nonconformal hydrodynamic (different colors) simulations of OO collisions at LHC (empty circles) and RHIC (filled triangles) as well as PbPb collisions (empty squares) and AuAu collisions (filled diamonds) at different values of the shear viscosity (different colors) compared to the conformal kinetic theory flow response curve $\kappa(\hat{\gamma}$ (black solid) and its ideal hydrodynamic limit (black dashed) as well as a rescaling of the same curve by a factor of $0.8$ (grey solid and dashed).}
    \label{fig:kappa_fit_switch}
\end{figure}

We further analyze the effects of a nonconformal equation of state in Fig.~\ref{fig:kappa_fit_switch}, where we compare the opacity dependence of the average response coefficients in conformal and nonconformal hydrodynamics for OO, AuAu and PbPb collisions at RHIC and LHC at two different values of the specific shear viscosity.  Since the nonconformal equation of state introduces an additional dependence on the overall energy scale of the collision, we find that for the nonconformal equation of state, the response of individual systems shows a somewhat larger spread. While at LHC energies the effects of a nonconformal equation of state can actually be rather well described by a constant scaling $\sim 0.8$ of the conformal response, at RHIC energies the nonconformal equation of state introduces an additional centrality dependence of the flow response, where more central/peripheral collisions exhibit a somewhat stronger/weaker suppression. Notably, this difference can also already be observed for ideal hydrodynamics, where nonconformal results at RHIC energies show a stronger centrality dependence than at LHC energies.

\section{Conclusion}\label{sec:conclusion}

We extended earlier investigations of the (in)applicability of hydrodynamics in small systems~\cite{Ambrus:2022koq,Ambrus:2022qya}, to study the development of collective flow based on event-by-event initial simulations of PbPb and AuAu collisions and most importantly OO collisions at RHIC and LHC, which are expected to cover the transition regime to hydrodynamic behavior. By comparing simulation results from a macroscopic description in hydrodynamics to a more microscopic description in conformal kinetic theory, we assessed the impact of non-equilibrium effects beyond hydrodynamics.

We discussed the geometrical properties of our ensembles of initial states generated using the \trento -code with sets of precalculated nucleon positions in the colliding nuclei. When compared to AuAu and PbPb, fluctuations in OO collisions create a larger spread of geometric quantities in central multiplicity classes, most notably in the initial ellipticity $\epsilon_2$. The less dominant role of the mean geometry also eliminates some correlations between initial state observables that are present in AuAu and PbPb.

Strikingly, we find that the event-by-event elliptic flow $\varepsilon_p$ in conformal RTA kinetic theory and hydrodynamics exhibits a remarkable degree of universality, where results for the flow response coefficient $\kappa=\varepsilon_p/\epsilon_2$ in different systems closely line up along a universal curve, characterizing the flow response as a function of the opacity $\hat{\gamma}$, which encodes the dependence on the systems size, energy scale and QCD transport properties. By comparing the collective flow response in conformal kinetic theory and hydrodynamics, we find that hydrodynamics provides an accurate description of flow for large collision systems up to peripheral centrality classes, but shows deviations in small systems, which restricts its range of applicability to opacities $\hat{\gamma}\gtrsim 3$. Specifically, for OO collision at RHIC and LHC,  the sensitivity to the underlying microscopic dynamics is typically at the $10\%$ level, indicating that a precise understanding of the initial state geometry is required to probe the nonequilibrium dynamcis of the QGP beyond hydrodynamics, and we refer to our companion paper~\cite{Ambrus:2024eqa} for further discussions.

When considering the centrality dependence of cumulants of the elliptic flow, we found that there are two competing effects dictating the shape of these curves. One is the centrality dependence of the mean initial state ellipticity which as a function of centrality first increases, then peaks in the mid-central to peripheral regime and then drops off. The other is the centrality dependence of the opacity causing more central events to have a larger flow response. Interestingly, we find that in OO collisions, the competition of the two effects causes the second order flow cumulant to vary only little with centrality up to peripheral classes. Another important finding was that within the centrality classes, moments of the elliptic flow distribution can be approximated quite well by multiplying moments of the initial ellipticity distribution with the mean flow response in the respective centrality class. This factorization motivates the construction of further observables. Our results show that appropriate ratios of flow cumulants become entirely insensitive to the dynamical response and directly probe the initial state geometry.

Finally, we performed a first test of how our conclusions might be affected when going towards more realistic descriptions by including nonconformal effects. We found that at high energies -- like at LHC --  conformal and nonconformal elliptic flow results mostly differ by a constant factor of $0.8$, which would not affect our discussion. However, at RHIC energies an additional centrality dependent effect is present, which introduces a spread around the universal flow response curve.

\begin{acknowledgments}
This work is supported by the Deutsche Forschungsgemeinschaft (DFG, German Research Foundation)
through the CRC-TR 211 ’Strong-interaction matter under extreme conditions’– project
number 315477589 – TRR 211.
V.E.A. gratefully acknowledges support by the European Union - NextGenerationEU through grant No. 760079/23.05.2023, funded by the Romanian ministry of research, innovation and digitization through Romania’s National Recovery and Resilience Plan, call no.~PNRR-III-C9-2022-I8.
C.W. was supported by the program Excellence Initiative–Research University of the University of Wrocław of the Ministry of Education and Science.
C.W. has also received funding from the European Research Council (ERC) under the European Union’s Horizon 2020 research and innovation programme (grant number: 101089093 / project acronym: High-TheQ). Views
and opinions expressed are however those of the authors
only and do not necessarily reflect those of the European
Union or the European Research Council. Neither the
European Union nor the granting authority can be held
responsible for them.
Numerical calculations presented in this work were performed at the Paderborn Center for Parallel Computing (PC2), the Center for Scientific Computing (CSC) at the Goethe-University of Frankfurt and 
the National Energy Research Scientific Computing Center (NERSC), a Department of Energy Office of Science User Facility, under the group name m2443 and we gratefully acknowledge their support. 
\end{acknowledgments}

\appendix

\section{LOebe}\label{app:LOebe}

The general setup of the Linear Order event-by-event (LOebe) code computing results for flow observables from an input of an initial energy density profile to linear order in opacity $\hat{\gamma}$ has already been laid out in our previous publication~\cite[App. B]{Ambrus:2022koq}. However, the earlier version of the code solved the relevant integrals by Monte Carlo sampling of an integrand that was computed from an interpolated initial condition and therefore required several hours of runtime for a single result. In order to produce event-by-event results, we changed the integration method to improve the runtime by a factor $\mathcal{O}(1000)$. 

Given an initial energy density profile $e(\tau_0,\xT)$, we need to compute the first order correction to transverse integrals of the spatial parts of the energy momentum tensor, which enters elliptic flow $\varepsilon_p$ [c.f. Eq.~\eqref{eq:epsp_def}] and transverse energy $\d E_\perp / \d\eta$ (cf. Eq.~\eqref{eq:Eperp_def}). At time $\tau$, they are obtained as
\begin{align}
\int_{\xT} T^{(1)ij} &= \frac{1}{\tau}\int_{\tau_0}^{\tau} d{\tau'} \int_{\xT} \frac{1}{\tau_R} \Bigg[
\tau'e^{(0)}
F^{ij}_{\rm eq} -
\tau_0 F_0^{ij}\Bigg]\;,
\label{eq:LO_T1ij}\\
F_0^{ij}&= \gamma\int \frac{d\phi_p}{2\pi}  e(\tau_0, \xT - \vT \Delta \tau') \nonumber\\
&\quad\times\mathbf{v}_\perp^i \mathbf{v}_\perp^j \left(1 - \vT \cdot \bm{\beta}\right) \;,\label{eq:LO_F0ij}
\end{align}
where $\vT=(\cos(\phi_p),\sin(\phi_p))$ and the $\gamma$- and  $\bm{\beta}$-factors of the zeroth order collective flow velocity as well as the relaxation time $\tau_R$, the zeroth order energy density $\epsilon^{(0)}$ and the equilibrium contribution $F^{ij}_{\rm eq}$ are evaluated at $(\tau', \xT)$. Since the computation of zeroth order quantities, as well as $F_{\rm eq}^{ij}$, remain unchanged, we refer for their precise form to our previous publication \cite{Ambrus:2022koq}.

The new integration method computes the integrals appearing in Eq. \eqref{eq:LO_T1ij}, as well as the $\phi_p$-integral in Eq. \eqref{eq:LO_F0ij}, one-by-one, based on two major simplifications. The first is the realization that assuming a bilinear interpolation of an initial energy density distribution given on a transverse grid, the contribution to the $\phi_p$-integral from each cell in transverse space can be computed analytically, turning the integral into a sum of contributions from all relevant cells. The second is the choice to compute this $\phi_p$-integral on a grid of transverse positions $\xT$ that is identical to the one that the initial condition was provided in. This means that the integral over $\xT$ becomes a sum over grid points. As we will see, this also introduces another simplification to the $\phi_p$-integration. Finally, the integral over $\tau'$ is computed via a trapezoid rule. In this case, there is a significant difference to a simple Riemann sum because the times $\tau'$ at which the integrand is evaluated are not equidistant. They are chosen to be denser in regions with bigger contributions.

The only remaining problem is the evaluation of the $\phi_p$-integral, which will now be explained. The integration procedure is to be repeated for each time $\tau'$ and each position in transverse space $\xT$ appearing in the discretization of the respective integrals in \eqref{eq:LO_F0ij}. The $\phi_p$-integral can be understood as a path at initial time along the circle consisting of points with lightlike distance to the considered position $\xT$ with a time difference $\Delta\tau'=\tau'-\tau_0$. This integral is computed in the following three steps.

\textbf{1. Identify cells in the initial state grid that intersect with the circle}

The algorithm starts in positive x-direction from the center of the circle, i.e. from the cell whose lower left corner is the grid point at $\xT+(a\lfloor\Delta\tau'/a\rfloor,0)$, where $a$ is the grid spacing, and follows the circle counterclockwise. The simplification coming from the grid in $\xT$ being chosen to be identical to the grid of the initial condition is that the circle center is itself a grid point. In the following we will write $\xT=(x_C,y_C)$. Now it can easily be determined through which of the sides the circle exits the cell by checking which of the cell corners lie inside the circle.  The next cell to check is the one in the direction in which the circle exited the previous cell. This procedure is repeated until the algorithm arrives back at the starting point.

\textbf{2. Find the angles at which the circle enters and exits each cell}

When the circle enters or exits a cell, one of the cartesian coordinates is exactly equal to a grid coordinate. Now we can draw a right-angled triangle with the difference of this coordinate to that of the circle center being a cathetus and the radius of the circle $\Delta\tau$ being the hypotenuse. The corresponding angle can then be determined via a fitting branch of the $\arccos$ or $\arcsin$ function. The correct branch can be determined via the signs of the differences $x-x_C$ and $y-y_C$.

\textbf{3. Compute the integral contribution from each cell and sum}

We now consider the contribution of a single cell. We denote the coordinates of its corners by $x_0<x_1$ and $y_0<y_1$ and the values of the input energy density profile as
\begin{align}
    e_{00}&=e(\tau_0,x_0,y_0)\;,&\quad e_{10}&=e(\tau_0,x_1,y_0)\;,\nonumber\\
    \quad e_{01}&=e(\tau_0,x_0,y_1)\;,&\quad e_{11}&=e(\tau_0,x_1,y_1)\;.
\end{align}

The bilinear interpolation of $e(\tau',\xT-\vT \Delta\tau')=e(\tau_0,x_C-\Delta\tau'\mathrm{cos}(\phi_p),y_C-\Delta\tau'\mathrm{sin}(\phi_p))$ on the part of the circle that intersects the cell is then given by the following expression.
\begin{align}
    e(\tau',\xT-\vT \Delta\tau')=\qquad\qquad\qquad\qquad\qquad\qquad\qquad\quad\nonumber\\
    \quad\frac{1}{a^2}\Big[e_{00}~(x_1-x_c+\Delta\tau'\cos(\phi_p))(y_1-y_c+\Delta\tau'\sin(\phi_p))\nonumber\\
    +e_{01}~(x_1-x_c+\Delta\tau'\cos(\phi_p))(y_c-\Delta\tau'\sin(\phi_p)-y_0)\nonumber\\
    +e_{10}~(x_c-\Delta\tau'\cos(\phi_p)-x_0)(y_1-y_C+\Delta\tau'\sin(\phi_p)\nonumber\\
    +e_{11}~(x_c-\Delta\tau'\cos(\phi_p)-x_0)(y_C-\Delta\tau'\sin(\phi_p)-y_0)\Big]
\end{align}

Now it becomes clear that the integrand in the contributions to Eq. \eqref{eq:LO_F0ij} from this cell are sums of terms containing different powers of $\cos(\phi_p)$ and $\sin(\phi_p)$. In order to express our result for the $\phi_p$-integral, we define
\begin{align}
    I_{nm}=\int_{\phi_{\rm min}}^{\phi_{\rm max}}\frac{\d \phi_p}{2\pi} \sin^n(\phi_p)\cos^m(\phi_p)\;,
\end{align}
where the angles $\phi_{\rm min}$ and $\phi_{\rm max}$ are where the circle enters and exits the cell. These integrals can easily be computed analytically. Now, the contributions $F_0^{ij}(\phi_{\rm min},\phi_{\rm max})$ to the three components $F_0^{ij}$ coming from this cell are given as follows. The total quantities $F_0^{ij}$ are obtained as a sum of the $F_0^{ij}(\phi_{\rm min},\phi_{\rm max})$ from all cells that the circle crosses.

\begin{widetext}
\begin{align}
    F_0^{11}(\phi_{\rm min},\phi_{\rm max})&=\gamma\int_{\phi_{\rm min}}^{\phi_{\rm max}}\frac{\d \phi_p}{2\pi} \cos^2(\phi_p)(1-\beta_1\cos(\phi_p)-\beta_2\sin(\phi_p)) e(\tau',\xT-\vT \Delta\tau')\\
    &=\frac{\gamma}{2\pi a^2}\Big\lbrace(I_{02}-\beta_1I_{03}-\beta_2I_{12})[e_{00}(x_1-x_C)(y_1-y_C)+e_{01}(x_1-x_C)(y_C-y_0)\nonumber\\
    &\qquad\qquad\qquad\qquad\qquad\qquad+e_{10}(x_C-x_0)(y_1-y_C)+e_{11}(x_C-x_0)(y_C-y_0)]\nonumber\\
    &\qquad\qquad +\Delta\tau'(I_{12}-\beta_1I_{13}-\beta_2I_{22})[(e_{00}-e_{01})(x_1-x_C)+(e_{10}-e_{11})(x_C-x_0)]\nonumber\\
    &\qquad\qquad +\Delta\tau'(I_{03}-\beta_1I_{04}-\beta_2I_{13})[(e_{00}-e_{10})(y_1-y_C)+(e_{01}-e_{11})(y_C-y_0)]\nonumber\\
    &\qquad\qquad +\Delta\tau'^2(I_{13}-\beta_1I_{14}-\beta_2I_{23})[e_{00}-e_{01}-e_{10}+e_{11})] \Big\rbrace \nonumber\\
    F_0^{12}(\phi_{\rm min},\phi_{\rm max})&=\gamma\int_{\phi_{\rm min}}^{\phi_{\rm max}}\frac{\d \phi_p}{2\pi} \cos(\phi_p)\sin(\phi_p)(1-\beta_1\cos(\phi_p)-\beta_2\sin(\phi_p)) e(\tau',\xT-\vT \Delta\tau')\\
    &=\frac{\gamma}{2\pi a^2}\Big\lbrace(I_{11}-\beta_1I_{12}-\beta_2I_{21})[e_{00}(x_1-x_C)(y_1-y_C)+e_{01}(x_1-x_C)(y_C-y_0)\nonumber\\
    &\qquad\qquad\qquad\qquad\qquad\qquad+e_{10}(x_C-x_0)(y_1-y_C)+e_{11}(x_C-x_0)(y_C-y_0)]\nonumber\\
    &\qquad\qquad +\Delta\tau'(I_{21}-\beta_1I_{22}-\beta_2I_{31})[(e_{00}-e_{01})(x_1-x_C)+(e_{10}-e_{11})(x_C-x_0)]\nonumber\\
    &\qquad\qquad +\Delta\tau'(I_{12}-\beta_1I_{13}-\beta_2I_{22})[(e_{00}-e_{10})(y_1-y_C)+(e_{01}-e_{11})(y_C-y_0)]\nonumber\\
    &\qquad\qquad +\Delta\tau'^2(I_{22}-\beta_1I_{23}-\beta_2I_{32})[e_{00}-e_{01}-e_{10}+e_{11})] \Big\rbrace \nonumber\\
    F_0^{22}(\phi_{\rm min},\phi_{\rm max})&=\gamma\int_{\phi_{\rm min}}^{\phi_{\rm max}}\frac{\d \phi_p}{2\pi} \sin^2(\phi_p)(1-\beta_1\cos(\phi_p)-\beta_2\sin(\phi_p)) e(\tau',\xT-\vT \Delta\tau')\\
    &=\frac{\gamma}{2\pi a^2}\Big\lbrace(I_{20}-\beta_1I_{21}-\beta_2I_{30})[e_{00}(x_1-x_C)(y_1-y_C)+e_{01}(x_1-x_C)(y_C-y_0)\nonumber\\
    &\qquad\qquad\qquad\qquad\qquad\qquad+e_{10}(x_C-x_0)(y_1-y_C)+e_{11}(x_C-x_0)(y_C-y_0)]\nonumber\\
    &\qquad\qquad +\Delta\tau'(I_{30}-\beta_1I_{31}-\beta_2I_{40})[(e_{00}-e_{01})(x_1-x_C)+(e_{10}-e_{11})(x_C-x_0)]\nonumber\\
    &\qquad\qquad +\Delta\tau'(I_{21}-\beta_1I_{22}-\beta_2I_{31})[(e_{00}-e_{10})(y_1-y_C)+(e_{01}-e_{11})(y_C-y_0)]\nonumber\\
    &\qquad\qquad +\Delta\tau'^2(I_{31}-\beta_1I_{32}-\beta_2I_{41})[e_{00}-e_{01}-e_{10}+e_{11})] \Big\rbrace \nonumber
\end{align}
\end{widetext}

\section{Flow cumulants at fixed final state transverse energy}\label{app:scaling_details}

\begin{figure*}
    \centering
    \includegraphics[width=.49\textwidth]{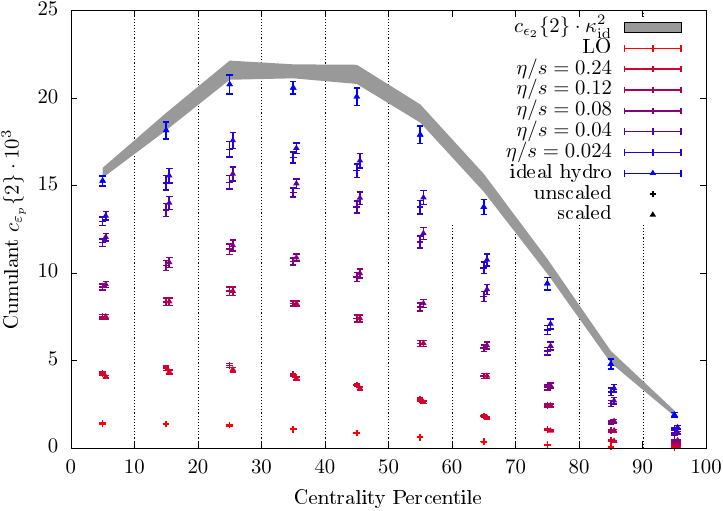}
    \includegraphics[width=.49\textwidth]{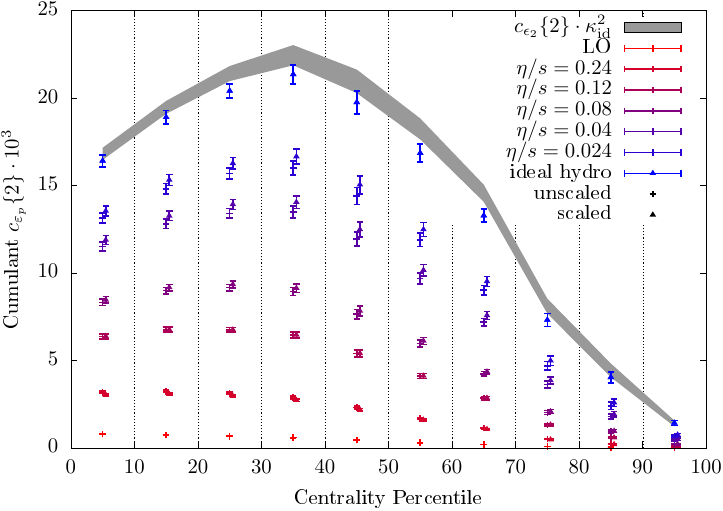}
    \includegraphics[width=.49\textwidth]{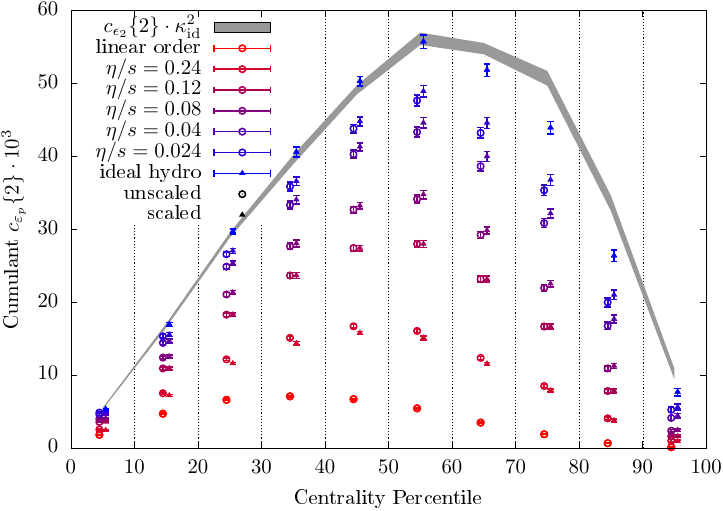}
    \includegraphics[width=.49\textwidth]{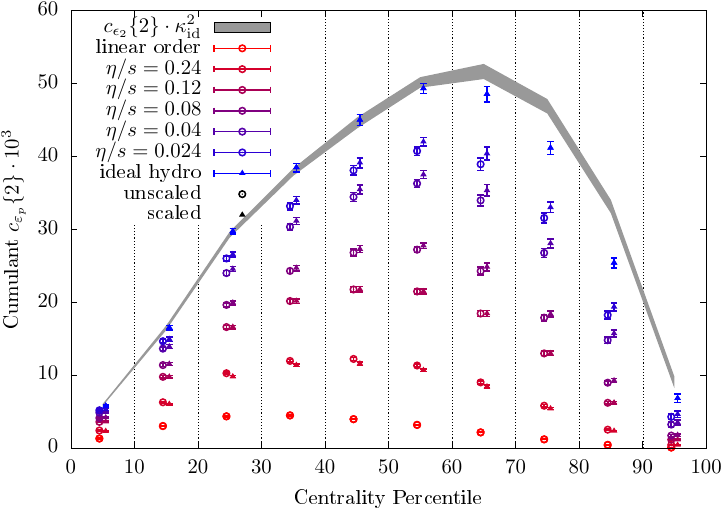}
    \caption{Second order elliptic flow cumulants $c_{\varepsilon_p}\{2\}$ as a function of centrality classes of OO collisions at LHC (top left) and RHIC (top right) as well as PbPb collisions (bottom left) and AuAu collisions (bottom right) from simulations in kinetic theory with unchanged energy (pluses) and scaled energy to match experimental results (triangles) covering a range in shear viscosity from the linear order result (red) to ideal hydrodynamics (blue) and various values of $\eta/s$ in between (color gradient from red to blue). Scaled results for the two limiting cases are trivial and therefore omitted. Error bars were obtained via jackknife resampling. Also shown are cumulants $c_{\epsilon_2}\{2\}$ of the initial state ellipticity scaled by a power of the mean ideal hydrodynamic flow response coefficient $\kappa_{\rm id}^{2k}$ (grey bands).}
    \label{fig:scaled_cumulants_vs_centrality_second}
\end{figure*}

\begin{figure*}
    \centering
    \includegraphics[width=.49\textwidth]{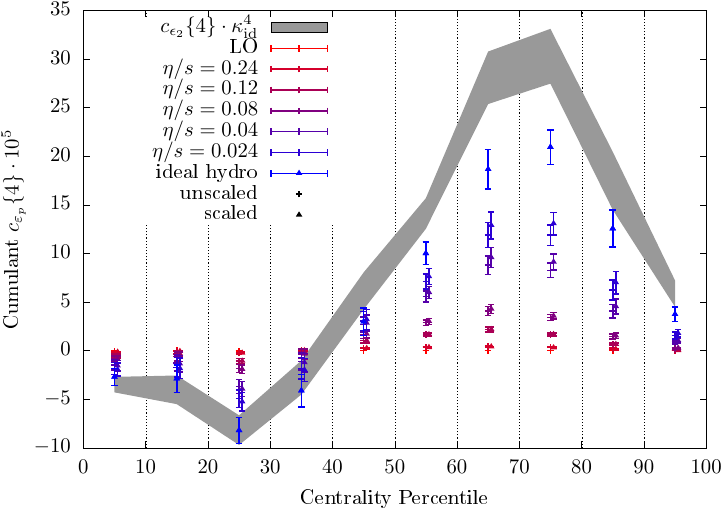}
    \includegraphics[width=.49\textwidth]{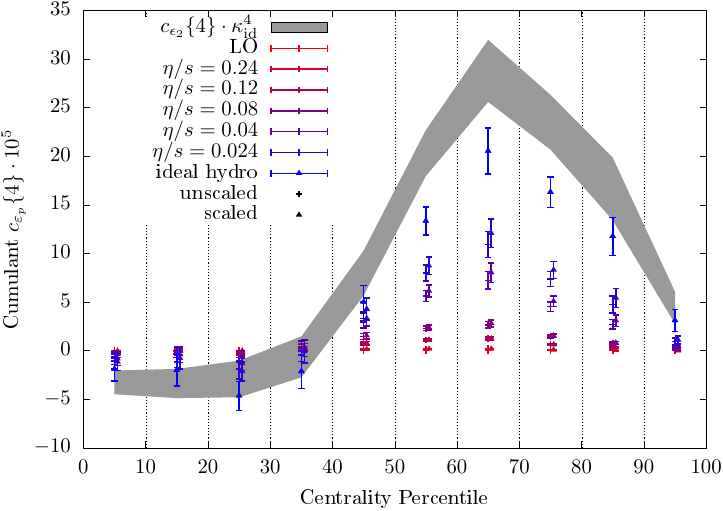}
    \includegraphics[width=.49\textwidth]{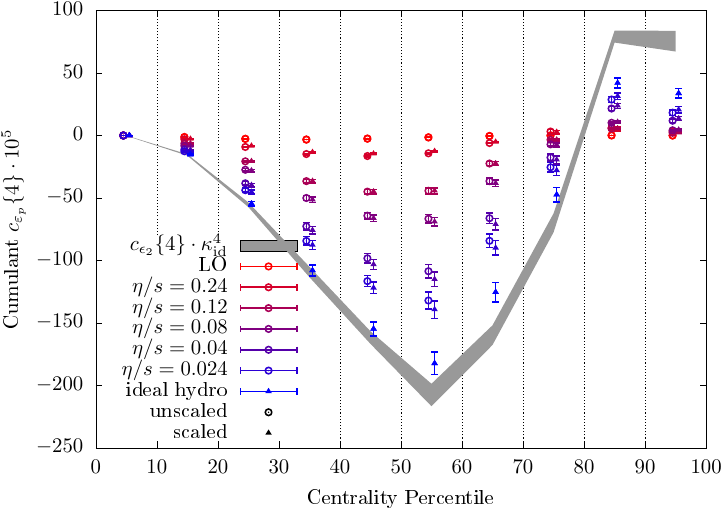}
    \includegraphics[width=.49\textwidth]{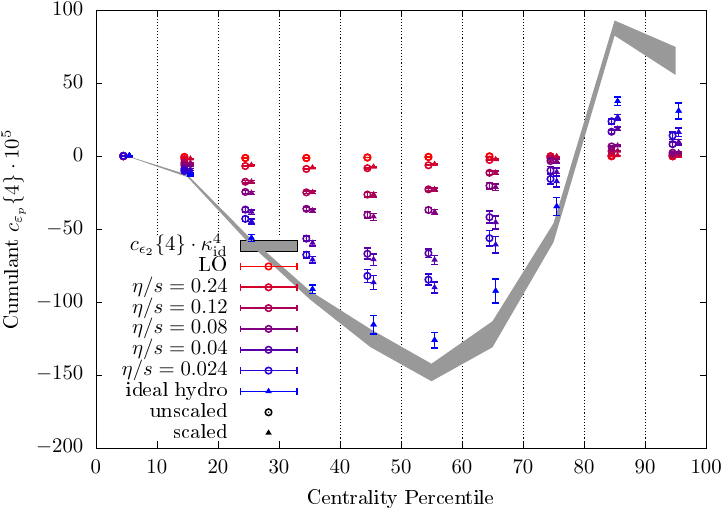}
    \caption{
    Same as Fig~\ref{fig:scaled_cumulants_vs_centrality_second}, but for the fourth order elliptic flow cumulant $c_{\varepsilon_p}\{4\}$.}
    \label{fig:scaled_cumulants_vs_centrality_fourth}
\end{figure*}

\begin{figure*}
    \centering
    \includegraphics[width=.49\textwidth]{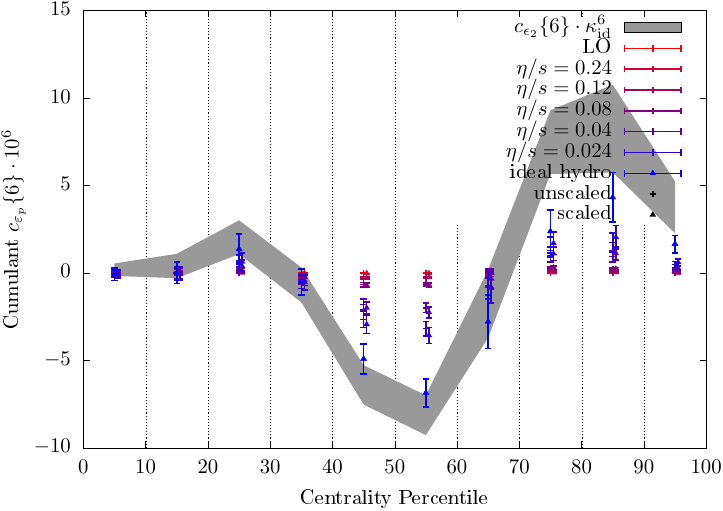}
    \includegraphics[width=.49\textwidth]{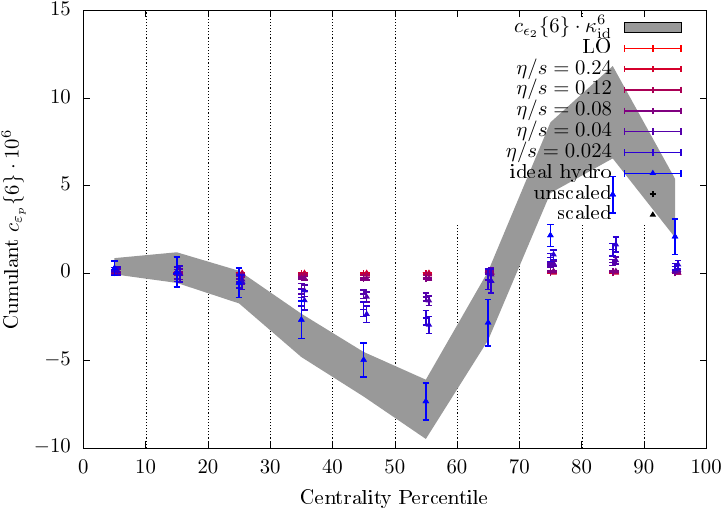}
    \includegraphics[width=.49\textwidth]{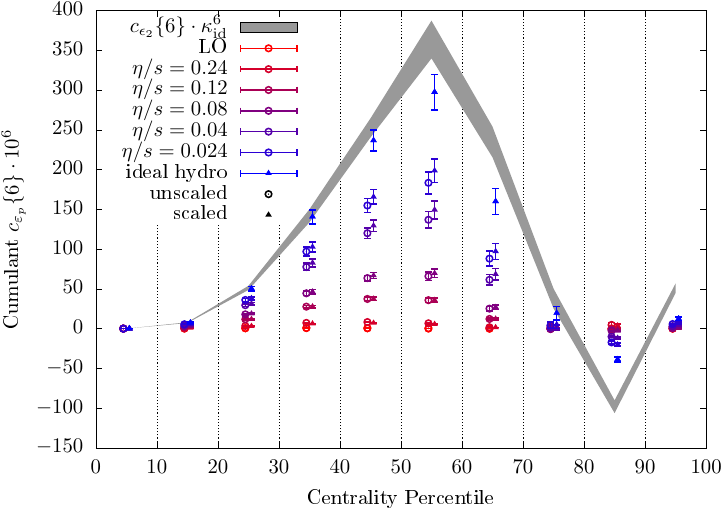}
    \includegraphics[width=.49\textwidth]{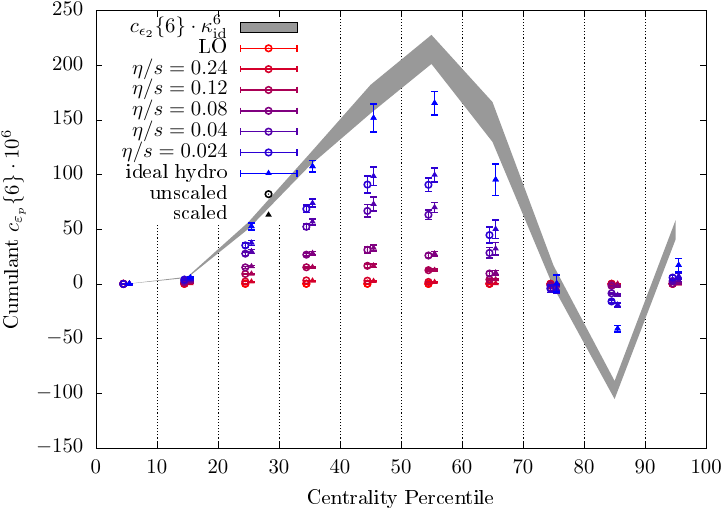}
    \caption{
    Same as Fig~\ref{fig:scaled_cumulants_vs_centrality_second}, but for the sixth order elliptic flow cumulant $c_{\varepsilon_p}\{6\}$.}
    \label{fig:scaled_cumulants_vs_centrality_sixth}
\end{figure*}

\begin{figure*}
    \centering
    \includegraphics[width=.49\textwidth]{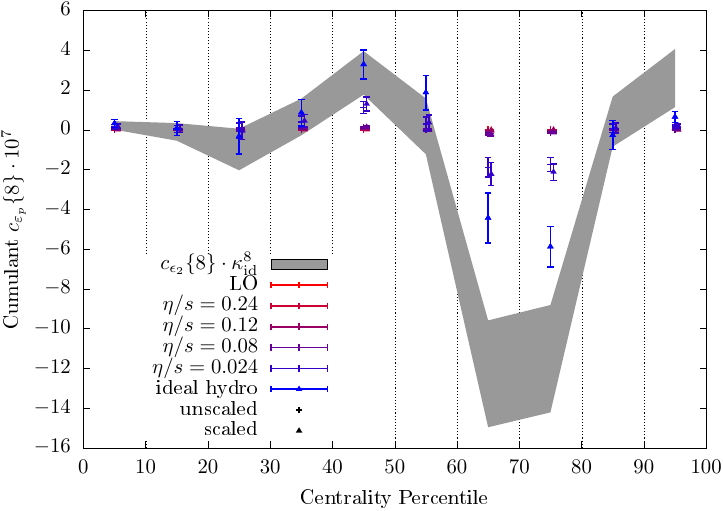}
    \includegraphics[width=.49\textwidth]{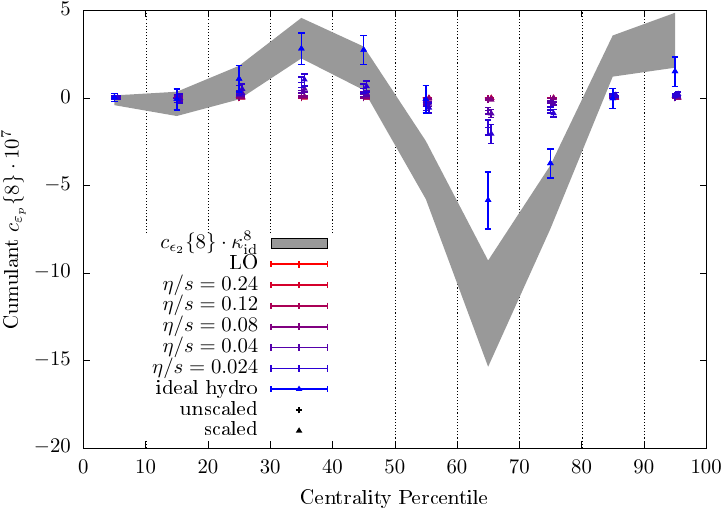}
    \includegraphics[width=.49\textwidth]{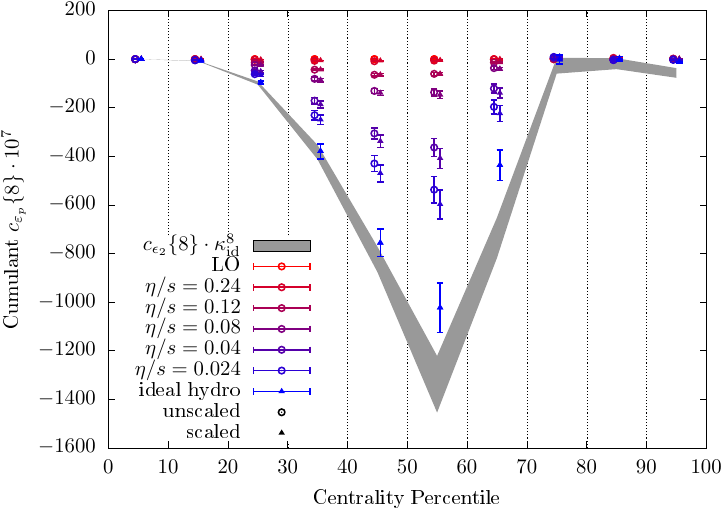}
    \includegraphics[width=.49\textwidth]{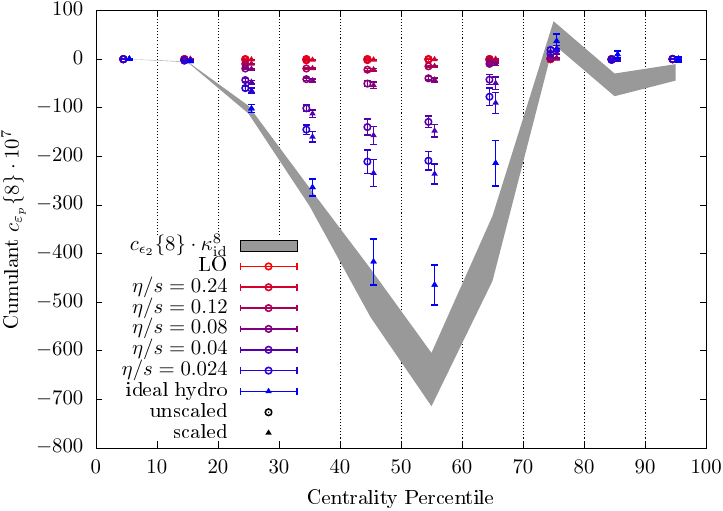}
    \caption{
    Same as Fig~\ref{fig:scaled_cumulants_vs_centrality_second}, but for the eighth order elliptic flow cumulant $c_{\varepsilon_p}\{8\}$.}
    \label{fig:scaled_cumulants_vs_centrality_eighth}
\end{figure*}

For most of our simulation results, we used the same initial conditions for different values of $\eta/s$. For a more realistic setup, the overall normalization should be adjusted in order to match our final state energy to experimental data. From a theoretical point of view, this is not problematic, if we view this as a scan of the opacity dependence at fixed initial state geometry. To the dynamics in conformal RTA, both the shear viscosity $\eta/s$ and the initial energy scale $\epni$ only matter via their influence in the opacity $\hat{\gamma}$, and varying one is equivalent to varying the other. Since $\hat{\gamma}$ depends on $\epni$ only via its fourth root, scaling the initial condition to match the final state energy actually has little effect. This is what we want to demonstrate now by comparing the scaled and unscaled cases.

These results were obtained not from additional simulations but from a well-motivated extrapolation procedure from the results for fixed initial transverse energy that were discussed in Section~\ref{sec:elliptic} in the following way. Since we have demonstrated already that the work and flow response curves (see Appendix A of our companion paper~\cite{Ambrus:2024eqa}) are universal in conformal RTA, instead of rerunning the simulations with varying initial state normalizations until the final state energy matches, we simply extrapolate the scaled results from the data we have already obtained. For each collision system and value of the shear viscosity, we determine a single global scaling parameter to match the global mean transverse energy. We first use the $f_{\rm work}(\hat{\gamma})=\Big\langle\frac{\d E_\perp/\d\eta}{\d E_\perp^{(0)}/\d\eta}\Big\rangle$ curve to determine what scaling factor would be needed for the event-by-event initial energy $\epni$ in order to arrive at the desired mean final state energy $\d E_\perp / \d\eta$. Due to the dependence of $\hat{\gamma}$ on $\epni$, this is an implicit equation and we find the solution numerically by iteration. Then we compute for each $\eta/s$ the change in the event-by-event flow response due to the scaling of the initial condition according to the flow response curve $\kappa(\hat{\gamma})$. The cumulants are then computed for the set of rescaled event-by-event flow responses that were obtained in this way. 

For the results obtained in this way we present the second order flow cumulants $c_{\varepsilon_p}\{2\}$ in Fig.~\ref{fig:scaled_cumulants_vs_centrality_second}, the fourth order flow cumulants $c_{\varepsilon_p}\{4\}$ in Fig.~\ref{fig:scaled_cumulants_vs_centrality_fourth}, the sixth order flow cumulants $c_{\varepsilon_p}\{6\}$ in Fig.~\ref{fig:scaled_cumulants_vs_centrality_sixth} and the eighth order flow cumulants $c_{\varepsilon_p}\{8\}$ in Fig.~\ref{fig:scaled_cumulants_vs_centrality_eighth}. In each case, the top left plot shows OO at LHC, the top right one shows OO at RHIC, the bottom left one shows PbPb and the bottom right one shows AuAu. The scaled results plotted with triangles are compared to the unscaled results plotted with pluses. Results for different values of the shear viscosity $\eta/s$ are plotted in different colors. No scaled result was computed in the cases of ideal hydrodynamics and opacity linearization. In the former case scaling does not change the result as the response coefficient is the same and in the latter the final state energy, while technically not fully determined, will not change much. Comparing the scaled and unscaled results, for $\eta/s=0.12$ they remain unchanged, since the transverse energy was already matching. For higher $\eta/s$, we have to compensate for the lower amount of work performed in the evolution by decreasing the initial energy, yielding a decreased opacity and flow response. Accordingly, for lower $\eta/s$, scaling gives higher results. Of course this was entirely expected, since we performed the scaling according to the universal curves, so we get out what we put in. The point of this exercise was to demonstrate that the change due to this scaling is negligible, especially in the qualitative form of the curves, which the results do support.

\section{Problems with the eos switching setup}\label{app:switching_problems}

\begin{figure*}
    \centering
    \includegraphics[width=0.49\linewidth]{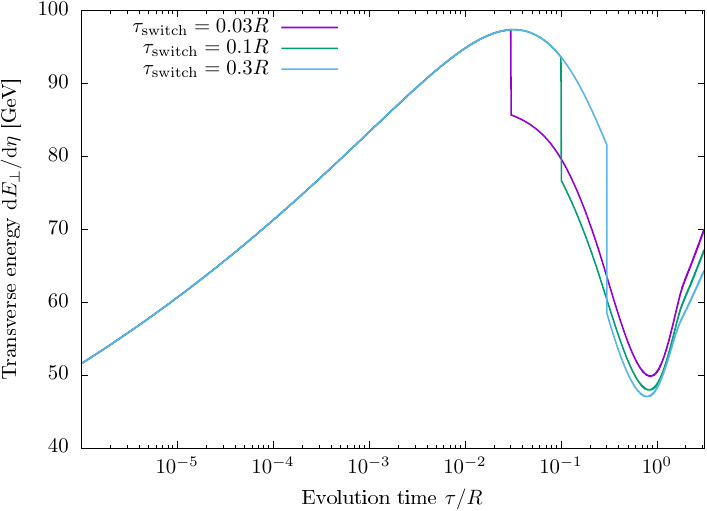}
    \includegraphics[width=0.49\linewidth]{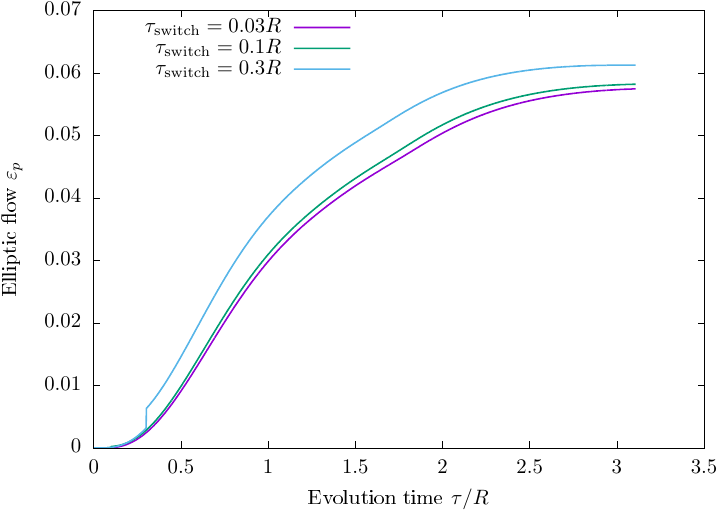}
    \caption{
    Time evolution of transverse energy $\d E_\perp/\d\eta$ (left) and elliptic flow $\varepsilon_p$ (right) from nonconformal hydrodynamic simulations of the evolution of an OO collision event at LHC from the 30-40\% centrality class, where the equation of state was switched at varying times $\tau_{\rm switch}=0.03R$ (purple), $0.1R$ (green) and $0.3R$ (light blue).}
    \label{fig:tdep_switch}
\end{figure*}

Figure~\ref{fig:tdep_switch} shows the time evolution of transverse energy on the left and elliptic flow on the right for an example profile from the 30-40\% centrality class of OO collisions at LHC in the nonconformal simulation setup. The three different colors show results from switching at $\tau/R=0.03$ (cyan), $0.1$ (dark violet) and $0.3$ (green). Varying the switching times on this scale can cause the final state results to differ on the order of 10\%. Since we scale the initial condition to match nonconformal results for the final state transverse energy, for the final state results that we are tracking, this difference affects only elliptic flow. What is most striking in these plots is the jump in transverse energy at the moment of switching. Since we chose to match the energy but change the pressure, some of the components of the energy-momentum tensor will jump. This type of discontinuity will necessarily happen in some way or another when switching between equations of state and this is the biggest source of theoretical uncertainty in our setup. In fact, the direction of the jump is not fixed either. In some events, transverse energy jumps up, in others it jumps down.

\bibliography{main}

\begin{thebibliography}{66}%
\makeatletter
\providecommand \@ifxundefined [1]{%
 \@ifx{#1\undefined}
}%
\providecommand \@ifnum [1]{%
 \ifnum #1\expandafter \@firstoftwo
 \else \expandafter \@secondoftwo
 \fi
}%
\providecommand \@ifx [1]{%
 \ifx #1\expandafter \@firstoftwo
 \else \expandafter \@secondoftwo
 \fi
}%
\providecommand \natexlab [1]{#1}%
\providecommand \enquote  [1]{``#1''}%
\providecommand \bibnamefont  [1]{#1}%
\providecommand \bibfnamefont [1]{#1}%
\providecommand \citenamefont [1]{#1}%
\providecommand \href@noop [0]{\@secondoftwo}%
\providecommand \href [0]{\begingroup \@sanitize@url \@href}%
\providecommand \@href[1]{\@@startlink{#1}\@@href}%
\providecommand \@@href[1]{\endgroup#1\@@endlink}%
\providecommand \@sanitize@url [0]{\catcode `\\12\catcode `\$12\catcode `\&12\catcode `\#12\catcode `\^12\catcode `\_12\catcode `\%12\relax}%
\providecommand \@@startlink[1]{}%
\providecommand \@@endlink[0]{}%
\providecommand \url  [0]{\begingroup\@sanitize@url \@url }%
\providecommand \@url [1]{\endgroup\@href {#1}{\urlprefix }}%
\providecommand \urlprefix  [0]{URL }%
\providecommand \Eprint [0]{\href }%
\providecommand \doibase [0]{http://dx.doi.org/}%
\providecommand \selectlanguage [0]{\@gobble}%
\providecommand \bibinfo  [0]{\@secondoftwo}%
\providecommand \bibfield  [0]{\@secondoftwo}%
\providecommand \translation [1]{[#1]}%
\providecommand \BibitemOpen [0]{}%
\providecommand \bibitemStop [0]{}%
\providecommand \bibitemNoStop [0]{.\EOS\space}%
\providecommand \EOS [0]{\spacefactor3000\relax}%
\providecommand \BibitemShut  [1]{\csname bibitem#1\endcsname}%
\let\auto@bib@innerbib\@empty
\bibitem [{\citenamefont {Heinz}(2013)}]{Heinz:2013wva}%
  \BibitemOpen
  \bibfield  {author} {\bibinfo {author} {\bibfnamefont {U.~W.}\ \bibnamefont {Heinz}},\ }\href {\doibase 10.1088/1742-6596/455/1/012044} {\bibfield  {journal} {\bibinfo  {journal} {J. Phys. Conf. Ser.}\ }\textbf {\bibinfo {volume} {455}},\ \bibinfo {pages} {012044} (\bibinfo {year} {2013})},\ \Eprint {http://arxiv.org/abs/1304.3634} {arXiv:1304.3634 [nucl-th]} \BibitemShut {NoStop}%
\bibitem [{\citenamefont {Shen}\ \emph {et~al.}(2016)\citenamefont {Shen}, \citenamefont {Qiu}, \citenamefont {Song}, \citenamefont {Bernhard}, \citenamefont {Bass},\ and\ \citenamefont {Heinz}}]{Shen:2014vra}%
  \BibitemOpen
  \bibfield  {author} {\bibinfo {author} {\bibfnamefont {C.}~\bibnamefont {Shen}}, \bibinfo {author} {\bibfnamefont {Z.}~\bibnamefont {Qiu}}, \bibinfo {author} {\bibfnamefont {H.}~\bibnamefont {Song}}, \bibinfo {author} {\bibfnamefont {J.}~\bibnamefont {Bernhard}}, \bibinfo {author} {\bibfnamefont {S.}~\bibnamefont {Bass}}, \ and\ \bibinfo {author} {\bibfnamefont {U.}~\bibnamefont {Heinz}},\ }\href {\doibase 10.1016/j.cpc.2015.08.039} {\bibfield  {journal} {\bibinfo  {journal} {Comput. Phys. Commun.}\ }\textbf {\bibinfo {volume} {199}},\ \bibinfo {pages} {61} (\bibinfo {year} {2016})},\ \Eprint {http://arxiv.org/abs/1409.8164} {arXiv:1409.8164 [nucl-th]} \BibitemShut {NoStop}%
\bibitem [{\citenamefont {McDonald}\ \emph {et~al.}(2017)\citenamefont {McDonald}, \citenamefont {Shen}, \citenamefont {Fillion-Gourdeau}, \citenamefont {Jeon},\ and\ \citenamefont {Gale}}]{McDonald:2016vlt}%
  \BibitemOpen
  \bibfield  {author} {\bibinfo {author} {\bibfnamefont {S.}~\bibnamefont {McDonald}}, \bibinfo {author} {\bibfnamefont {C.}~\bibnamefont {Shen}}, \bibinfo {author} {\bibfnamefont {F.}~\bibnamefont {Fillion-Gourdeau}}, \bibinfo {author} {\bibfnamefont {S.}~\bibnamefont {Jeon}}, \ and\ \bibinfo {author} {\bibfnamefont {C.}~\bibnamefont {Gale}},\ }\href {\doibase 10.1103/PhysRevC.95.064913} {\bibfield  {journal} {\bibinfo  {journal} {Phys. Rev. C}\ }\textbf {\bibinfo {volume} {95}},\ \bibinfo {pages} {064913} (\bibinfo {year} {2017})},\ \Eprint {http://arxiv.org/abs/1609.02958} {arXiv:1609.02958 [hep-ph]} \BibitemShut {NoStop}%
\bibitem [{\citenamefont {Putschke}\ \emph {et~al.}(2019)\citenamefont {Putschke} \emph {et~al.}}]{Putschke:2019yrg}%
  \BibitemOpen
  \bibfield  {author} {\bibinfo {author} {\bibfnamefont {J.~H.}\ \bibnamefont {Putschke}} \emph {et~al.},\ }\href@noop {} {\  (\bibinfo {year} {2019})},\ \bibinfo {note} {arXiv:1903.07706},\ \Eprint {http://arxiv.org/abs/1903.07706} {arXiv:1903.07706 [nucl-th]} \BibitemShut {NoStop}%
\bibitem [{\citenamefont {Nijs}\ \emph {et~al.}(2021)\citenamefont {Nijs}, \citenamefont {van~der Schee}, \citenamefont {G\"ursoy},\ and\ \citenamefont {Snellings}}]{Nijs:2020roc}%
  \BibitemOpen
  \bibfield  {author} {\bibinfo {author} {\bibfnamefont {G.}~\bibnamefont {Nijs}}, \bibinfo {author} {\bibfnamefont {W.}~\bibnamefont {van~der Schee}}, \bibinfo {author} {\bibfnamefont {U.}~\bibnamefont {G\"ursoy}}, \ and\ \bibinfo {author} {\bibfnamefont {R.}~\bibnamefont {Snellings}},\ }\href {\doibase 10.1103/PhysRevC.103.054909} {\bibfield  {journal} {\bibinfo  {journal} {Phys. Rev. C}\ }\textbf {\bibinfo {volume} {103}},\ \bibinfo {pages} {054909} (\bibinfo {year} {2021})},\ \Eprint {http://arxiv.org/abs/2010.15134} {arXiv:2010.15134 [nucl-th]} \BibitemShut {NoStop}%
\bibitem [{\citenamefont {Everett}\ \emph {et~al.}(2021)\citenamefont {Everett} \emph {et~al.}}]{JETSCAPE:2020mzn}%
  \BibitemOpen
  \bibfield  {author} {\bibinfo {author} {\bibfnamefont {D.}~\bibnamefont {Everett}} \emph {et~al.} (\bibinfo {collaboration} {JETSCAPE}),\ }\href {\doibase 10.1103/PhysRevC.103.054904} {\bibfield  {journal} {\bibinfo  {journal} {Phys. Rev. C}\ }\textbf {\bibinfo {volume} {103}},\ \bibinfo {pages} {054904} (\bibinfo {year} {2021})},\ \Eprint {http://arxiv.org/abs/2011.01430} {arXiv:2011.01430 [hep-ph]} \BibitemShut {NoStop}%
\bibitem [{\citenamefont {Xu}\ and\ \citenamefont {Greiner}(2005)}]{Xu:2004mz}%
  \BibitemOpen
  \bibfield  {author} {\bibinfo {author} {\bibfnamefont {Z.}~\bibnamefont {Xu}}\ and\ \bibinfo {author} {\bibfnamefont {C.}~\bibnamefont {Greiner}},\ }\href {\doibase 10.1103/PhysRevC.71.064901} {\bibfield  {journal} {\bibinfo  {journal} {Phys. Rev. C}\ }\textbf {\bibinfo {volume} {71}},\ \bibinfo {pages} {064901} (\bibinfo {year} {2005})},\ \Eprint {http://arxiv.org/abs/hep-ph/0406278} {arXiv:hep-ph/0406278} \BibitemShut {NoStop}%
\bibitem [{\citenamefont {Xu}\ \emph {et~al.}(2008)\citenamefont {Xu}, \citenamefont {Greiner},\ and\ \citenamefont {Stocker}}]{Xu:2007jv}%
  \BibitemOpen
  \bibfield  {author} {\bibinfo {author} {\bibfnamefont {Z.}~\bibnamefont {Xu}}, \bibinfo {author} {\bibfnamefont {C.}~\bibnamefont {Greiner}}, \ and\ \bibinfo {author} {\bibfnamefont {H.}~\bibnamefont {Stocker}},\ }\href {\doibase 10.1103/PhysRevLett.101.082302} {\bibfield  {journal} {\bibinfo  {journal} {Phys. Rev. Lett.}\ }\textbf {\bibinfo {volume} {101}},\ \bibinfo {pages} {082302} (\bibinfo {year} {2008})},\ \Eprint {http://arxiv.org/abs/0711.0961} {arXiv:0711.0961 [nucl-th]} \BibitemShut {NoStop}%
\bibitem [{\citenamefont {Lin}\ \emph {et~al.}(2005)\citenamefont {Lin}, \citenamefont {Ko}, \citenamefont {Li}, \citenamefont {Zhang},\ and\ \citenamefont {Pal}}]{Lin:2004en}%
  \BibitemOpen
  \bibfield  {author} {\bibinfo {author} {\bibfnamefont {Z.-W.}\ \bibnamefont {Lin}}, \bibinfo {author} {\bibfnamefont {C.~M.}\ \bibnamefont {Ko}}, \bibinfo {author} {\bibfnamefont {B.-A.}\ \bibnamefont {Li}}, \bibinfo {author} {\bibfnamefont {B.}~\bibnamefont {Zhang}}, \ and\ \bibinfo {author} {\bibfnamefont {S.}~\bibnamefont {Pal}},\ }\href {\doibase 10.1103/PhysRevC.72.064901} {\bibfield  {journal} {\bibinfo  {journal} {Phys. Rev. C}\ }\textbf {\bibinfo {volume} {72}},\ \bibinfo {pages} {064901} (\bibinfo {year} {2005})},\ \Eprint {http://arxiv.org/abs/nucl-th/0411110} {arXiv:nucl-th/0411110} \BibitemShut {NoStop}%
\bibitem [{\citenamefont {Bratkovskaya}\ \emph {et~al.}(2011)\citenamefont {Bratkovskaya}, \citenamefont {Cassing}, \citenamefont {Konchakovski},\ and\ \citenamefont {Linnyk}}]{Bratkovskaya:2011wp}%
  \BibitemOpen
  \bibfield  {author} {\bibinfo {author} {\bibfnamefont {E.~L.}\ \bibnamefont {Bratkovskaya}}, \bibinfo {author} {\bibfnamefont {W.}~\bibnamefont {Cassing}}, \bibinfo {author} {\bibfnamefont {V.~P.}\ \bibnamefont {Konchakovski}}, \ and\ \bibinfo {author} {\bibfnamefont {O.}~\bibnamefont {Linnyk}},\ }\href {\doibase 10.1016/j.nuclphysa.2011.03.003} {\bibfield  {journal} {\bibinfo  {journal} {Nucl. Phys. A}\ }\textbf {\bibinfo {volume} {856}},\ \bibinfo {pages} {162} (\bibinfo {year} {2011})},\ \Eprint {http://arxiv.org/abs/1101.5793} {arXiv:1101.5793 [nucl-th]} \BibitemShut {NoStop}%
\bibitem [{\citenamefont {Linnyk}\ \emph {et~al.}(2016)\citenamefont {Linnyk}, \citenamefont {Bratkovskaya},\ and\ \citenamefont {Cassing}}]{Linnyk:2015rco}%
  \BibitemOpen
  \bibfield  {author} {\bibinfo {author} {\bibfnamefont {O.}~\bibnamefont {Linnyk}}, \bibinfo {author} {\bibfnamefont {E.~L.}\ \bibnamefont {Bratkovskaya}}, \ and\ \bibinfo {author} {\bibfnamefont {W.}~\bibnamefont {Cassing}},\ }\href {\doibase 10.1016/j.ppnp.2015.12.003} {\bibfield  {journal} {\bibinfo  {journal} {Prog. Part. Nucl. Phys.}\ }\textbf {\bibinfo {volume} {87}},\ \bibinfo {pages} {50} (\bibinfo {year} {2016})},\ \Eprint {http://arxiv.org/abs/1512.08126} {arXiv:1512.08126 [nucl-th]} \BibitemShut {NoStop}%
\bibitem [{\citenamefont {Ambru\cb{s}}\ \emph {et~al.}(2022)\citenamefont {Ambru\cb{s}}, \citenamefont {Schlichting},\ and\ \citenamefont {Werthmann}}]{Ambrus:2021fej}%
  \BibitemOpen
  \bibfield  {author} {\bibinfo {author} {\bibfnamefont {V.~E.}\ \bibnamefont {Ambru\cb{s}}}, \bibinfo {author} {\bibfnamefont {S.}~\bibnamefont {Schlichting}}, \ and\ \bibinfo {author} {\bibfnamefont {C.}~\bibnamefont {Werthmann}},\ }\href {\doibase 10.1103/PhysRevD.105.014031} {\bibfield  {journal} {\bibinfo  {journal} {Phys. Rev. D}\ }\textbf {\bibinfo {volume} {105}},\ \bibinfo {pages} {014031} (\bibinfo {year} {2022})},\ \Eprint {http://arxiv.org/abs/2109.03290} {arXiv:2109.03290 [hep-ph]} \BibitemShut {NoStop}%
\bibitem [{\citenamefont {Heller}\ \emph {et~al.}(2018)\citenamefont {Heller}, \citenamefont {Kurkela}, \citenamefont {Spali\'nski},\ and\ \citenamefont {Svensson}}]{Heller:2016rtz}%
  \BibitemOpen
  \bibfield  {author} {\bibinfo {author} {\bibfnamefont {M.~P.}\ \bibnamefont {Heller}}, \bibinfo {author} {\bibfnamefont {A.}~\bibnamefont {Kurkela}}, \bibinfo {author} {\bibfnamefont {M.}~\bibnamefont {Spali\'nski}}, \ and\ \bibinfo {author} {\bibfnamefont {V.}~\bibnamefont {Svensson}},\ }\href {\doibase 10.1103/PhysRevD.97.091503} {\bibfield  {journal} {\bibinfo  {journal} {Phys. Rev. D}\ }\textbf {\bibinfo {volume} {97}},\ \bibinfo {pages} {091503} (\bibinfo {year} {2018})},\ \Eprint {http://arxiv.org/abs/1609.04803} {arXiv:1609.04803 [nucl-th]} \BibitemShut {NoStop}%
\bibitem [{\citenamefont {Spali\'nski}(2018)}]{Spalinski:2017mel}%
  \BibitemOpen
  \bibfield  {author} {\bibinfo {author} {\bibfnamefont {M.}~\bibnamefont {Spali\'nski}},\ }\href {\doibase 10.1016/j.physletb.2017.11.059} {\bibfield  {journal} {\bibinfo  {journal} {Phys. Lett. B}\ }\textbf {\bibinfo {volume} {776}},\ \bibinfo {pages} {468} (\bibinfo {year} {2018})},\ \Eprint {http://arxiv.org/abs/1708.01921} {arXiv:1708.01921 [hep-th]} \BibitemShut {NoStop}%
\bibitem [{\citenamefont {Romatschke}(2017)}]{Romatschke:2017acs}%
  \BibitemOpen
  \bibfield  {author} {\bibinfo {author} {\bibfnamefont {P.}~\bibnamefont {Romatschke}},\ }\href {\doibase 10.1007/JHEP12(2017)079} {\bibfield  {journal} {\bibinfo  {journal} {JHEP}\ }\textbf {\bibinfo {volume} {12}},\ \bibinfo {pages} {079} (\bibinfo {year} {2017})},\ \Eprint {http://arxiv.org/abs/1710.03234} {arXiv:1710.03234 [hep-th]} \BibitemShut {NoStop}%
\bibitem [{\citenamefont {Du}\ and\ \citenamefont {Schlichting}(2021{\natexlab{a}})}]{Du:2020dvp}%
  \BibitemOpen
  \bibfield  {author} {\bibinfo {author} {\bibfnamefont {X.}~\bibnamefont {Du}}\ and\ \bibinfo {author} {\bibfnamefont {S.}~\bibnamefont {Schlichting}},\ }\href {\doibase 10.1103/PhysRevD.104.054011} {\bibfield  {journal} {\bibinfo  {journal} {Phys. Rev. D}\ }\textbf {\bibinfo {volume} {104}},\ \bibinfo {pages} {054011} (\bibinfo {year} {2021}{\natexlab{a}})},\ \Eprint {http://arxiv.org/abs/2012.09079} {arXiv:2012.09079 [hep-ph]} \BibitemShut {NoStop}%
\bibitem [{\citenamefont {Du}\ and\ \citenamefont {Schlichting}(2021{\natexlab{b}})}]{Du:2020zqg}%
  \BibitemOpen
  \bibfield  {author} {\bibinfo {author} {\bibfnamefont {X.}~\bibnamefont {Du}}\ and\ \bibinfo {author} {\bibfnamefont {S.}~\bibnamefont {Schlichting}},\ }\href {\doibase 10.1103/PhysRevLett.127.122301} {\bibfield  {journal} {\bibinfo  {journal} {Phys. Rev. Lett.}\ }\textbf {\bibinfo {volume} {127}},\ \bibinfo {pages} {122301} (\bibinfo {year} {2021}{\natexlab{b}})},\ \Eprint {http://arxiv.org/abs/2012.09068} {arXiv:2012.09068 [hep-ph]} \BibitemShut {NoStop}%
\bibitem [{\citenamefont {Kurkela}\ \emph {et~al.}(2019{\natexlab{a}})\citenamefont {Kurkela}, \citenamefont {Mazeliauskas}, \citenamefont {Paquet}, \citenamefont {Schlichting},\ and\ \citenamefont {Teaney}}]{Kurkela:2018wud}%
  \BibitemOpen
  \bibfield  {author} {\bibinfo {author} {\bibfnamefont {A.}~\bibnamefont {Kurkela}}, \bibinfo {author} {\bibfnamefont {A.}~\bibnamefont {Mazeliauskas}}, \bibinfo {author} {\bibfnamefont {J.-F.}\ \bibnamefont {Paquet}}, \bibinfo {author} {\bibfnamefont {S.}~\bibnamefont {Schlichting}}, \ and\ \bibinfo {author} {\bibfnamefont {D.}~\bibnamefont {Teaney}},\ }\href {\doibase 10.1103/PhysRevLett.122.122302} {\bibfield  {journal} {\bibinfo  {journal} {Phys. Rev. Lett.}\ }\textbf {\bibinfo {volume} {122}},\ \bibinfo {pages} {122302} (\bibinfo {year} {2019}{\natexlab{a}})},\ \Eprint {http://arxiv.org/abs/1805.01604} {arXiv:1805.01604 [hep-ph]} \BibitemShut {NoStop}%
\bibitem [{\citenamefont {Kurkela}\ \emph {et~al.}(2018)\citenamefont {Kurkela}, \citenamefont {Mazeliauskas}, \citenamefont {Paquet}, \citenamefont {Schlichting},\ and\ \citenamefont {Teaney}}]{Kurkela:2018gitrep}%
  \BibitemOpen
  \bibfield  {author} {\bibinfo {author} {\bibfnamefont {A.}~\bibnamefont {Kurkela}}, \bibinfo {author} {\bibfnamefont {A.}~\bibnamefont {Mazeliauskas}}, \bibinfo {author} {\bibfnamefont {J.-F.}\ \bibnamefont {Paquet}}, \bibinfo {author} {\bibfnamefont {S.}~\bibnamefont {Schlichting}}, \ and\ \bibinfo {author} {\bibfnamefont {D.}~\bibnamefont {Teaney}},\ }\href@noop {} {\enquote {\bibinfo {title} {K{\o}{MP}{\o}{ST}},}\ }\bibinfo {howpublished} {\url{https://github.com/KMPST/KoMPoST}} (\bibinfo {year} {2018})\BibitemShut {NoStop}%
\bibitem [{\citenamefont {Liyanage}\ \emph {et~al.}(2022)\citenamefont {Liyanage}, \citenamefont {Everett}, \citenamefont {Chattopadhyay},\ and\ \citenamefont {Heinz}}]{Liyanage:2022nua}%
  \BibitemOpen
  \bibfield  {author} {\bibinfo {author} {\bibfnamefont {D.}~\bibnamefont {Liyanage}}, \bibinfo {author} {\bibfnamefont {D.}~\bibnamefont {Everett}}, \bibinfo {author} {\bibfnamefont {C.}~\bibnamefont {Chattopadhyay}}, \ and\ \bibinfo {author} {\bibfnamefont {U.}~\bibnamefont {Heinz}},\ }\href {\doibase 10.1103/PhysRevC.105.064908} {\bibfield  {journal} {\bibinfo  {journal} {Phys. Rev. C}\ }\textbf {\bibinfo {volume} {105}},\ \bibinfo {pages} {064908} (\bibinfo {year} {2022})},\ \Eprint {http://arxiv.org/abs/2205.00964} {arXiv:2205.00964 [nucl-th]} \BibitemShut {NoStop}%
\bibitem [{\citenamefont {M\"antysaari}\ \emph {et~al.}(2017)\citenamefont {M\"antysaari}, \citenamefont {Schenke}, \citenamefont {Shen},\ and\ \citenamefont {Tribedy}}]{Mantysaari:2017cni}%
  \BibitemOpen
  \bibfield  {author} {\bibinfo {author} {\bibfnamefont {H.}~\bibnamefont {M\"antysaari}}, \bibinfo {author} {\bibfnamefont {B.}~\bibnamefont {Schenke}}, \bibinfo {author} {\bibfnamefont {C.}~\bibnamefont {Shen}}, \ and\ \bibinfo {author} {\bibfnamefont {P.}~\bibnamefont {Tribedy}},\ }\href {\doibase 10.1016/j.physletb.2017.07.038} {\bibfield  {journal} {\bibinfo  {journal} {Phys. Lett. B}\ }\textbf {\bibinfo {volume} {772}},\ \bibinfo {pages} {681} (\bibinfo {year} {2017})},\ \Eprint {http://arxiv.org/abs/1705.03177} {arXiv:1705.03177 [nucl-th]} \BibitemShut {NoStop}%
\bibitem [{\citenamefont {Zhao}\ \emph {et~al.}(2023)\citenamefont {Zhao}, \citenamefont {Ryu}, \citenamefont {Shen},\ and\ \citenamefont {Schenke}}]{Zhao:2022ugy}%
  \BibitemOpen
  \bibfield  {author} {\bibinfo {author} {\bibfnamefont {W.}~\bibnamefont {Zhao}}, \bibinfo {author} {\bibfnamefont {S.}~\bibnamefont {Ryu}}, \bibinfo {author} {\bibfnamefont {C.}~\bibnamefont {Shen}}, \ and\ \bibinfo {author} {\bibfnamefont {B.}~\bibnamefont {Schenke}},\ }\href {\doibase 10.1103/PhysRevC.107.014904} {\bibfield  {journal} {\bibinfo  {journal} {Phys. Rev. C}\ }\textbf {\bibinfo {volume} {107}},\ \bibinfo {pages} {014904} (\bibinfo {year} {2023})},\ \Eprint {http://arxiv.org/abs/2211.16376} {arXiv:2211.16376 [nucl-th]} \BibitemShut {NoStop}%
\bibitem [{\citenamefont {Werner}\ \emph {et~al.}(2011)\citenamefont {Werner}, \citenamefont {Karpenko},\ and\ \citenamefont {Pierog}}]{Werner:2010ss}%
  \BibitemOpen
  \bibfield  {author} {\bibinfo {author} {\bibfnamefont {K.}~\bibnamefont {Werner}}, \bibinfo {author} {\bibfnamefont {I.}~\bibnamefont {Karpenko}}, \ and\ \bibinfo {author} {\bibfnamefont {T.}~\bibnamefont {Pierog}},\ }\href {\doibase 10.1103/PhysRevLett.106.122004} {\bibfield  {journal} {\bibinfo  {journal} {Phys. Rev. Lett.}\ }\textbf {\bibinfo {volume} {106}},\ \bibinfo {pages} {122004} (\bibinfo {year} {2011})},\ \Eprint {http://arxiv.org/abs/1011.0375} {arXiv:1011.0375 [hep-ph]} \BibitemShut {NoStop}%
\bibitem [{\citenamefont {Werner}\ \emph {et~al.}(2014)\citenamefont {Werner}, \citenamefont {Bleicher}, \citenamefont {Guiot}, \citenamefont {Karpenko},\ and\ \citenamefont {Pierog}}]{Werner:2013ipa}%
  \BibitemOpen
  \bibfield  {author} {\bibinfo {author} {\bibfnamefont {K.}~\bibnamefont {Werner}}, \bibinfo {author} {\bibfnamefont {M.}~\bibnamefont {Bleicher}}, \bibinfo {author} {\bibfnamefont {B.}~\bibnamefont {Guiot}}, \bibinfo {author} {\bibfnamefont {I.}~\bibnamefont {Karpenko}}, \ and\ \bibinfo {author} {\bibfnamefont {T.}~\bibnamefont {Pierog}},\ }\href {\doibase 10.1103/PhysRevLett.112.232301} {\bibfield  {journal} {\bibinfo  {journal} {Phys. Rev. Lett.}\ }\textbf {\bibinfo {volume} {112}},\ \bibinfo {pages} {232301} (\bibinfo {year} {2014})},\ \Eprint {http://arxiv.org/abs/1307.4379} {arXiv:1307.4379 [nucl-th]} \BibitemShut {NoStop}%
\bibitem [{\citenamefont {Schenke}\ and\ \citenamefont {Venugopalan}(2014)}]{Schenke:2014zha}%
  \BibitemOpen
  \bibfield  {author} {\bibinfo {author} {\bibfnamefont {B.}~\bibnamefont {Schenke}}\ and\ \bibinfo {author} {\bibfnamefont {R.}~\bibnamefont {Venugopalan}},\ }\href {\doibase 10.1103/PhysRevLett.113.102301} {\bibfield  {journal} {\bibinfo  {journal} {Phys. Rev. Lett.}\ }\textbf {\bibinfo {volume} {113}},\ \bibinfo {pages} {102301} (\bibinfo {year} {2014})},\ \Eprint {http://arxiv.org/abs/1405.3605} {arXiv:1405.3605 [nucl-th]} \BibitemShut {NoStop}%
\bibitem [{\citenamefont {Schenke}(2021)}]{Schenke:2021mxx}%
  \BibitemOpen
  \bibfield  {author} {\bibinfo {author} {\bibfnamefont {B.}~\bibnamefont {Schenke}},\ }\href {\doibase 10.1088/1361-6633/ac14c9} {\bibfield  {journal} {\bibinfo  {journal} {Rept. Prog. Phys.}\ }\textbf {\bibinfo {volume} {84}},\ \bibinfo {pages} {082301} (\bibinfo {year} {2021})},\ \Eprint {http://arxiv.org/abs/2102.11189} {arXiv:2102.11189 [nucl-th]} \BibitemShut {NoStop}%
\bibitem [{\citenamefont {Demirci}\ \emph {et~al.}(2021)\citenamefont {Demirci}, \citenamefont {Lappi},\ and\ \citenamefont {Schlichting}}]{Demirci:2021kya}%
  \BibitemOpen
  \bibfield  {author} {\bibinfo {author} {\bibfnamefont {S.}~\bibnamefont {Demirci}}, \bibinfo {author} {\bibfnamefont {T.}~\bibnamefont {Lappi}}, \ and\ \bibinfo {author} {\bibfnamefont {S.}~\bibnamefont {Schlichting}},\ }\href {\doibase 10.1103/PhysRevD.103.094025} {\bibfield  {journal} {\bibinfo  {journal} {Phys. Rev. D}\ }\textbf {\bibinfo {volume} {103}},\ \bibinfo {pages} {094025} (\bibinfo {year} {2021})},\ \Eprint {http://arxiv.org/abs/2101.03791} {arXiv:2101.03791 [hep-ph]} \BibitemShut {NoStop}%
\bibitem [{\citenamefont {Schenke}\ \emph {et~al.}(2015)\citenamefont {Schenke}, \citenamefont {Schlichting},\ and\ \citenamefont {Venugopalan}}]{Schenke:2015aqa}%
  \BibitemOpen
  \bibfield  {author} {\bibinfo {author} {\bibfnamefont {B.}~\bibnamefont {Schenke}}, \bibinfo {author} {\bibfnamefont {S.}~\bibnamefont {Schlichting}}, \ and\ \bibinfo {author} {\bibfnamefont {R.}~\bibnamefont {Venugopalan}},\ }\href {\doibase 10.1016/j.physletb.2015.05.051} {\bibfield  {journal} {\bibinfo  {journal} {Phys. Lett. B}\ }\textbf {\bibinfo {volume} {747}},\ \bibinfo {pages} {76} (\bibinfo {year} {2015})},\ \Eprint {http://arxiv.org/abs/1502.01331} {arXiv:1502.01331 [hep-ph]} \BibitemShut {NoStop}%
\bibitem [{\citenamefont {McLerran}\ and\ \citenamefont {Skokov}(2016)}]{McLerran:2015sva}%
  \BibitemOpen
  \bibfield  {author} {\bibinfo {author} {\bibfnamefont {L.}~\bibnamefont {McLerran}}\ and\ \bibinfo {author} {\bibfnamefont {V.}~\bibnamefont {Skokov}},\ }\href {\doibase 10.1016/j.nuclphysa.2015.12.005} {\bibfield  {journal} {\bibinfo  {journal} {Nucl. Phys. A}\ }\textbf {\bibinfo {volume} {947}},\ \bibinfo {pages} {142} (\bibinfo {year} {2016})},\ \Eprint {http://arxiv.org/abs/1510.08072} {arXiv:1510.08072 [hep-ph]} \BibitemShut {NoStop}%
\bibitem [{\citenamefont {Schenke}\ \emph {et~al.}(2022)\citenamefont {Schenke}, \citenamefont {Schlichting},\ and\ \citenamefont {Singh}}]{Schenke:2022mjv}%
  \BibitemOpen
  \bibfield  {author} {\bibinfo {author} {\bibfnamefont {B.}~\bibnamefont {Schenke}}, \bibinfo {author} {\bibfnamefont {S.}~\bibnamefont {Schlichting}}, \ and\ \bibinfo {author} {\bibfnamefont {P.}~\bibnamefont {Singh}},\ }\href {\doibase 10.1103/PhysRevD.105.094023} {\bibfield  {journal} {\bibinfo  {journal} {Phys. Rev. D}\ }\textbf {\bibinfo {volume} {105}},\ \bibinfo {pages} {094023} (\bibinfo {year} {2022})},\ \Eprint {http://arxiv.org/abs/2201.08864} {arXiv:2201.08864 [nucl-th]} \BibitemShut {NoStop}%
\bibitem [{\citenamefont {Kurkela}\ \emph {et~al.}(2019{\natexlab{b}})\citenamefont {Kurkela}, \citenamefont {Wiedemann},\ and\ \citenamefont {Wu}}]{Kurkela:2019kip}%
  \BibitemOpen
  \bibfield  {author} {\bibinfo {author} {\bibfnamefont {A.}~\bibnamefont {Kurkela}}, \bibinfo {author} {\bibfnamefont {U.~A.}\ \bibnamefont {Wiedemann}}, \ and\ \bibinfo {author} {\bibfnamefont {B.}~\bibnamefont {Wu}},\ }\href {\doibase 10.1140/epjc/s10052-019-7428-6} {\bibfield  {journal} {\bibinfo  {journal} {Eur. Phys. J. C}\ }\textbf {\bibinfo {volume} {79}},\ \bibinfo {pages} {965} (\bibinfo {year} {2019}{\natexlab{b}})},\ \Eprint {http://arxiv.org/abs/1905.05139} {arXiv:1905.05139 [hep-ph]} \BibitemShut {NoStop}%
\bibitem [{\citenamefont {Kurkela}\ \emph {et~al.}(2020)\citenamefont {Kurkela}, \citenamefont {Taghavi}, \citenamefont {Wiedemann},\ and\ \citenamefont {Wu}}]{Kurkela:2020wwb}%
  \BibitemOpen
  \bibfield  {author} {\bibinfo {author} {\bibfnamefont {A.}~\bibnamefont {Kurkela}}, \bibinfo {author} {\bibfnamefont {S.~F.}\ \bibnamefont {Taghavi}}, \bibinfo {author} {\bibfnamefont {U.~A.}\ \bibnamefont {Wiedemann}}, \ and\ \bibinfo {author} {\bibfnamefont {B.}~\bibnamefont {Wu}},\ }\href {\doibase 10.1016/j.physletb.2020.135901} {\bibfield  {journal} {\bibinfo  {journal} {Phys. Lett. B}\ }\textbf {\bibinfo {volume} {811}},\ \bibinfo {pages} {135901} (\bibinfo {year} {2020})},\ \Eprint {http://arxiv.org/abs/2007.06851} {arXiv:2007.06851 [hep-ph]} \BibitemShut {NoStop}%
\bibitem [{\citenamefont {Ambrus}\ \emph {et~al.}(2023{\natexlab{a}})\citenamefont {Ambrus}, \citenamefont {Schlichting},\ and\ \citenamefont {Werthmann}}]{Ambrus:2022koq}%
  \BibitemOpen
  \bibfield  {author} {\bibinfo {author} {\bibfnamefont {V.~E.}\ \bibnamefont {Ambrus}}, \bibinfo {author} {\bibfnamefont {S.}~\bibnamefont {Schlichting}}, \ and\ \bibinfo {author} {\bibfnamefont {C.}~\bibnamefont {Werthmann}},\ }\href {\doibase 10.1103/PhysRevD.107.094013} {\bibfield  {journal} {\bibinfo  {journal} {Phys. Rev. D}\ }\textbf {\bibinfo {volume} {107}},\ \bibinfo {pages} {094013} (\bibinfo {year} {2023}{\natexlab{a}})},\ \Eprint {http://arxiv.org/abs/2211.14379} {arXiv:2211.14379 [hep-ph]} \BibitemShut {NoStop}%
\bibitem [{\citenamefont {Ambrus}\ \emph {et~al.}(2023{\natexlab{b}})\citenamefont {Ambrus}, \citenamefont {Schlichting},\ and\ \citenamefont {Werthmann}}]{Ambrus:2022qya}%
  \BibitemOpen
  \bibfield  {author} {\bibinfo {author} {\bibfnamefont {V.~E.}\ \bibnamefont {Ambrus}}, \bibinfo {author} {\bibfnamefont {S.}~\bibnamefont {Schlichting}}, \ and\ \bibinfo {author} {\bibfnamefont {C.}~\bibnamefont {Werthmann}},\ }\href {\doibase 10.1103/PhysRevLett.130.152301} {\bibfield  {journal} {\bibinfo  {journal} {Phys. Rev. Lett.}\ }\textbf {\bibinfo {volume} {130}},\ \bibinfo {pages} {152301} (\bibinfo {year} {2023}{\natexlab{b}})},\ \Eprint {http://arxiv.org/abs/2211.14356} {arXiv:2211.14356 [hep-ph]} \BibitemShut {NoStop}%
\bibitem [{\citenamefont {Ambrus}\ \emph {et~al.}(2024)\citenamefont {Ambrus}, \citenamefont {Schlichting},\ and\ \citenamefont {Werthmann}}]{Ambrus:2024eqa}%
  \BibitemOpen
  \bibfield  {author} {\bibinfo {author} {\bibfnamefont {V.~E.}\ \bibnamefont {Ambrus}}, \bibinfo {author} {\bibfnamefont {S.}~\bibnamefont {Schlichting}}, \ and\ \bibinfo {author} {\bibfnamefont {C.}~\bibnamefont {Werthmann}},\ }\href@noop {} {\  (\bibinfo {year} {2024})},\ \Eprint {http://arxiv.org/abs/2411.19709} {arXiv:2411.19709 [hep-ph]} \BibitemShut {NoStop}%
\bibitem [{\citenamefont {Werthmann}\ \emph {et~al.}(2025)\citenamefont {Werthmann}, \citenamefont {Ambrus},\ and\ \citenamefont {Schlichting}}]{werthmann_2025_14849750}%
  \BibitemOpen
  \bibfield  {author} {\bibinfo {author} {\bibfnamefont {C.}~\bibnamefont {Werthmann}}, \bibinfo {author} {\bibfnamefont {V.~E.}\ \bibnamefont {Ambrus}}, \ and\ \bibinfo {author} {\bibfnamefont {S.}~\bibnamefont {Schlichting}},\ }\href {\doibase 10.5281/zenodo.14849750} {\enquote {\bibinfo {title} {Plot data for "collective dynamics in heavy and light-ion collisions -- i) kinetic theory vs. hydrodynamics"},}\ } (\bibinfo {year} {2025})\BibitemShut {NoStop}%
\bibitem [{\citenamefont {Bazavov}\ \emph {et~al.}(2014)\citenamefont {Bazavov} \emph {et~al.}}]{HotQCD:2014kol}%
  \BibitemOpen
  \bibfield  {author} {\bibinfo {author} {\bibfnamefont {A.}~\bibnamefont {Bazavov}} \emph {et~al.} (\bibinfo {collaboration} {HotQCD}),\ }\href {\doibase 10.1103/PhysRevD.90.094503} {\bibfield  {journal} {\bibinfo  {journal} {Phys. Rev. D}\ }\textbf {\bibinfo {volume} {90}},\ \bibinfo {pages} {094503} (\bibinfo {year} {2014})},\ \Eprint {http://arxiv.org/abs/1407.6387} {arXiv:1407.6387 [hep-lat]} \BibitemShut {NoStop}%
\bibitem [{\citenamefont {Borsanyi}\ \emph {et~al.}(2016)\citenamefont {Borsanyi} \emph {et~al.}}]{Borsanyi:2016ksw}%
  \BibitemOpen
  \bibfield  {author} {\bibinfo {author} {\bibfnamefont {S.}~\bibnamefont {Borsanyi}} \emph {et~al.},\ }\href {\doibase 10.1038/nature20115} {\bibfield  {journal} {\bibinfo  {journal} {Nature}\ }\textbf {\bibinfo {volume} {539}},\ \bibinfo {pages} {69} (\bibinfo {year} {2016})},\ \Eprint {http://arxiv.org/abs/1606.07494} {arXiv:1606.07494 [hep-lat]} \BibitemShut {NoStop}%
\bibitem [{\citenamefont {Mueller}(2000)}]{Mueller:1999pi}%
  \BibitemOpen
  \bibfield  {author} {\bibinfo {author} {\bibfnamefont {A.~H.}\ \bibnamefont {Mueller}},\ }\href {\doibase 10.1016/S0370-2693(00)00084-8} {\bibfield  {journal} {\bibinfo  {journal} {Phys. Lett. B}\ }\textbf {\bibinfo {volume} {475}},\ \bibinfo {pages} {220} (\bibinfo {year} {2000})},\ \Eprint {http://arxiv.org/abs/hep-ph/9909388} {arXiv:hep-ph/9909388} \BibitemShut {NoStop}%
\bibitem [{\citenamefont {Schlichting}(2024)}]{Schlichting:2024uok}%
  \BibitemOpen
  \bibfield  {author} {\bibinfo {author} {\bibfnamefont {S.}~\bibnamefont {Schlichting}},\ }\href {\doibase 10.1051/epjconf/202429601020} {\bibfield  {journal} {\bibinfo  {journal} {EPJ Web Conf.}\ }\textbf {\bibinfo {volume} {296}},\ \bibinfo {pages} {01020} (\bibinfo {year} {2024})},\ \Eprint {http://arxiv.org/abs/2401.03841} {arXiv:2401.03841 [hep-ph]} \BibitemShut {NoStop}%
\bibitem [{\citenamefont {Heiselberg}\ and\ \citenamefont {Levy}(1999)}]{Heiselberg:1998es}%
  \BibitemOpen
  \bibfield  {author} {\bibinfo {author} {\bibfnamefont {H.}~\bibnamefont {Heiselberg}}\ and\ \bibinfo {author} {\bibfnamefont {A.-M.}\ \bibnamefont {Levy}},\ }\href {\doibase 10.1103/PhysRevC.59.2716} {\bibfield  {journal} {\bibinfo  {journal} {Phys. Rev. C}\ }\textbf {\bibinfo {volume} {59}},\ \bibinfo {pages} {2716} (\bibinfo {year} {1999})},\ \Eprint {http://arxiv.org/abs/nucl-th/9812034} {arXiv:nucl-th/9812034} \BibitemShut {NoStop}%
\bibitem [{\citenamefont {Borghini}\ and\ \citenamefont {Gombeaud}(2011)}]{Borghini:2010hy}%
  \BibitemOpen
  \bibfield  {author} {\bibinfo {author} {\bibfnamefont {N.}~\bibnamefont {Borghini}}\ and\ \bibinfo {author} {\bibfnamefont {C.}~\bibnamefont {Gombeaud}},\ }\href {\doibase 10.1140/epjc/s10052-011-1612-7} {\bibfield  {journal} {\bibinfo  {journal} {Eur. Phys. J. C}\ }\textbf {\bibinfo {volume} {71}},\ \bibinfo {pages} {1612} (\bibinfo {year} {2011})},\ \Eprint {http://arxiv.org/abs/1012.0899} {arXiv:1012.0899 [nucl-th]} \BibitemShut {NoStop}%
\bibitem [{\citenamefont {Denicol}\ \emph {et~al.}(2012)\citenamefont {Denicol}, \citenamefont {Niemi}, \citenamefont {Molnar},\ and\ \citenamefont {Rischke}}]{Denicol:2012cn}%
  \BibitemOpen
  \bibfield  {author} {\bibinfo {author} {\bibfnamefont {G.~S.}\ \bibnamefont {Denicol}}, \bibinfo {author} {\bibfnamefont {H.}~\bibnamefont {Niemi}}, \bibinfo {author} {\bibfnamefont {E.}~\bibnamefont {Molnar}}, \ and\ \bibinfo {author} {\bibfnamefont {D.~H.}\ \bibnamefont {Rischke}},\ }\href {\doibase 10.1103/PhysRevD.85.114047} {\bibfield  {journal} {\bibinfo  {journal} {Phys. Rev. D}\ }\textbf {\bibinfo {volume} {85}},\ \bibinfo {pages} {114047} (\bibinfo {year} {2012})},\ \bibinfo {note} {[Erratum: Phys.Rev.D 91, 039902 (2015)]},\ \Eprint {http://arxiv.org/abs/1202.4551} {arXiv:1202.4551 [nucl-th]} \BibitemShut {NoStop}%
\bibitem [{\citenamefont {Jaiswal}(2013)}]{Jaiswal:2013npa}%
  \BibitemOpen
  \bibfield  {author} {\bibinfo {author} {\bibfnamefont {A.}~\bibnamefont {Jaiswal}},\ }\href {\doibase 10.1103/PhysRevC.87.051901} {\bibfield  {journal} {\bibinfo  {journal} {Phys. Rev. C}\ }\textbf {\bibinfo {volume} {87}},\ \bibinfo {pages} {051901} (\bibinfo {year} {2013})},\ \Eprint {http://arxiv.org/abs/1302.6311} {arXiv:1302.6311 [nucl-th]} \BibitemShut {NoStop}%
\bibitem [{\citenamefont {Moln\'ar}\ \emph {et~al.}(2014)\citenamefont {Moln\'ar}, \citenamefont {Niemi}, \citenamefont {Denicol},\ and\ \citenamefont {Rischke}}]{Molnar:2013lta}%
  \BibitemOpen
  \bibfield  {author} {\bibinfo {author} {\bibfnamefont {E.}~\bibnamefont {Moln\'ar}}, \bibinfo {author} {\bibfnamefont {H.}~\bibnamefont {Niemi}}, \bibinfo {author} {\bibfnamefont {G.~S.}\ \bibnamefont {Denicol}}, \ and\ \bibinfo {author} {\bibfnamefont {D.~H.}\ \bibnamefont {Rischke}},\ }\href {\doibase 10.1103/PhysRevD.89.074010} {\bibfield  {journal} {\bibinfo  {journal} {Phys. Rev. D}\ }\textbf {\bibinfo {volume} {89}},\ \bibinfo {pages} {074010} (\bibinfo {year} {2014})},\ \Eprint {http://arxiv.org/abs/1308.0785} {arXiv:1308.0785 [nucl-th]} \BibitemShut {NoStop}%
\bibitem [{\citenamefont {Ambrus}\ \emph {et~al.}(2022)\citenamefont {Ambrus}, \citenamefont {Moln\'ar},\ and\ \citenamefont {Rischke}}]{Ambrus:2022vif}%
  \BibitemOpen
  \bibfield  {author} {\bibinfo {author} {\bibfnamefont {V.~E.}\ \bibnamefont {Ambrus}}, \bibinfo {author} {\bibfnamefont {E.}~\bibnamefont {Moln\'ar}}, \ and\ \bibinfo {author} {\bibfnamefont {D.~H.}\ \bibnamefont {Rischke}},\ }\href {\doibase 10.1103/PhysRevD.106.076005} {\bibfield  {journal} {\bibinfo  {journal} {Phys. Rev. D}\ }\textbf {\bibinfo {volume} {106}},\ \bibinfo {pages} {076005} (\bibinfo {year} {2022})},\ \Eprint {http://arxiv.org/abs/2207.05670} {arXiv:2207.05670 [nucl-th]} \BibitemShut {NoStop}%
\bibitem [{\citenamefont {Karpenko}\ \emph {et~al.}(2014)\citenamefont {Karpenko}, \citenamefont {Huovinen},\ and\ \citenamefont {Bleicher}}]{Karpenko:2013wva}%
  \BibitemOpen
  \bibfield  {author} {\bibinfo {author} {\bibfnamefont {I.}~\bibnamefont {Karpenko}}, \bibinfo {author} {\bibfnamefont {P.}~\bibnamefont {Huovinen}}, \ and\ \bibinfo {author} {\bibfnamefont {M.}~\bibnamefont {Bleicher}},\ }\href {\doibase 10.1016/j.cpc.2014.07.010} {\bibfield  {journal} {\bibinfo  {journal} {Comput. Phys. Commun.}\ }\textbf {\bibinfo {volume} {185}},\ \bibinfo {pages} {3016} (\bibinfo {year} {2014})},\ \Eprint {http://arxiv.org/abs/1312.4160} {arXiv:1312.4160 [nucl-th]} \BibitemShut {NoStop}%
\bibitem [{\citenamefont {Moreland}\ \emph {et~al.}(2015)\citenamefont {Moreland}, \citenamefont {Bernhard},\ and\ \citenamefont {Bass}}]{Moreland:2014oya}%
  \BibitemOpen
  \bibfield  {author} {\bibinfo {author} {\bibfnamefont {J.~S.}\ \bibnamefont {Moreland}}, \bibinfo {author} {\bibfnamefont {J.~E.}\ \bibnamefont {Bernhard}}, \ and\ \bibinfo {author} {\bibfnamefont {S.~A.}\ \bibnamefont {Bass}},\ }\href {\doibase 10.1103/PhysRevC.92.011901} {\bibfield  {journal} {\bibinfo  {journal} {Phys. Rev. C}\ }\textbf {\bibinfo {volume} {92}},\ \bibinfo {pages} {011901} (\bibinfo {year} {2015})},\ \Eprint {http://arxiv.org/abs/1412.4708} {arXiv:1412.4708 [nucl-th]} \BibitemShut {NoStop}%
\bibitem [{\citenamefont {Nijs}\ and\ \citenamefont {van~der Schee}(2022)}]{Nijs:2021clz}%
  \BibitemOpen
  \bibfield  {author} {\bibinfo {author} {\bibfnamefont {G.}~\bibnamefont {Nijs}}\ and\ \bibinfo {author} {\bibfnamefont {W.}~\bibnamefont {van~der Schee}},\ }\href {\doibase 10.1103/PhysRevC.106.044903} {\bibfield  {journal} {\bibinfo  {journal} {Phys. Rev. C}\ }\textbf {\bibinfo {volume} {106}},\ \bibinfo {pages} {044903} (\bibinfo {year} {2022})},\ \Eprint {http://arxiv.org/abs/2110.13153} {arXiv:2110.13153 [nucl-th]} \BibitemShut {NoStop}%
\bibitem [{\citenamefont {Liyanage}\ \emph {et~al.}(2023)\citenamefont {Liyanage}, \citenamefont {S\"urer}, \citenamefont {Plumlee}, \citenamefont {Wild},\ and\ \citenamefont {Heinz}}]{Liyanage:2023nds}%
  \BibitemOpen
  \bibfield  {author} {\bibinfo {author} {\bibfnamefont {D.}~\bibnamefont {Liyanage}}, \bibinfo {author} {\bibfnamefont {O.}~\bibnamefont {S\"urer}}, \bibinfo {author} {\bibfnamefont {M.}~\bibnamefont {Plumlee}}, \bibinfo {author} {\bibfnamefont {S.~M.}\ \bibnamefont {Wild}}, \ and\ \bibinfo {author} {\bibfnamefont {U.}~\bibnamefont {Heinz}},\ }\href {\doibase 10.1103/PhysRevC.108.054905} {\bibfield  {journal} {\bibinfo  {journal} {Phys. Rev. C}\ }\textbf {\bibinfo {volume} {108}},\ \bibinfo {pages} {054905} (\bibinfo {year} {2023})},\ \Eprint {http://arxiv.org/abs/2302.14184} {arXiv:2302.14184 [nucl-th]} \BibitemShut {NoStop}%
\bibitem [{\citenamefont {Alvioli}\ \emph {et~al.}(2009)\citenamefont {Alvioli}, \citenamefont {Drescher},\ and\ \citenamefont {Strikman}}]{Alvioli:2009ab}%
  \BibitemOpen
  \bibfield  {author} {\bibinfo {author} {\bibfnamefont {M.}~\bibnamefont {Alvioli}}, \bibinfo {author} {\bibfnamefont {H.~J.}\ \bibnamefont {Drescher}}, \ and\ \bibinfo {author} {\bibfnamefont {M.}~\bibnamefont {Strikman}},\ }\href {\doibase 10.1016/j.physletb.2009.08.067} {\bibfield  {journal} {\bibinfo  {journal} {Phys. Lett. B}\ }\textbf {\bibinfo {volume} {680}},\ \bibinfo {pages} {225} (\bibinfo {year} {2009})},\ \Eprint {http://arxiv.org/abs/0905.2670} {arXiv:0905.2670 [nucl-th]} \BibitemShut {NoStop}%
\bibitem [{\citenamefont {Alvioli}\ \emph {et~al.}(2012)\citenamefont {Alvioli}, \citenamefont {Holopainen}, \citenamefont {Eskola},\ and\ \citenamefont {Strikman}}]{Alvioli:2011sk}%
  \BibitemOpen
  \bibfield  {author} {\bibinfo {author} {\bibfnamefont {M.}~\bibnamefont {Alvioli}}, \bibinfo {author} {\bibfnamefont {H.}~\bibnamefont {Holopainen}}, \bibinfo {author} {\bibfnamefont {K.~J.}\ \bibnamefont {Eskola}}, \ and\ \bibinfo {author} {\bibfnamefont {M.}~\bibnamefont {Strikman}},\ }\href {\doibase 10.1103/PhysRevC.85.034902} {\bibfield  {journal} {\bibinfo  {journal} {Phys. Rev. C}\ }\textbf {\bibinfo {volume} {85}},\ \bibinfo {pages} {034902} (\bibinfo {year} {2012})},\ \Eprint {http://arxiv.org/abs/1112.5306} {arXiv:1112.5306 [hep-ph]} \BibitemShut {NoStop}%
\bibitem [{\citenamefont {Lim}\ \emph {et~al.}(2019)\citenamefont {Lim}, \citenamefont {Carlson}, \citenamefont {Loizides}, \citenamefont {Lonardoni}, \citenamefont {Lynn}, \citenamefont {Nagle}, \citenamefont {Orjuela~Koop},\ and\ \citenamefont {Ouellette}}]{Lim:2018huo}%
  \BibitemOpen
  \bibfield  {author} {\bibinfo {author} {\bibfnamefont {S.~H.}\ \bibnamefont {Lim}}, \bibinfo {author} {\bibfnamefont {J.}~\bibnamefont {Carlson}}, \bibinfo {author} {\bibfnamefont {C.}~\bibnamefont {Loizides}}, \bibinfo {author} {\bibfnamefont {D.}~\bibnamefont {Lonardoni}}, \bibinfo {author} {\bibfnamefont {J.~E.}\ \bibnamefont {Lynn}}, \bibinfo {author} {\bibfnamefont {J.~L.}\ \bibnamefont {Nagle}}, \bibinfo {author} {\bibfnamefont {J.~D.}\ \bibnamefont {Orjuela~Koop}}, \ and\ \bibinfo {author} {\bibfnamefont {J.}~\bibnamefont {Ouellette}},\ }\href {\doibase 10.1103/PhysRevC.99.044904} {\bibfield  {journal} {\bibinfo  {journal} {Phys. Rev. C}\ }\textbf {\bibinfo {volume} {99}},\ \bibinfo {pages} {044904} (\bibinfo {year} {2019})},\ \Eprint {http://arxiv.org/abs/1812.08096} {arXiv:1812.08096 [nucl-th]} \BibitemShut {NoStop}%
\bibitem [{TGl()}]{TGlauberMC}%
  \BibitemOpen
  \href@noop {} {\enquote {\bibinfo {title} {{TGlauberMC on HepForge}.}}\ }\bibinfo {howpublished} {\url{https://tglaubermc.hepforge.org}},\ \bibinfo {note} {v3.2}\BibitemShut {NoStop}%
\bibitem [{\citenamefont {Rybczy\'nski}\ and\ \citenamefont {Broniowski}(2019)}]{Rybczynski:2019adt}%
  \BibitemOpen
  \bibfield  {author} {\bibinfo {author} {\bibfnamefont {M.}~\bibnamefont {Rybczy\'nski}}\ and\ \bibinfo {author} {\bibfnamefont {W.}~\bibnamefont {Broniowski}},\ }\href {\doibase 10.1103/PhysRevC.100.064912} {\bibfield  {journal} {\bibinfo  {journal} {Phys. Rev. C}\ }\textbf {\bibinfo {volume} {100}},\ \bibinfo {pages} {064912} (\bibinfo {year} {2019})},\ \Eprint {http://arxiv.org/abs/1910.09489} {arXiv:1910.09489 [hep-ph]} \BibitemShut {NoStop}%
\bibitem [{\citenamefont {Giacalone}\ \emph {et~al.}(2019)\citenamefont {Giacalone}, \citenamefont {Mazeliauskas},\ and\ \citenamefont {Schlichting}}]{Giacalone:2019ldn}%
  \BibitemOpen
  \bibfield  {author} {\bibinfo {author} {\bibfnamefont {G.}~\bibnamefont {Giacalone}}, \bibinfo {author} {\bibfnamefont {A.}~\bibnamefont {Mazeliauskas}}, \ and\ \bibinfo {author} {\bibfnamefont {S.}~\bibnamefont {Schlichting}},\ }\href {\doibase 10.1103/PhysRevLett.123.262301} {\bibfield  {journal} {\bibinfo  {journal} {Phys. Rev. Lett.}\ }\textbf {\bibinfo {volume} {123}},\ \bibinfo {pages} {262301} (\bibinfo {year} {2019})},\ \Eprint {http://arxiv.org/abs/1908.02866} {arXiv:1908.02866 [hep-ph]} \BibitemShut {NoStop}%
\bibitem [{\citenamefont {Voloshin}\ and\ \citenamefont {Zhang}(1996)}]{Voloshin:1994mz}%
  \BibitemOpen
  \bibfield  {author} {\bibinfo {author} {\bibfnamefont {S.}~\bibnamefont {Voloshin}}\ and\ \bibinfo {author} {\bibfnamefont {Y.}~\bibnamefont {Zhang}},\ }\href {\doibase 10.1007/s002880050141} {\bibfield  {journal} {\bibinfo  {journal} {Z. Phys. C}\ }\textbf {\bibinfo {volume} {70}},\ \bibinfo {pages} {665} (\bibinfo {year} {1996})},\ \Eprint {http://arxiv.org/abs/hep-ph/9407282} {arXiv:hep-ph/9407282} \BibitemShut {NoStop}%
\bibitem [{\citenamefont {Borghini}\ \emph {et~al.}(2001{\natexlab{a}})\citenamefont {Borghini}, \citenamefont {Dinh},\ and\ \citenamefont {Ollitrault}}]{Borghini:2000sa}%
  \BibitemOpen
  \bibfield  {author} {\bibinfo {author} {\bibfnamefont {N.}~\bibnamefont {Borghini}}, \bibinfo {author} {\bibfnamefont {P.~M.}\ \bibnamefont {Dinh}}, \ and\ \bibinfo {author} {\bibfnamefont {J.-Y.}\ \bibnamefont {Ollitrault}},\ }\href {\doibase 10.1103/PhysRevC.63.054906} {\bibfield  {journal} {\bibinfo  {journal} {Phys. Rev. C}\ }\textbf {\bibinfo {volume} {63}},\ \bibinfo {pages} {054906} (\bibinfo {year} {2001}{\natexlab{a}})},\ \Eprint {http://arxiv.org/abs/nucl-th/0007063} {arXiv:nucl-th/0007063} \BibitemShut {NoStop}%
\bibitem [{\citenamefont {Borghini}\ \emph {et~al.}(2001{\natexlab{b}})\citenamefont {Borghini}, \citenamefont {Dinh},\ and\ \citenamefont {Ollitrault}}]{Borghini:2001vi}%
  \BibitemOpen
  \bibfield  {author} {\bibinfo {author} {\bibfnamefont {N.}~\bibnamefont {Borghini}}, \bibinfo {author} {\bibfnamefont {P.~M.}\ \bibnamefont {Dinh}}, \ and\ \bibinfo {author} {\bibfnamefont {J.-Y.}\ \bibnamefont {Ollitrault}},\ }\href {\doibase 10.1103/PhysRevC.64.054901} {\bibfield  {journal} {\bibinfo  {journal} {Phys. Rev. C}\ }\textbf {\bibinfo {volume} {64}},\ \bibinfo {pages} {054901} (\bibinfo {year} {2001}{\natexlab{b}})},\ \Eprint {http://arxiv.org/abs/nucl-th/0105040} {arXiv:nucl-th/0105040} \BibitemShut {NoStop}%
\bibitem [{\citenamefont {Bhalerao}\ \emph {et~al.}(2011)\citenamefont {Bhalerao}, \citenamefont {Luzum},\ and\ \citenamefont {Ollitrault}}]{Bhalerao:2011yg}%
  \BibitemOpen
  \bibfield  {author} {\bibinfo {author} {\bibfnamefont {R.~S.}\ \bibnamefont {Bhalerao}}, \bibinfo {author} {\bibfnamefont {M.}~\bibnamefont {Luzum}}, \ and\ \bibinfo {author} {\bibfnamefont {J.-Y.}\ \bibnamefont {Ollitrault}},\ }\href {\doibase 10.1103/PhysRevC.84.034910} {\bibfield  {journal} {\bibinfo  {journal} {Phys. Rev. C}\ }\textbf {\bibinfo {volume} {84}},\ \bibinfo {pages} {034910} (\bibinfo {year} {2011})},\ \Eprint {http://arxiv.org/abs/1104.4740} {arXiv:1104.4740 [nucl-th]} \BibitemShut {NoStop}%
\bibitem [{\citenamefont {Giacalone}\ \emph {et~al.}(2017)\citenamefont {Giacalone}, \citenamefont {Yan}, \citenamefont {Noronha-Hostler},\ and\ \citenamefont {Ollitrault}}]{Giacalone:2016eyu}%
  \BibitemOpen
  \bibfield  {author} {\bibinfo {author} {\bibfnamefont {G.}~\bibnamefont {Giacalone}}, \bibinfo {author} {\bibfnamefont {L.}~\bibnamefont {Yan}}, \bibinfo {author} {\bibfnamefont {J.}~\bibnamefont {Noronha-Hostler}}, \ and\ \bibinfo {author} {\bibfnamefont {J.-Y.}\ \bibnamefont {Ollitrault}},\ }\href {\doibase 10.1103/PhysRevC.95.014913} {\bibfield  {journal} {\bibinfo  {journal} {Phys. Rev. C}\ }\textbf {\bibinfo {volume} {95}},\ \bibinfo {pages} {014913} (\bibinfo {year} {2017})},\ \Eprint {http://arxiv.org/abs/1608.01823} {arXiv:1608.01823 [nucl-th]} \BibitemShut {NoStop}%
\bibitem [{\citenamefont {Aad}\ \emph {et~al.}(2014)\citenamefont {Aad} \emph {et~al.}}]{ATLAS:2014qxy}%
  \BibitemOpen
  \bibfield  {author} {\bibinfo {author} {\bibfnamefont {G.}~\bibnamefont {Aad}} \emph {et~al.} (\bibinfo {collaboration} {ATLAS}),\ }\href {\doibase 10.1140/epjc/s10052-014-3157-z} {\bibfield  {journal} {\bibinfo  {journal} {Eur. Phys. J. C}\ }\textbf {\bibinfo {volume} {74}},\ \bibinfo {pages} {3157} (\bibinfo {year} {2014})},\ \Eprint {http://arxiv.org/abs/1408.4342} {arXiv:1408.4342 [hep-ex]} \BibitemShut {NoStop}%
\bibitem [{\citenamefont {Noronha-Hostler}\ \emph {et~al.}(2016)\citenamefont {Noronha-Hostler}, \citenamefont {Yan}, \citenamefont {Gardim},\ and\ \citenamefont {Ollitrault}}]{Noronha-Hostler:2015dbi}%
  \BibitemOpen
  \bibfield  {author} {\bibinfo {author} {\bibfnamefont {J.}~\bibnamefont {Noronha-Hostler}}, \bibinfo {author} {\bibfnamefont {L.}~\bibnamefont {Yan}}, \bibinfo {author} {\bibfnamefont {F.~G.}\ \bibnamefont {Gardim}}, \ and\ \bibinfo {author} {\bibfnamefont {J.-Y.}\ \bibnamefont {Ollitrault}},\ }\href {\doibase 10.1103/PhysRevC.93.014909} {\bibfield  {journal} {\bibinfo  {journal} {Phys. Rev. C}\ }\textbf {\bibinfo {volume} {93}},\ \bibinfo {pages} {014909} (\bibinfo {year} {2016})},\ \Eprint {http://arxiv.org/abs/1511.03896} {arXiv:1511.03896 [nucl-th]} \BibitemShut {NoStop}%
\bibitem [{\citenamefont {Sambataro}\ \emph {et~al.}(2022)\citenamefont {Sambataro}, \citenamefont {Sun}, \citenamefont {Minissale}, \citenamefont {Plumari},\ and\ \citenamefont {Greco}}]{Sambataro:2022sns}%
  \BibitemOpen
  \bibfield  {author} {\bibinfo {author} {\bibfnamefont {M.~L.}\ \bibnamefont {Sambataro}}, \bibinfo {author} {\bibfnamefont {Y.}~\bibnamefont {Sun}}, \bibinfo {author} {\bibfnamefont {V.}~\bibnamefont {Minissale}}, \bibinfo {author} {\bibfnamefont {S.}~\bibnamefont {Plumari}}, \ and\ \bibinfo {author} {\bibfnamefont {V.}~\bibnamefont {Greco}},\ }\href {\doibase 10.1140/epjc/s10052-022-10802-2} {\bibfield  {journal} {\bibinfo  {journal} {Eur. Phys. J. C}\ }\textbf {\bibinfo {volume} {82}},\ \bibinfo {pages} {833} (\bibinfo {year} {2022})},\ \Eprint {http://arxiv.org/abs/2206.03160} {arXiv:2206.03160 [hep-ph]} \BibitemShut {NoStop}%
\bibitem [{\citenamefont {Steinheimer}\ \emph {et~al.}(2011)\citenamefont {Steinheimer}, \citenamefont {Schramm},\ and\ \citenamefont {Stocker}}]{Steinheimer:2010ib}%
  \BibitemOpen
  \bibfield  {author} {\bibinfo {author} {\bibfnamefont {J.}~\bibnamefont {Steinheimer}}, \bibinfo {author} {\bibfnamefont {S.}~\bibnamefont {Schramm}}, \ and\ \bibinfo {author} {\bibfnamefont {H.}~\bibnamefont {Stocker}},\ }\href {\doibase 10.1088/0954-3899/38/3/035001} {\bibfield  {journal} {\bibinfo  {journal} {J. Phys. G}\ }\textbf {\bibinfo {volume} {38}},\ \bibinfo {pages} {035001} (\bibinfo {year} {2011})},\ \Eprint {http://arxiv.org/abs/1009.5239} {arXiv:1009.5239 [hep-ph]} \BibitemShut {NoStop}%
\bibitem [{\citenamefont {Krupczak}\ \emph {et~al.}(2024)\citenamefont {Krupczak} \emph {et~al.}}]{ExTrEMe:2023nhy}%
  \BibitemOpen
  \bibfield  {author} {\bibinfo {author} {\bibfnamefont {R.}~\bibnamefont {Krupczak}} \emph {et~al.} (\bibinfo {collaboration} {ExTrEMe}),\ }\href {\doibase 10.1103/PhysRevC.109.034908} {\bibfield  {journal} {\bibinfo  {journal} {Phys. Rev. C}\ }\textbf {\bibinfo {volume} {109}},\ \bibinfo {pages} {034908} (\bibinfo {year} {2024})},\ \Eprint {http://arxiv.org/abs/2311.02210} {arXiv:2311.02210 [nucl-th]} \BibitemShut {NoStop}%
\end{thebibliography}%

\end{document}